\documentstyle[prb,eqsecnum,aps,epsf,epsfig,floats]{revtex}
\begin{document}

\title{THEORY OF THE SPIN BATH}

\author{N. V. Prokof'ev$^{1,2,3}$  and P. C. E. Stamp$^{3,4}$}
\address{
$^{1}$ Russian Science Centre "Kurchatov Institute", Moscow 123182, Russia\\
$^{2}$ Physics and Astronomy, Hasbrouck Laboratory, University of
Massachussetts, Amherst, MA 01003, USA  \\
$^{3}$ Physics Department, and Canadian Institute for Advanced Research,\\ 
University of British Columbia, 6224 Agricultural
Rd.,
Vancouver B.C., Canada V6T 1Z1 \\
$^{4}$ Spinoza Institute, and Institute for Theoretical Physics,
Minnaert Building,\\
University of Utrecht, Leuvenlaan 4, 3508 TD Utrecht,
the Netherlands. }
\maketitle

\vspace{0.7cm}
\begin{abstract}

The quantum dynamics of mesoscopic or macroscopic systems is always
complicated by their coupling to many "environmental" modes. 
At low $T$ these environmental effects are dominated by 
{\it localised} modes, such as nuclear and paramagnetic spins, and
defects (which also dominate the entropy and specific heat).
This environment, at low energies, maps onto a "spin bath" model. 
This contrasts with "oscillator bath" models (originated by Feynman
and Vernon) which describe {\it delocalised} environmental modes such
as electrons, phonons, photons, magnons, etc. 
The couplings to $N$ spin bath modes are 
{\it independent} of $N$ (rather than the  $\sim O(1/\sqrt{N})$
dependence typical of oscillator baths), and often strong.
One cannot in generalmap a spin bath to an oscillator bath (or vice-versa); they
constitute distinct "universality classes" of quantum environment.

We show how the mapping to spin bath models is made, and then  
discuss several examples in
detail, including moving particles, magnetic solitons,
nanomagnets, and SQUIDs, coupled to nuclear and paramagnetic spin
environments. 

We then focus on the "Central Spin" model,
which couples a central 2-level system to a background spin
bath. It is the spin bath analogue of the famous
"spin-boson" oscillator model, and describes, eg., 
the tunneling dynamics of nanoscopic and
mesoscopic magnets and superconductors. 
We show how to average over (or "integrate out")  
spin bath modes, using an operator instanton technique, to find 
the Central spin dynamics. The formal
manouevres involve 4 separate averages- 
each average corresponds
physically to a different "decoherence" mechanism acting on the
central spin dynamics. Each environmental spin has its own topological
"spin phase", which by interacting with the phase of the central
system, decoheres it- this can happen even
without dissipation. We give analytic results for the
central spin correlation functions, under various conditions.

We then describe the 
application of this theory to magnetic and superconducting systems.
Particular attention is given to recent
work on tunneling magnetic macromolecules, where the role of the 
nuclear spin bath in controlling the tunneling is very clear; we also 
discuss other magnetic systems in the quantum regime,
and the influence of nuclear and paramagnetic spins on flux dynamics in 
SQUIDs. 

Finally, we discuss decoherence mechanisms and  
coherence experiments in superconductors and magnets.
We show that a spin bath environment causes decoherence even  
in the $T \rightarrow 0$ limit. Control of this decoherence will be
essential in the effort to construct "qubits" for quantum computers.

\end{abstract}

\vspace{0.5cm}
\pacs{PACS numbers: }

\section{Introduction}
\label{sec:1}

Many problems in quantum physics can be discussed using a model in which 
one or more mesoscopic or even macroscopic coordinates $Q, Q'$, etc.,
interact with a background environment (one 
coordinate might also 
represent an experimental probe, or even an observer). 
In such models (which have a long history 
\cite{oschist,LZhist,SBhist}) all
variables, including the environmental ones, 
are treated {\it quantum-mechanically}. The aim is 
to find the behaviour 
of $Q, Q'$, etc., after {\it averaging} over the environmental variables in 
some way.

It is certainly not obvious that one can discuss the real world in this 
way, given the complexity of $N$-body systems. However we now know that
many (but not all) 
mesoscopic or macroscopic systems can
be described {\it at low energies} 
by a few "canonical models", where a simple "central system"
(eg., a 2-level system, or an oscillator) couples to an environment. 
Remarkably, there appear to be only two types of environment, describable 
as baths of either oscillators or spins.
One way of trying to justify such 
models is the "renormalisation group" 
viewpoint \cite{RG}, which
maintains that 
most physical systems fall into a few "universality classes", each scaling to its
own "fixed point" in the space of possible Hamiltonians. 
All systems in a given university class use the same canonical 
Hamiltonian- the differences between different systems lying in 
the different values of the relevant couplings in this Hamiltonian. 

Although this "hard RG" philosophy clearly fails in some 
cases, it is a useful starting point for the present article, in which 
the quantum environment is modelled by a "spin
bath" (usually of 2-level systems, or "spin-$1/2$" systems). The finite
Hilbert space of each bath spin makes the spin bath 
appropriate for describing the low
energy dynamics of a set of {\it localised} environmental modes. 
We concentrate
on one particular "central spin" model \cite{PS93,sta94,PRB,PSchi95,PS96} 
in which the central system 
itself reduces to 
a 2-level system; but we also discuss cases where  
the central system is a one-dimensional coordinate (a "particle") 
moving through a field of spins. 

Another well-known set of canonical models describes the environment 
as a set of uncoupled oscillators- these include  the "spin-
boson" model \cite{ajl87,weiss} and the 
"Caldeira-Leggett" model \cite{weiss,cal83}. The spin-boson model couples
a central 2-level system to the oscillators, and is thus the  
analogue of the central spin model; and the Caldeira-Leggett model couples a
tunneling particle to the oscillators. These oscillator models all 
derive from a 
scheme proposed by Feynman and Vernon \cite{feyv63}, to describe a central
system coupled {\it weakly} to $N$ environmental modes; as they showed, 
the mapping to an oscillator bath can only be made rigourously {\it if the 
coupling is weak}. Oscillator 
models are thus best
adapted to $N$ {\it delocalised} environmental modes (where the
coupling is automatically $\sim 1/N^{1/2}$, and thus small for large $N$).

However, readers familiar with low-temperature physics will know that at 
low energies, the entropy and heat 
capacity of almost all real physical systems 
are dominated instead by {\it local} modes such as defects,
impurity spins, and nuclear spins \cite{loun}. 
Typically these relax very slowly 
at low $T$ because little phase space is available in their coupling to  
any delocalised modes (or to each 
other). However they often couple 
strongly to any mesoscopic or macroscopic collective coordinate, which then
easily perturbs them.
This coupling is of course 
{\it independent} of $N$. 

Unfortunately, even though spin bath models have been
studied sporadically for many years \cite{SBhist}, the results have  
often been misleading, either because they 
treat some weak-coupling limit 
(sometimes made \cite{mermin,hanggi98} 
by arbitrarily multiplying the coupling to each of the $N$ bath spins by
a factor $1/N$, for no good physical reason), or because they 
drop some of the important couplings to the bath spins, in order to 
solve the model. In the weak-coupling limit, spin bath models 
can be mapped to oscillator baths \cite{sta88,neto93,seattlePS} (in
accordance with the original remarks of Feynman and Vernon \cite{feyv63}).
However one is often nowhere near the weak-coupling limit, and the
mapping to the oscillator bath then {\it fails} in general 
\cite{PS93,PRB,PS96,seattlePS}. This demands a new approach, which is the 
subject of this review. 

It may be useful to mention {\it why} many physicists are 
interested in models of this kind. 
Here are some of the reasons:

(i) Very rapid progress in work on intrinsically quantum processes 
(interference,
tunneling, etc.) occurring at the nanoscopic and mesoscopic scales 
\cite{dube97,imry,kagajl92,toms98,IJMPB92,QTM94},
plus speculations about the coming "nanotechnological revolution".
Perhaps the most exciting idea in this area is that of making "quantum  computers" using nanoscopic 
superconductors 
\cite{qubitS}, semiconductors \cite{qubitN}, or nanomagnets.
Needless to say, the technological 
repercussions of this work will be enormous, provided the crucial problem of 
decoherence can be overcome.

(ii) Physicists need to understand the mechanisms of 
decoherence and quantum dissipation \cite{weiss}
in nature, and
the crossover to (or "emergence" of) classical behaviour from
quantum
physics as either size, temperature $T$, external fields, or couplings to the
environment are increased. These issues are not only relevant
to quantum device design, but also to problems in 
quantum gravity, and to the infamous "quantum measurement"
problem \cite{qmmt}.
The existence of low-$T$ canonical models,  
going beyond the phenomenology 
of stochastic or master equations \cite{master} to work with 
closed Hamiltonians, is invaluable here. Recent examples include 
the
analysis of quantum spin glass relaxation \cite{cugl98},
quantum relaxation in nanomagnets 
\cite
{PS93,PRB,PS96,IJMPB92,QTM94,garg95,rose98,copp95,sang97,PS98,ohm98,luc98,ww1,ww2,ww3}
and tunneling superconductors 
\cite{cal83,kagajl92,voss,washburn,clarke,lukens,naka,amb85,SQUIDth}),
and the study of 
quantum chaotic systems,  
stochastic
resonance \cite{hanggi98a} and dissipative tunneling in AC fields 
\cite{hanggi98b}. Earlier such models have   
been used for decades to discuss relaxation 
in fields like quantum chemistry \cite{hanggi89} or nuclear 
physics.
 
(iii) Both oscillator and spin bath models map to many important models in 
quantum field theory. Thus the "spin-boson"
model mentioned above \cite{ajl87} 
maps, for specific parameter values, to the Kondo model, 
the 
Thirring and Sine-Gordon models, and various other 2-dimensional field
theories. Although similar mappings have yet to be exploited in great detail 
for the spin bath, they will obviously be very useful for, eg., 
lattice spin models. 

Most work in these areas has used oscillator bath 
representations of the environment, with the tacit assumption that 
delocalised environmental modes dominate the physics. However, 
experiments on quantum nanomagnetic systems
\cite{copp95,sang97,ohm98,luc98,ww1,ww2,ww3}, on glasses 
\cite{glass}, 
and on mesoscopic conductors \cite{TLS,mohanty}, 
as well as theoretical debate about the mechanisms of 
decoherence in nature, clearly require a more general point of view.
In fact we shall see that spin baths behave very differently from  
oscillator baths. For 
example, the oft-discussed connection between dissipation and decoherence
which exists for oscillator bath environments \cite{cal83,ajl84} 
is absent here- one can even have
decoherence with no dissipation at all \cite{PS93}, because of the quantum 
phase associated with the spin bath dynamics. 

Although the existence of a quantum phase associated with spin is 
obvious (it is a quantum variable), it was not until Haldane \cite{fdmh83}
and Berry \cite{mvb84} discussed its topological properties that physicists
realised its practical importance, 
in terms of the path traced out on the spin sphere. Just as in the
usual Aharonov-Bohm effect, the "flux" enclosed by a path (given 
here by $\omega S$, where $S$ is the spin and  
$\omega$ is the enclosed solid angle on the unit spin sphere) is equal to a
dynamical phase- but now these are both in {\it spin space}, not in real
space. These ideas (and related experiments) 
were the centre of enormous interest in the mid-late 
1980's, in almost all fields of physics (and were extensively reviewed 
then \cite{spinphase}).

In this article we will be interested in
the spin phase of the 
{\it environment}. 
We stress here that the environmental "spin" variables may not 
necessarily refer to real spins 
(they can describe defects, or other such "2 level systems"), but they 
will still have an associated dynamic topological phase, which can be 
described by an effective spin bath variable.
The environmental spin phase interacts with the phase of the
central system, causing phase decoherence in its dynamics 
\cite{PS93,sta94,sta88}.
From the point of
view of measurement theory, this environmental phase decoherence 
comes from a 
"phase measurement" made 
by the spin environment \cite{sta88}, 
in a kind of "inverse Stern-Gerlach" setup (where the spins, instead of being
measured, are themselves 
doing the measuring!). Such phase decoherence also arises from oscillator
baths, in a rather different (and much less effective) way \cite{stern}. 
In fact, the relevant phases 
involve both an adiabatic "Berry" term and
a second term coming from transitions between different bath states 
(section 3.A).  
There are also other decoherence mechanisms
associated with the bath spins, coming from both the temporal fluctuations
of the bias on the central system caused by the spin bath (section 3.D), and
from the precession of the spins in the spin bath (with their associated phase)
{\it in between} transitions
of the central system (section 3.C). Thus the 
question of how the bath spin dynamics influences the central system is not
simply a question of looking at Berry phases. 
 
Practical application
of the theory to, eg., SQUIDs, or nanomagnets, or "qubits" (section 5), 
must include 
all mechanisms properly (section 3.E). The tactic adopted in this article 
is to
focus on a "Central Spin" model \cite{PS93,sta94,PRB,PSchi95,PS96} 
(sections 3 and 4), in which 
the role of each term is exposed rather clearly; after this one  
sees how things generalise to other models. This model is directly relevant 
to qubits, and to the observation of mesoscopic or "macroscopic"
quantum coherence- indeed we maintain that any practical design of 
such devices must involve the elimination, by one means or another, of 
{\it all} decoherence mechanisms from the relevant spin bath \cite{seattlePS}.   

We begin the article (section 2) by showing how both oscillator and spin bath 
models arise as the low-energy truncated versions of higher energy 
Hamiltonians. We give several examples, both magnetic and superconducting, to
illustrate this.
Then, in section 3 we explain how, mathematically, one averages over 
the spin bath variables to find the 
behaviour of the central system. This is done pedagogically- 
we use the example of  
the Central Spin model (and compare it with the spin-boson model). Various 
simple limits are introduced, and solved, before the general technique is 
given at the end of section 3.
In Section 4 we give some results \cite{PRB} for the dynamics of
the Central Spin model, in various regimes, to show how the physics
is influenced by the spin bath; and we also show how the system reduces to 
the spin-boson system in the weak-coupling limit.
For those readers interested in the mathematical details,
these are sketched in 2 Appendices.

Finally, in section 5 we return to physical applications,
particularly to quantum magnetic systems and 
superconductors.
We then discuss decoherence, and show how 
this should persist
even in the $T \rightarrow 0$ limit. We finish by discussing the important
application of the Central Spin model to decoherence in qubits and in 
quantum computation.


\section{The Low-Energy Effective Hamiltonian}
\label{sec:H-EFF}

In studying the low-energy
dynamics of a central quantum system coupled to an environment, we begin  
by ``truncating out" the unwanted high-energy physics, 
to produce a low-energy effective Hamiltonian. This is of
course a quite general technique in physics, and 
one way to approach it is illustrated in Fig.1. Typically one has 
a reasonably accurately known ``high-energy" or
``bare" Hamiltonian (or Lagrangian) for a quantum system, valid
below some ``ultraviolet" upper energy cut-off energy $E_c$, and
having the form
\begin{equation}
\tilde{H}_{\mbox{\scriptsize Bare}} (E_c)\;\; = \;\;\tilde{H}_o (\tilde{P},
\tilde{Q} ) \;+\; \tilde{H}_{\mbox{\scriptsize int}} (\tilde{P},
\tilde{Q}; \tilde{p}, \tilde{q})\; +\;\tilde{H}_{\mbox{\scriptsize
env}} (\tilde{p}, \tilde{q}) \; \;\;\;\;\;\;\;\;(E <E_c)\;,
\label{2.1}
\end{equation}
where $\tilde{Q}$ is an $\tilde{M}$-dimensional coordinate
describing that part of the system we are interested in (with
$\tilde{P}$ the corresponding conjugate momentum), and $(\tilde{p},
\tilde{q})$ are $\tilde{N}$-dimensional coordinates describing all
other degrees of freedom which may couple to $(\tilde{P},
\tilde{Q} )$. Conventionally one refers to $(\tilde{p},
\tilde{q})$  as environmental coordinates.
$\tilde{H}_{\mbox{\scriptsize Bare}}$ 
is of course a
low-energy form of some other even higher-energy Hamiltonian, in a
chain extending ultimately back to quarks, leptons, and perhaps strings.

If, however, one is only interested in physics below a much lower
energy scale $\Omega_o$, then the question is - can we find a
new effective Hamiltonian, of form
\begin{equation}
H_{\mbox{\scriptsize eff}}(\Omega_o) \;\;=\;\; 
H_o (P,Q ) + H_{\mbox{\scriptsize int}} 
(P,Q; p, q) +H_{\mbox{\scriptsize env}} (p, q) \; \;\;\;\;\;\;(E <\Omega_o)\;,
\label{2.2}
\end{equation}
in the truncated Hilbert space of energies below $\Omega_o$? In
this $H_{\mbox{\scriptsize eff}}$, $P$ and $Q$ are generalised
$m$-dimensional coordinates of interest, and $p,q$ are
$N$-dimensional environmental coordinates coupled to them. Since we
have truncated the total Hilbert space, we have in general that
$M<\tilde{M}$ and $N<\tilde{N}$.

Why do we make this truncation (after all, its inevitable effect will be to 
generate new couplings between the low-energy modes)? 
Essentially because
in many cases the truncation pushes the new $H_{\mbox{\scriptsize eff}}$
towards some low-energy ``fixed point" Hamiltonian; and many
different physical systems may flow to the same fixed point.
This allows us to speak of ``universality classes" of quantum
environment, and of a small number of "canonical" effective Hamiltonians.
All physical systems in the same universality class  
will be described by the same form for $H_{\mbox{\scriptsize
eff}}$,
albeit with different values for the couplings.
As one varies the UV cut-off $\Omega_o$, the
couplings change and any
given system moves in the "coupling" or "effective Hamiltonian"
space; but they all move 
towards the same fixed point (or fixed line, in some cases) 
as $\Omega_0$ is reduced.
The various coupling terms in 
$H_{eff}(\Omega_o)$,
simply parametrise the path it takes as $\Omega_o \to 0$
(Fig.1).

Explicit derivations of $H_{\mbox{\scriptsize
eff}}$ for particular systems are lengthy; 
see  eg. \cite{feyv63,seattlePS,kagajl92,QTM94,ajl84,SQUIDth} 
for general discussions, ref. \cite{seattlePS} for comparision of spin and 
oscillator bath systems, 
and \cite{cal83,dube97,tup97,amb85,ajl88,rose99,tatara} for specific examples. 
In this article 
we will go directly to the canonical models, giving some 
examples of each so that readers can see how the high-energy Hamiltonians 
are related to the models for some real systems. To warm up we recall 
the basic structure of the
oscillator bath effective Hamiltonians, and then move on immediately to
discuss various canonical models involving spin baths.

\subsection{Oscillator Bath models}
\label{sec:2A}

For models in the general "universality class" 
of oscillator bath environments, 
$H_{\mbox{\scriptsize eff}}$ takes the form \cite{feyv63,ajl84}: 
\begin{eqnarray}
H_{\mbox{\scriptsize eff}}(\Omega_o)
\;\; = \;\;  H_o (P,Q ) \;+\; \sum_{q=1}^{N} \bigg[ F_q(P,Q)x_q +G_q(P,Q)p_q
 \bigg] 
\;\;  + \;\; {1 \over 2} \sum_{q=1}^{N} \left(  {p_q^2 \over m_q} +
 m_q \omega_q^2 x_q^2 \right)   \;,
\label{2.3}
\end{eqnarray}
where the generalised bath coordinates $(q_k,p_k)$ are now oscillator 
displacements $x_q$ and momenta $p_q$; these describe {\it delocalised} 
modes. The couplings $F_q(P,Q)$ and $G_q(P,Q)$ are
$\sim O(N^{-1/2})$, so that in the "thermodynamic limit" $N \gg 1$, 
appropriate to 
a macroscopic environment of delocalised oscillators, these couplings are 
small \cite{cal83,feyv63,ajl84} (a number of studies have also shown how 
higher-order couplings can be absorbed into linear but $T$-dependent couplings 
\cite{nonlin}). A special
case of (\ref{2.3}) is the Feynman-Vernon
bilinear coupling form \cite{feyv63}:
\begin{eqnarray}
H_{\mbox{\scriptsize eff}}(\Omega_o)
 =  H_o (P,Q ) + \sum_{q=1}^{N}c_q x_q Q 
 + {1 \over 2} \sum_{q=1}^{N} \left(  {p_q^2 \over m_q} +
m_q \omega_q^2 x_q^2 \right)   \;,
\label{2.3a}
\end{eqnarray}
where the couplings $c_q \sim O(N^{-1/2})$ as well.
In recent years great
attention has been given to problems where $H_o(P,Q)$
describes a tunneling system (the
"Caldeira-Leggett" model \cite{weiss,cal83}); there have also been
extensive studies of an oscillator coupled to oscillators 
\cite{grabert88}, of free particles coupled to oscillators 
\cite{grabert88,hakim} and of band particles coupled to oscillators
\cite{weiss,schmid83}.

Suppose now the potential $V(Q)$ has a 2-well form, with small oscillation 
frequencies $\sim \Omega_o$, and a "bias" energy difference between the  
two minima $< \Omega_o$.  Then for energies $< \Omega_o$,
one further truncates
to the celebrated ``spin-boson" model \cite{ajl87}: 
\begin{eqnarray}
H_{\mbox{\scriptsize SB}}(\Omega_o)
\; = \;   \Delta(\Phi_o) \hat{\tau}_x +\xi_H \hat{\tau}_z +
 \sum_{q=1}^{N} \bigg[ c_q^{\parallel} \hat{\tau}_z + 
(c_q^{\perp} \hat{\tau}_- + h.c. ) \bigg] x_q 
\;\;  +\;\; {1 \over 2} \sum_{q=1}^{N} \left(  {p_q^2 \over m_q} +
 m_q \omega_q^2 x_q^2 \right)   \;,
\label{2.4}
\end{eqnarray}
where the two-level central system (with tunneling amplitude $\Delta(\Phi_o)$
and longitudinal bias $\xi_H$)
is described by the
Pauli matrix vector $\hat{\vec{\tau}}$, coupled to background
oscillators having energies $\omega_q < \Omega_o$. We have introduced a 
topological phase
$\Phi_o$ for the central system, which depends in general on an external
field; the simplest and best-known example is a form 
$\Delta(\Phi_o)= 2\Delta_o \cos \Phi_o$, arising from the interference between 
2 paths of amplitude $ \Delta_o e^{\pm i \Phi_o}$ in the motion of 
the central system. This kind of "Aharonov-Bohm" interference is well-known in 
superconductors (where the phase is just the 
superconducting order parameter phase), and in magnets (where it is the 
topological spin phase \cite{fdmh83,mvb84,spinphase}). 
One can have a more complicated
dependence of $\Delta(\Phi_o)$ on $\Phi_o$ (eg., using multiple SQUIDs, or a
nanomagnet with $m$-fold rotation symmetry), 
but we will stick with the simple $2\cos \Phi_o$ dependence 
in this article.

For consistency we must assume $\xi_H < \Omega_o$, 
otherwise higher levels will be involved.
Typically 
$c_q^{\perp }$ is dropped, because its effects are down on those of
$c_q^{\parallel}$ by a factor $\sim (\Delta_o/\Omega_o)^2$ in tunneling
rates; but sometimes  $c_q^{\parallel} =0$
(for reasons of
symmetry), and then $c_q^{\perp}$ must be retained. 
The tunneling amplitude $\Delta_o \sim \Omega_o e^{-A_o}$, where
$A_o$ is the tunneling
action. 

The fame of the spin-boson model partly arises because many
well-known problems in condensed matter physics can be mapped to
it- this is a good example of the ``universality" mentioned above. 
Because the effect of each oscillator on the central system (and
vice-versa) is 
very small, it may be entirely incorporated in second-order perturbation
theory (ie., to order $\sim (F_q^2/\omega_q), (G_q^2/\omega_q)$) for the 
general form (\ref{2.3}), or to order $\sim (c_q^2/\omega_q)$ for the 
bilinear forms (\ref{2.3a}),(\ref{2.4}). In the latter case 
this immediately 
encapsulates all environmental effects in the spectral function
\cite{cal83,feyv63}
\begin{equation}
J_{\alpha}(\omega ) = {\pi \over 2} \sum_{q=1}^N 
{\vert c_q^{\alpha} \vert^2 \over
m_q\omega_q }~ \delta (\omega -\omega_q) \;,
\label{3.6}
\end{equation}
where $\alpha = \perp, \parallel$. In general $J_{\alpha}(\omega)$ also
depends on $T$, even in the low-$\omega$ limit \cite{PS96,IJMPB92,ajl84}. The
case where $J_{\alpha}(\omega) \sim \omega$ is referred to as "Ohmic"
\cite{cal83}. Because the $c_q^{\alpha} \sim N^{-1/2}$, the
$J_{\alpha}(\omega)$ are independent of $N$ and have the usual "response
function" form.

\subsection{Two examples of Spin-Boson systems}
\label{sec:2B}

We give just 2 examples here of how a spin-boson model can arise, 
in the description of mesoscopic systems at low energies.
Both truncations ignore the presence of spin bath modes (for which see
sections 2.E and 2.F, where we return to these 2 examples).

\vspace{4mm}

{\bf (i) Nanomagnet coupled to phonons or electrons}:
The electronic spin dynamics of
nanomagnets are often described by a "giant spin"
Hamiltonian $H_o(\vec{S})$, describing a
quantum rotator with spin quantum number  $S=\vert \vec{S} \vert
\gg 1$. 
This model \cite{vHem86,Enz86} assumes the
individual electronic moments are locked together by strong 
exchange interactions $J_{ij}$ into a monodomain
giant spin, with
$\vec{S} = \sum_j \vec{s}_j$ (summed over local moment sites).
This only only works \cite{seattlePS} below
a UV cut-off energy $E_c$ considerably less than $J_{ij}$. However we are 
interested in the quantum dynamics for energies $< \Omega_o$, where  
$\Omega_o$ is controlled mostly by the single-ion magnetic anisotropy;
in real nanomagnets $\Omega_o \sim 0.1-10 ~K$. 

Any real nanomagnet has couplngs to 
a spin bath of nuclear and paramagnetic spins,
and to oscillator baths of phonons and electrons, which we now describe. 

The "high-energy"coupling between phonons and $\vec{S}$ is 
described by terms like \cite{AbraBl,trem}:
\begin{equation}
{\cal H}^{\phi}_2 \sim \Omega_o U(\hat{\vec{S}}) \left(
{m_e \over M_a } \right)^{1/4} \sum_{\vec{q}} \left(
{\omega_q \over \Theta_D }  \right)^{1/2}
[b_{\vec{q}}+b_{\vec{q}}^{\dag}] \;,
\label{1.ph}
\end{equation}
where m$_e$ is the electron mass, $M_a$ the mass of the molecule, and  
$\Theta_D \sim c_s a^{-1}$ is
the Debye temperature (with $a$ the relevant lattice spacing, 
and $c_s$ the sound velocity).  The interaction $U(\vec{S}) \sim S$ and 
dimensionless; a typical example is the non-diagonal term 
$(\hat{S}_x\hat{S}_z /S)$, which causes 
phonon
emission or absorption with a change $\pm 1$ in $S_z$
(since $S_x = \frac{1}{2}(S_{+} + S_{-}))$.
One also has diagonal terms in which $S_z S_x$ is replaced
by, eg. $S^{2}_{z}$;
and there are also higher couplings to, eg, pairs of phonons. 

Truncation to the "quantum regime" \cite{PS96} 
then gives the spin-boson model
(\ref{2.4}), with a dominant 
non-diagonal coupling $c_q^{\perp} \sim S \Omega_o 
\vert \vec{q} \vert ^{1/2}$, coming from terms like (\ref{1.ph}). 
In the absence of external fields in 
Hamiltonian $H_o^{eff}(\vec{S})$, the diagonal coupling $c_q^{\parallel}$
is actually {\it zero} 
(because of 
time-reversal symmetry). The 
Caldeira-Leggett spectral function for the system has the form
$J_{\perp}(\omega) \sim S^2 (\Omega_o^2/\rho c_s^5) \omega^3$
where $B_{\perp} \sim (S^2 \Omega_o^2/\Theta_D)$; here $\rho$ is the 
density of the medium supporting Debye phonons, and 
$\Theta_D^4 \sim \rho c_s^5$.

If now we couple {\it electrons} to the giant spin, it is the {\it diagonal}
coupling which dominates \cite{PS96}. The electronic 
coupling to $\vec{S}$ depends 
on the type of magnetism.
Some details have been  
worked out for 
Kondo interactions with conduction electrons- the 
coupling to $\vec{S}$ is 
\begin{equation}
H_{int}^{GK} \ = \ \frac{1}{2} \bar{J} \hat{\vec{S}}.
\hat{\vec{\sigma}}^{\alpha \beta} \
\sum_{\vec{k}\vec{q}}
F_q \ c^{+}_{\vec{k}+\vec{q}_i \alpha} \ c_{\vec{k}_i \beta}
\label{1.35}
\end{equation}
where $\bar{J}$ is the mean value of the 
Kondo couplings to each individual electronic spin in the 
nanomagnet, and $\vec{S} F_q = \int(d^{3}r/V_o)
\vec{s}(\vec{r})
e^{i\vec{q}.\vec{r}}$ is a "form factor" integrating the localised
electron spin 
density $\vec{s}(\vec{r})$ 
over the nanomagnetic volume $V_o$. At low energies
the corresponding spin-boson model has an "Ohmic" 
diagonal spectral function
$J_{\parallel}(\omega) \ = \ \pi \alpha_{\kappa} \omega$.
The size of $\alpha_{\kappa}$
depends on how the electrons permeate the nanomagnet
\cite{PS96}; if they permeate freely,
$\alpha_{\kappa} 
\sim \ g^2S^{4/3}$,
where $g = \bar{J}N(0)$ is 
the mean dimensionless Kondo coupling, 
and
$N(0)$ the
Fermi surface density of states. Typically $g \sim  0.1$, 
so $\alpha_{\kappa}$ can be big.

\vspace{4mm}

{\bf (ii) RF SQUID (Flux coupled to electrons)}:
We briefly recall one well-known application of 
the spin-boson model, to an RF SQUID coupled to both 
normal electrons (in shunts, etc.), and Bogoliubov quasiparticles
\cite{kagajl92,amb85,ajl88,SQUIDth}. 
The flux $\phi$ passing through a superconducting ring with a weak
link moves in a multiwell potential, which can be adjusted so that the lowest 
2 wells (each with small oscillation or "Josephson plasma" frequency 
$\Omega_o \sim 2\pi[E_J/\pi C]^{1/2}/\phi_o$, where $E_J$ is the 
Josephson weak link coupling energy) are almost degenerate, and
dominate the low-energy properties. The high-energy coupling between the flux
and the electronic quasiparticles has the form \cite{amb85,ajl88,SQUIDth} 
\begin{equation}
H_{int} = \{ \cos (\pi \phi/\phi_o) \sum_q t_q U_q^S (a_q + a_q^{\dagger})
\; + \; i \sin(\pi \phi/\phi_o) \sum_q t_q U_q^A  (a_q - a_q^{\dagger}) \}
\label{SQUIDcplg}
\end{equation}
where $q \equiv (\vec{k}, \vec{k}')$ labels oscillator states describing
a quasiparticle pair $\vert \vec{k} \vec{k}' \rangle$ with energy 
$\omega_q = E_k + E_{k'}$, $t_q$ is the relevant junction tunneling matrix 
element, $U_q^{S/A}$ the symmetric/antisymmetric 
BCS coherence factor, and  
$\phi_o$ is the flux quantum. Thus we have a coupling to both the momenta and 
coordinates of the oscillators, which can also be written as a coupling to 
2 independent oscillator baths \cite{amb85}. 
The $T$-dependence of the coherence 
factors (coming from the BCS gap dependence) as well as the gap structure
in their energy dependence, gives a complex 
structure in $J(\omega,T)$. The reduction to the spin-boson model is now
trivial \cite{ajl87}, the minima in $\phi$-space of the 
effective potential corresponding 
to the 2 eigenstates of $\hat{\tau}_z$.

\subsection{Spin Bath Environments}
\label{sec:2C}

Now suppose we have a high-energy
Hamiltonian of form (\ref{2.1}), but where the environmental
coordinates $(\tilde{p}, \tilde{q})$ are a set of
$N$ spin-$1/2$ variables $\{ \hat{\vec{\sigma}}_k \}$, (i.e.,
two-level systems); and we assume the interspin couplings to be  
weak. Then, instead of (\ref{2.3}), we have  
\begin{equation}
H = H_o (P,Q ) + H_{\mbox{\scriptsize int}} 
(P,Q;\{ \hat{\vec{\sigma}} \} ) +H_{\mbox{\scriptsize env}} 
(\{ \hat{\vec{\sigma}} \}) \;;
\label{2.5}
\end{equation}
\begin{equation}
H_{\mbox{\scriptsize int}} (P,Q;\{ \hat{\vec{\sigma}} \} ) =
\sum_{k=1}^N \bigg[ F_k^{\parallel}(P,Q) \hat{\sigma}_k^z +
[ F_k^{\perp}(P,Q) \hat{\sigma}_k^- +h.c. ] \bigg] \;;
\label{2.6}
\end{equation}
\begin{equation}
H_{\mbox{\scriptsize env}} (\{ \hat{\vec{\sigma}} \}) =\sum_{k=1}^N
\vec{h}_k \hat{\vec{\sigma}}_k +  \sum_{k=1}^N\sum_{k'=1}^N
V_{kk'}^{\alpha \beta} \hat{\sigma}_k^\alpha
\hat{\sigma}_{k'}^\beta \;,
\label{2.7}
\end{equation}
for energy scales $E<E_c$. Thus we now have a central system coupled to  
a "spin bath", described by 
$H_{\mbox{\scriptsize env}} (\{ \hat{\vec{\sigma}} \})$ in (\ref{2.7}). 
The couplings $ F_k^{\parallel}(P,Q)$ and 
$F_k^{\perp}(P,Q)$, between the central system and the bath spins, are usually 
much greater than the interspin couplings 
$V_{kk'}^{\alpha \beta}$; this means that the dynamics of 
each spin is largely "slaved" to that of the central system. 

Unlike oscillator baths (whose modes typically represent delocalised
environmental degrees of freedom), the $\{ \hat{\vec{\sigma}}_k \}$
represent {\it localised} modes (whose weak spatial overlap explains why
the $V_{kk'}^{\alpha \beta}$ are small).
This fact underlies a crucial difference between 
oscillator and spin bath environments- the couplings 
$F_k^{\parallel}(P,Q)$ and 
$F_k^{\perp}(P,Q)$ are {\it independent
of the number} $N$ {\it of bath spins}. 
Thus the larger is $N$, the larger is 
the total effect of the spin bath on the central system- there is no 
strict thermodynamic limit in the system, and it is not meaningful to 
let $N \to \infty$. We emphasize also that we see no justification
in general for spin bath models in which 
$F_k^{\parallel},F_k^{\perp} \sim O(N^{-1/2})$, or even $\sim O(1/N)$ 
(although one can certainly invent artificial models 
of this kind). Thus, if we add more localised environmental 
modes to our environment, it is clear that the different modes are 
approximately independent (as they will be if quasi-localised), so that  
their individual couplings to the central system will be hardly
affected, ie., will depend only weakly on $N$. 

There is nothing to stop generalisation of this model to include bath spins
$\{ \vec{I}_k \}$, with $I_k = \vert \vec{I}_k \vert > 1/2$; the $(2I_k + 1)$
states then represent the degrees of freedom of, eg., a defect, 
or a spin (again, localised). This introduces
tensor (eg., quadrupolar) couplings to the bath spins 
\cite{rose99,rose99a}, and
thereby complicates the mathematics- but does not alter the basic physics.
We will not discuss this here (for the relevant formalism, and its application
to the $Fe$-8 molecular nanomagnet, see refs. \cite{rose99,rose99a}).

\subsection{Particle moving through a spin bath}
\label{sec:2D}

A particle moving through a spin bath is 
described by (\ref{2.5}), in which $P$ and $Q$ describe
the momentum and position of the moving particle. The
diagonal term $F_k^{\parallel}(P,Q)$ is analogous to the 
"position" oscillator coupling $F_q(P,Q)$ in (\ref{2.3}), and likewise
$F_k^{\perp}(P,Q)$ to corresponds to $G_q(P,Q)$. However both forms can 
be altered by canonical transformation, corresponding to a rotation 
between the different coordinates. The most common problems involve
a diagonal coordinate coupling $F_k^{\parallel}(Q)$ and a non-diagonal
momentum coupling $F_k^{\perp}(P)$. Then bath transitions (spin flips)
are induced by the motion of the particle, whereas a stationary particle
sees a "potential" 
$U(Q, \{ \sigma_k^z (t) \}) = \sum_k F_k^{\parallel}(Q) \sigma_k^z (t)$, 
in general time-dependent. 

A nice mesoscopic example of this is a large
magnetic soliton coupled to background spins \cite{dube97}. In
many realistic cases the most important such coupling will be to 
paramagnetic impurities, but here we consider the simpler case of
a hyperfine coupling to a set $\{ \hat{\vec{\sigma}} \}$ of
$N$ spin-$1/2$ nuclear spins. In this case $V_{kk'}^{\alpha
\beta}$ describes the extremely weak internuclear dipolar coupling;
typically $\vert V_{kk'}^{\alpha \beta} \vert \le 10^{-7} ~K$; and
$\vec{h}_k$ is any external field that might unfluence these
nuclei.

The "high-energy" Hamiltonian for such a wall is usually determined  
as an integral over the magnetisation density ${\bf M}({\bf r})$
and its spacetime gradients \cite{slonc}. 
>From this one 
eliminates the details of the wall profile altogether, to produce a ``bare''
Hamiltonian (ie., neglecting the environment) for the wall coordinate; in
simple cases where the wall demagnetisation field keeps the wall almost
flat, this gives
\cite{IJMPB92,sta91}
\begin{equation}
H_{w} = \frac{1}{2} M_{w} \dot{Q}^{2} - V(Q) - 2 S_{w} \mu_{B} M_{0}
H_{e} Q
\end{equation}
\label{Hwall}
for a wall with centre of mass coordinate $Q$ and surface area $S_w$. 
The "pinning potential" 
$V(Q) = V_{0} \; \mbox{sech}^{2} (Q/\lambda_{w})$, provided the pinning centre
is much smaller than the wall width $\lambda_w$.
The term linear in $Q$ comes from 
an external  magnetic field ${\bf H}_{e}$. 
 
What now of the environment? In the literature there is extensive
discussion
of the
effects of magnons (ie., spin waves)\cite{IJMPB92,sta91}, electrons 
\cite{tatara}, and
phonons \cite{dube97} on the
wall dynamics- these are all oscillator baths. 
However at low $T$ spin bath effects, 
coming from nuclear and paramagnetic spins, will completely dominate.
Even in Ni (where only 
$1 \%$ of the nuclei have spins,
with a tiny hyperfine coupling 
$\omega_{0} = 28.35 \, \mbox{MHz}
\sim 1.4 \, \mbox{mK}$), all real samples have an important concentration of
paramagnetic spins (caused by Oxygen in the sample) as well as many 
defects. In 
rare earths, the hyperfine coupling $\omega_k \sim 1-10 \, \mbox{GHz}$
($0.05- 0.5 \, \mbox{K}$), and hyperfine effects alone are quite massive.
Thus we must modify $H_w$ above to 
\begin{equation}
H = H_{w} + \sum_{k=1}^{N} \omega_{k}^{\alpha \beta} s_k^{\alpha} 
I_k^{\beta} +
 \sum_{k} \sum_{k'} V_{k k'}^{\alpha \beta}
I_{k}^{\alpha} I_{k'}^{\beta}
\label{DWH1}
\end{equation}
in which the electronic spins ${\bf s}_k$ couple locally to $N$ nuclear
spins
${\bf I}_k$
at positions ${\bf r}_{k}$ ($ k=1,2,3, ... \, N$), via a 
hyperfine coupling $\omega_k^{\alpha \beta}$ (and also in general to 
paramagnetic spins).
The internuclear coupling 
$|V_{k k'}^{\alpha \beta}| \sim 1-100 \, \mbox{kHz}$ ($0.05-5
\, \mu\mbox{K}$), ie., $\ll \omega_{k}$, but 
it gives the spin bath its own dynamics. 

To write the 
Hamiltonian in the form (\ref{2.6}), we write the continuum magnetisation 
${\bf M}({\bf r})= {\bf M}_o({\bf r}) + {\bf m}({\bf r})$, where 
${\bf M}_o({\bf r})$ is the slowly-varying part descibing the wall profile and 
${\bf m}({\bf r})$ describes fluctuations around this. Then
we rotate the spin quantisation axis to be locally
parallel to 
${\bf M}_o({\bf r})$, and get \cite{dube97} 
\begin{equation}
H=H_{w} + \sum_{k=1}^{N}  
\int \frac{d^{3}r}{\gamma_{g}} \delta ({\bf r}-{\bf r}_{k})\;
\bigg[  \omega_{k}^{\parallel}   M_{z}({\bf r}) I_{k}^{z} \;+\; 
 \omega_{k}^{\perp}[ m_{x}({\bf r}) I_{k}^{x}
+  m_{y}({\bf r}) I_{k}^{y}] \bigg] \;+\;
\frac{1}{2} \sum_{k} \sum_{k'} V_{k k'}^{\alpha \beta}
I_{k}^{\alpha} I_{k'}^{\beta}
\label{inutile}
\label{DWH2}
\end{equation}
displaying explicitly the longitudinal and transverse couplings. The "particle"
moves through a slowly fluctuating "random walk" potential field 
$U_{\parallel}(Q)\;$ (coming from the sum over  
couplings $ \omega_{k}^{\parallel}$ to randomly oriented spins). 
The transverse coupling (independent of $Q$ but not of $P$) 
causes "spin flip" transitions in the 
spin bath when the wall moves, even if the bath is at $T=0$.

One may also discuss problems in superconductors and normal metals involving
nuclear and paramagnetic spins, and other "defects", 
which can also be mapped to the same model of a
particle moving through a spin bath (sections 2.F, 5.B).

\subsection{The Central Spin Model}
\label{sec:22D}

Is there a low-energy effective Hamiltonian, analogous to
the spin-boson model, in which a "central" 2-level system
couples instead to a spin bath? 
The answer is yes, but the effective Hamiltonian
does not look quite so simple as the spin-boson one. In the {\it absence}
of any external field, the  
analogue of the spin-boson form in (\ref{2.4}) for a spin bath is actually 
\cite{PS93,PRB,PSchi95,PS96,tup97}
\begin{eqnarray}
H_{\mbox{\scriptsize CS}}(\Omega_o) 
& = & \left\{ 2\tilde{\Delta} \hat{\tau }_ -  \cos
\bigg[ \Phi - \sum_k \vec{V}_k \cdot \hat{\vec{\sigma }}_k  
 \bigg] + H.c. \right\} \nonumber \\
 & + & {\hat \tau }_z \sum_{k=1}^N \omega_k^{\parallel} \:
 {\vec l}_k \cdot {\hat {\vec \sigma }}_k  + \sum_{k=1}^N
 \omega_k^{\perp}\: {\vec m}_k \cdot {\hat {\vec \sigma }}_k
 +\sum_{k=1}^N \sum_{k'=1}^N V_{kk'}^{\alpha \beta } \hat{\sigma}_k^\alpha
 \hat{\sigma}_{k'}^\beta ~ \;.
 \label{1.24}
 \end{eqnarray}
where $\vec{\tau}$ describes the central spin, and the {$\sigma_k$} the
spin bath degrees of freedom. This form is not the most general one-
apart from dropping external field effects (for which see below) 
we have also restricted the central spin phase
in a simple $\cos \Phi$ form ({\it cf}. Introduction). As discussed below, 
both $\tilde{\Delta}$ and $\Phi$ incorporate spin bath renormalisation effects.
The factor of 2 in front of $\tilde{\Delta}$ is somewhat arbitrary (if the
cosine is one, then 
the actual "tunnel splitting" coming from (\ref{1.24}), in the absence of
spin bath effects, will be $4 \tilde{\Delta}$). 

The basic form of (\ref{1.24}) is actually fairly easy to understand. 
The extra phase in the first "non-diagonal" term (adding to $\Phi$)
comes from the topological phase of the bath spins as they make
transitions \cite{PS93}. There are also diagonal terms, and a weak
interaction $V_{kk'}$ between the bath spins. This form assumes the 
diagonal couplings $\omega_k^{\parallel}, \omega_k^{\perp}$ are $\ll$ 
the UV cutoff $\Omega_o$, and that 
$V_{kk'} \ll \omega_k^{\parallel}, \omega_k^{\perp}$. The ratio 
$V_{kk'}/\tilde \Delta$ is arbitrary. 
In sections 4 and 5 we shall see there is a weak coupling limit to
this model, in which it reduces to a spin-boson system.
 
Before this model was derived in this general form and then solved, a number 
of special cases had already been looked at \cite{SBhist}. In particular, 
Shimshoni and Gefen \cite{shim92}
included only the diagonal terms $\omega_k^{\parallel}$ and 
$\omega_k^{\perp}$, and examined the results in weak coupling in the presence 
of an AC field (see also \cite{rose98,rose99,rose99a}); they clearly 
recognised that the problem was different from an oscillator bath one. 

Let us now discuss the different
terms in (\ref{1.24}), in the order they appear (cf. Fig. 2).

(i) {\it Non-Diagonal terms}: That an extra phase term should exist, 
coming from the spin bath, 
is obvious on general
grounds (cf. Introduction). 
One can understand its algebraic form
in the following way. 
Notice that $H_{eff}$ in (\ref{1.24}) operates on both the central
spin {\it and} the spin bath;
and the effect of a single giant spin transition
on $\hat{\vec{\sigma}}_k$ can always be written as a transition between
initial and final nuclear states, in the form
$\mid \chi_k^{fin} \rangle = \hat{T}_k \mid \chi_k^{in} \rangle$, where 
\begin{equation}
\hat{T}_k = e^{-i\int d\tau H_{int} (\tau ) } = 
e^{-[\delta_k + i (\vec{V}_k \cdot
\vec{\sigma}_k + \phi_k)] }
\label{chi.2}
\end{equation}
The integral over $H_{int}$ is only defined once we know the
trajectory of the central system during the tunneling event.
It is in this sense
that we say \cite{PS93} that the instanton has become an "operator"
in the space of the spin bath modes. Notice that in general the 
central spin phase
$\Phi$ and splitting $\Delta$ are renormalised by the bath couplings 
\cite{PS93,PRB,PSchi95}:
\begin{equation}
\Phi = \Phi_o + \sum_k \phi_k  \;,\;\;\;\;\;\;\;\;\;\;\;\;\;\;\;
\tilde{\Delta} = \Delta_o \exp \{ -\sum_k \delta_k \}  \;.
\label{z.32}
\end{equation}
The $\{ \phi_k \}$ are "Berry phase" terms coming from the bath spin dynamics
during a central spin flip, and the $\{ \delta_k \}$ come from high-frequency
modifications of the original high-energy potential ("barrier 
fluctuations").
One can show that both $\phi_k$ and $\delta_k$ 
are $\sim O(\omega_k^2 /\Omega_o^2 )$, where $\omega_k$ is the larger of 
$\omega_k^{\parallel}, \omega_k^{\perp}$, and we will ignore 
these terms from now on.

Expanding out the cosine in (\ref{1.24}) gives a
series of terms like
$\hat{\tau}_{\pm}  \Gamma_{\alpha \beta \gamma \cdots }
\hat{\sigma}_{k_1}^{\alpha} \hat{\sigma}_{k_2}^{\beta}
\hat{\sigma}_{k_3}^{\gamma}  \cdots $ 
in which the instanton flip of the central spin couples simultaneously
to many {\it different} bath spins- a single central spin transition 
can cause multiple transitions in the bath (Fig. 2).
Later we will introduce a parameter 
$\lambda$
which measures the average number of bath spins 
flipping during each instanton.

(ii) {\it Diagonal terms}: These act {\it between} transitions of the central 
spin (Fig. 2), and are also easy to understand.
Formally, one starts by considering the "initial" and "final" fields 
(ie., before and after a 
transition of $\vec{\tau}$) acting on $\vec{\sigma}_k$. Calling these fields 
 $\vec{\gamma}_{k}^{(1)}$
and $\vec{\gamma}_{k}^{(2)}$ respectively (Fig. 3), we define the {\it
sum} and
the {\it difference} terms as
\begin{eqnarray}
\omega_k^{\parallel}{\vec l}_k & =& ({\vec \gamma }_k^{(1)} -
{\vec \gamma }_k^{(2)})/2 \nonumber \\
\omega_k^{\perp} {\vec m}_k & =& ({\vec \gamma }_k^{(1)} +
{\vec \gamma }_k ^{(2)})/2\;.
\label{1.25}
\end{eqnarray}
where ${\vec l}_k$ and ${\vec m}_k$ are unit vectors. Then the
truncated diagonal
interaction takes the form
\begin{equation}
H^D_{eff}\; =\;\; \sum_{k=1}^N \bigg\{ {\vec \gamma}_k^{(1)} {1+\hat{\tau}_z
\over 2} +
{\vec \gamma}_k^{(2)} {1-\hat{\tau}_z \over 2} \bigg\} \cdot {\hat {\vec
\sigma }}_k \;\;\;\equiv\;\;\;
{\hat \tau }_z \sum_{k=1}^N \omega_k^{\parallel} \:
{\vec l}_k \cdot
{\hat {\vec \sigma }}_k  + \sum_{k=1}^N \omega_k^{\perp}\: {\vec m}_k
\cdot
{\hat
{\vec \sigma }}_k  \;,
\label{1.26}
\end{equation}
i.e., one term which changes during a transition of the central system,
and one which does not.

The longitudinal coupling 
$ \hat{\tau}_z \sum \omega_k^{\parallel} \sigma_k^z$ 
determines the gross structure of 
the bath states in energy space: it also determines an "internal bias field"
$\epsilon (\{\sigma_k^z\}) = \sum \omega_k^{\parallel} \sigma_k^z$ acting on
$\tau_z$. The 2 levels of bath spin 
$\vec{\sigma_k}$ are split by 
energy $\omega_k^{\parallel}$, depending on whether $\sigma_k^z$ is parallel
or antiparallel to $\tau_z$. The effect of this on the $2^N$ fold multiplet of 
bath states surrounding each central spin state is shown in Fig. 4. Suppose 
we classify these states by their "polarisation group"; all bath states
whose total longitudinal polarisation $\sum \sigma_k^z = M$  are in 
polarisation group $M$. Since the $\omega_k^{\parallel}$ 
vary from one bath spin
to another, states in polarisation group $M$ are spread over an energy range 
$\tilde \Gamma_M$ ; and the 
entire manifold of states, comprising all polarisation groups, is spread 
over a larger energy range $E_o$. Let us define {\it normalised} densities 
of states $G_M(\epsilon)$ and $W(\epsilon)$ for these 2 distributions, so that
\begin{equation}
W(\epsilon) = (1/2^N)\sum_{M}C_N^{(N+M)/2} G_M(\epsilon)
\label{density}
\end{equation}
where $C_n^m = n!/m!(n-m)!$. In almost any physical case one will have 
strongly overlapping polarisation groups, so that 
for all but very small values of $N$, or except in the extreme wings of the 
distributions, one has
\begin{equation}
G_M (\epsilon) \sim (2/\pi \tilde \Gamma _M^2)^{1/2}
e^{-2\epsilon^2/\tilde \Gamma_M^2}.
\label{G_M}
\end{equation}
\begin{equation} 
W(\epsilon) \sim (2/\pi E_o^2)^{1/2}e^{-2\epsilon^2/E_o^2}.
\label{gauss}
\end{equation}
The simplest case is where the 
$\omega_k^{\parallel}$ cluster around a single 
central value $\omega_o$ with variance
\begin{equation}
\delta \omega_o = \sqrt{ {1 \over N}\sum_{k} (\omega_k^{\parallel}-\omega_o)^2 }
\label{4.28}
\end{equation}
For this case we define a parameter $\mu = N^{1/2}\delta \omega_o/\omega_o$,
characterising the degree of polarisation group overlap; overlap is complete 
when $\mu > 1$. 
Then $\tilde{\Gamma}_M \sim 2N^{1/2}\delta\omega_o$ and $E_o = 2N^{1/2}\omega_o$
(so that  $\tilde{\Gamma}_M/E_o \sim \delta\omega_o/\omega_o = N^{-1/2}\mu$). 
In the extremely unlikely 
case where $\mu \ll 1$, $W(\epsilon)$ can no longer be treated
as Gaussian- however there is an intrinsic lower limit to the linewidth
of each polarisation group set by the interspin interaction $V_{kk'}$. This
"intrinsic linewidth" $\Gamma_o \sim N^{1/2}V_o$, 
where $V_o$ is a typical value of $V_{kk'}$; 
in any physically realistic case this is usually enough 
by itself to cause complete overlap 
of all groups
(the essentially non-interacting 
case where $\tilde {\Gamma} = 0$, ie., $\omega_k = \omega_o$ 
for all bath spins, and $V_{kk'} = 0, \vec{V}_k= 0, \omega_k^{\perp} = 0$,
was actually studied by Garg \cite{garg95}).

The "transverse" couplings $\omega_k^{\perp}$ arise when the fields before/
after a transition, acting on the 
$\{ \vec{\sigma}_k \}$, are not exactly parallel or antiparallel. This can 
happen in many ways, either because of external fields which couple to the
bath spins, or because of a lack of symmetry in the 
underlying dynamics of the central system, or in its coupling to the bath
modes. Thus they are non-zero in any realistic situation. 

\vspace{3mm}

(iii) {\it Internal Spin Bath dynamics}: Finally, the interaction $V_{kk'}$ 
is usually 
so weak that it
does not change under truncation. If this term is absent, the spin bath will 
have no "intrinsic" dynamics, and remains inert between transitions of 
$\vec{\tau}$. Thus even if small, $V_{kk'}$ is important, since it allows the 
bath state to evolve during these intervals. The 
most important effect of $V_{kk'}$ is that it allows the longitudinal bias 
field $\epsilon (\{\sigma_k^z\})$ to fluctuate in time, between and during 
transitions of $\vec{\tau}$. 
Notice, however, that with the Hamiltonian in the form (\ref{1.24}), {\it only 
fluctuations within the same polarisation group are allowed}.
In NMR language, only $T_2$ processes occur in the intrinsic dynamics of the 
spin bath- changes in $M$ can only occur via the interaction with the central
spin. This will only be true at low $T$- at higher $T$ longitudinal relaxation
(ie., "$T_1$ processes", in NMR language) between different polarisation 
groups should be included in a realistic model. Such processes (which arise 
from the interaction of the spin bath modes with other environmental modes, or 
with thermally excited higher modes of the central system, above the UV cutoff 
energy $\Omega_o$) are almost always very slow when $kT \ll \Omega_o$. On the 
other hand $T_2$ fluctuations will persist until $kT \ll \Gamma_o$; in the 
physical examples studied so far this means they persist down to $\mu K$ 
temperatures or below. We will return briefly to this very low $T$ regime 
at the end of the article (section 5.C). 

\vspace{3mm}

{\bf External Field Effects}:  In general all of the parameters in (\ref{1.24})
will depend on any external field ${\bf H}_o$, because it changes the 
high-energy dynamics of both central system and bath- however we 
can make low-field expansions and separate out the most important terms. 
Defining the "Zeeman" coupling energies $\Omega_{H_o}$ and $\omega_k^{H_o}$ of 
central and bath spins to this field, it is easy to see that under the 
conditions $\Omega_{H_o}/\Omega_o < 1$ and $\omega_k^{H_o}/\omega_k < 1$,
the principal changes to (\ref{1.24}) will be (i) the addition of an obvious 
longitudinal coupling $\xi_{H_o} \hat{\tau}_z$ to the central spin, and (ii)
Qthe changes $\Phi_o \rightarrow \Phi_{H_o}$,
$\vec{V}_k \rightarrow \vec{V}_k^{H_o}$, and $\vec{ \omega}_k^{\perp} \equiv
\omega_k^{\perp}{\hat{\vec m}}_k \rightarrow 
\vec{\omega}_k^{\perp}({\bf H}_o)$,
where up to {\it linear order} in $H_o$ one has
\begin{eqnarray}
\Phi_{H_o} & =& \Phi_o + \psi(H_o) \;;\;\;\;\;\;\;\;\;\;\;\;\psi \sim 
2\pi \Omega_{H_o}/\Omega_o \nonumber \\
\vec{V}_k^{H_o} & =& \vec{V}_k + \vec{v}_k(H_o) \;;\;\;\;\;\;\;\;\;\;\;
\vert \vec{v}_k(H_o) \vert /\vert \vec{V}_k\vert 
\sim \omega_k^{H_o}/\omega_k \nonumber \\
\vec{\omega}_k^{\perp}({\bf H}_o)&=& 
\vec{\omega}_k^{\perp}+{\bf d}_k({\bf H}_o)
\;;\;\;\;\;\;\;\;\;\;\;\; \vert {\bf d}_k({\bf H}_o)\vert/\omega_k^{\perp}
\sim  \omega_k^{H_o}/\omega_k.
\label{field}
\end{eqnarray}
Thus even at low fields one has an important change in
all the topological phases in the problem, and also to the 
{\it transverse} diagonal coupling (which itself arises from internal fields).
In general ${\bf H}_0$ will also change the interspin bath couplings 
$V_{kk'}$; in a way which depends on the specific details of the problem.
In the next 2 sections we see how this works for both magnetic and 
superconducting systems.

\subsection{Nanomagnet coupled to nuclear and paramagnetic spins}
\label{sec:2E}

If we start from the "giant spin" model introduced above for a nanomagnet,
then a simple isotropic contact hyperfine coupling to nuclear 
spins will lead to a Hamiltonian 
(for $E < E_c$) like:
\begin{equation}
H(\vec{S};\{ \hat{\vec{\sigma}} \} ) = H_o (\vec{S} ) + {1 \over S}
\sum_{k=1}^N \omega_k \vec{S} \cdot \hat{\vec{\sigma}}_k 
+H_{\mbox{\scriptsize env}} (\{ \hat{\vec{\sigma}} \}) \;;
\label{2.8}
\end{equation}
where $H_o(\vec{S})$ is the  ``giant spin" Hamiltonian, 
and $H_{\mbox{\scriptsize env}} (\{ \hat{\vec{\sigma}} \})$
is the same as in (\ref{2.7}). 
The generalisation of this simple Hamiltonian to include dipolar hyperfine 
interactions, as well as to higher spin nuclei and to paramagnetic spins (with 
tensor and quadrupolar couplings) can be used if necessary 
\cite{rose99,rose99a}. 
Here we will assume for simplicity that
$\mid \vec{I}_k \mid = I = \frac{1}{2}$, and write $\vec{I}_k \rightarrow
\sigma_k$, i.e. the nuclear spins will be described by spin-$\frac{1}{2}$
Pauli
matrices.
In fact in many cases even if $I \neq \frac{1}{2}$, the low-energy nuclear spin
dynamics is well
described by a 2-level system. 
 
The truncation of 
$H(\vec{S};\{ \hat{\vec{\sigma}} \} )$ to a central spin Hamiltonian 
$H_{\mbox{\scriptsize eff}} (\vec{\tau};\{ \hat{\vec{\sigma}} \})$ has been
discussed in several ~
papers \cite{PS93,PRB,PS96,seattlePS,tup97,dauriac}. 
As an example we quote the result for a simple
easy axis-easy plane nanomagnet (for which $H_o(\vec{S}) = 
(1/S)[-K_2^{\parallel}\hat{S}_z^2 + K_2^{\perp}\hat{S}_y^2]$, and give it a
physical interpretation. In this case, assuming  
$\omega_k \ll \Omega_o$ and also a weak external field ${\bf H}_o$, the 
effective Hamiltonian is \cite{PS93,PRB,PS96,tup97}:
\begin{eqnarray}
H_{eff}(\Omega_o) 
& = & \left\{ 2 \Delta_o \hat{\tau }_ -  \cos \bigg[ \pi S -
i \sum_k \alpha_k \vec{n}_k \cdot \hat{\vec{\sigma }}_k 
- \beta_o {\bf n}_o . {\bf H}_o \bigg] + H.c. \right\} \nonumber \\
& + & {\hat \tau }_z \bigg[ \xi_H + 
                                   \sum_{k=1}^N \omega_k^{\parallel} \:
{\hat \sigma }_k^z \bigg] \; + \; 
\sum_{k=1}^N \sum_{k'=1}^N V_{kk'}^{\alpha \beta } \hat{\sigma}_k^\alpha
\hat{\sigma}_{k'}^\beta ~ \;.
\label{z.3301}
\end{eqnarray}
ie., a special case of the general form (\ref{1.24}), 
with the parameters $\xi_H = g \mu_B S_z H_o^z$, 
$\vec{l}_k =\hat{\vec{z}}$, $\omega_k^\parallel
 = \omega_k$, and $\omega_k^\perp =0$.
The vectors $\alpha_k\vec{n}_k$ and $\beta_o {\bf n}_o$ 
for this easy axis-easy plane case turn 
out to be (again assuming $\omega_k \ll \Omega_o$, and small ${\bf H}_o$): 
\begin{equation}
\alpha_k \vec{n}_k = {\pi \omega_k \over \Omega_o} \big(
\hat{\vec{x}},~
i \sqrt{K_{\parallel} / K_{\perp }}~ \hat{\vec{y}} \big)\;;\;
\;\;\;\;\;\;\;\;\;\;\;\;\;\;\;
\beta_o {\bf n}_o = {\pi g \mu_B S \over \Omega_o} \big(
\hat{\vec{x}},~
i \sqrt{K_{\parallel} / K_{\perp }}~ \hat{\vec{y}} \big)
\label{z.29}
\end{equation}
In this example there are 2 tunneling trajectories
(clockwise and counterclockwise in the easy plane), giving a result
$ e^{\pm i \alpha_k \vec{n}_k \cdot
\vec{\sigma}_k }$ for the "transfer matrix" $\hat{T}_k$ 
(in zero applied field). The resulting vector
$\alpha_k \vec{n}_k$ is the "average hyperfine field" acting on
$\vec{\sigma}_k$ during the tunneling event. To understand its
orientation (and why it is {\it complex}) we note that the nuclear spin
itself exercises a torque on $\vec{S}$ while it is tunneling, 
and this pushes $\vec{S}$
away from the easy plane. Consequently (a) the average field
acting on $\vec{\sigma}_k$ has a component out of the easy plane, in the
$y$-direction, and (b) $\vec{S}$ no longer moves exactly
along the easy-plane path, while tunneling, that it
would in the absence of  $\vec{\sigma}_k$ (and so its action increases,
via the imaginary part of $\alpha_k \vec{n}_k$).

The contribution $\beta_o {\bf n}_o . {\bf H}_o$ to the topological phase
\cite{tup97,boga92,garg93} comes from the area swept 
out by the giant spin on the spin sphere (cf. Introduction),
which changes as the field ${\bf H}_o$ changes;
it is essentially an "Aharonov-Bohm"
contribution to this phase from the external field \cite{boga92}, which
leads to spectacular oscillations (Fig. 5) in the effective tunneling amplitude
$\tilde \Delta \equiv 2 \Delta_o \cos [\pi S + i \beta_o {\bf n}_o.{\bf H}_o]$
for a field perpendicular to $\vec{z}$ \cite{rose99a,garg93}. Very
recently oscillations in the tunneling amplitude of $Fe$-8 magnetic
molecular crystals were seen which are related to this \cite{ww2}, 
although 
the presence of both nuclear spins and dipolar fields seriously complicates
their interpretation \cite{rose99a} (see section 5.A).

We recall from section II.B that the giant spin model truncates for nanomagnets
to a 2-level system for energies $\ll \Omega_o \sim 1-10~K$. Contact hyperfine 
couplings are in the range $1.3-30~mK$ (transition metals) or $40-500~mK$
(rare earths); on the other hand the internuclear couplings $V_{kk'}
\sim 10^{-8}-10^{-5}~K$. In the $Fe$-8 system just 
mentioned the hyperfine interactions are actually dominated by dipolar 
couplings between the 8 $Fe^{+3}$ ions and the 120 protons in the molecule; 
these couplings are in the range $\sim 1-100~ MHz$ ($0.05-5~mK$). When the 
hyperfine couplings are this weak we must also take into account the effect of 
external fields on the nuclear dynamics \cite{rose99a} (section 5.A). 
The values of $\Delta$ 
vary over a huge range, but typically $\Delta \ll \omega_k$ (in $Fe$-8, 
$\Delta \sim 10^{-7}~K$).

\subsection{SQUID coupled to nuclear and paramagnetic spins}
\label{sec:2F}

We consider again the RF SQUID, but now concentrate on the  
coupling of the flux $\Phi$ to the spin bath of nuclear and paramagnetic spins
which are within a penetration depth of the surface of the superconductor.
This example is very instructive in understanding the weak-coupling limit of
the central spin model (the following discussion is based on refs.
\cite{seattlePS,PSsquid}).

Suppose to start with we consider a "cubic geometry" \cite{seattlePS}, 
in which a cube of side $L = 1~ cm$ has a hole of radius $R = 0.2~cm$ through 
it, with in addition a slit connecting the hole to the exterior, spanned by
a cylindrical junction of length $l = 10^{-4}~cm$ and diameter $d= 2 \times
10^{-5}~cm$.
The magnetic field inside the hole corresponding to a
half-flux quantum is
$B_o=(\pi \hbar c /e \pi R^2)=2\times 10^{-6}\: G$,
whereas the magnetic field in the
junction is as high as
$B_j \sim B_o L/d = 10^{-1}\: G$.

There are both nuclear spins and paramagnetic impurities in the spin bath.
Consider first the
nuclear spins; assuming all nuclei have spins, we find that in the bulk of
the ring, within
a penetration depth of the surface, there are  
$N_r \sim 2\pi R \lambda_L L \times 10^{23}
\approx 5 \times 10^{17}$ 
nuclear spins coupling to the ring current; and in the junction itself,
a number more like
$N_j= (\pi d l \lambda_L ) \times 10^{23}
\approx 3 \times 10^{9}$. Thus each ring nuclear spin couples to the SQUID
with a diagonal coupling $\omega_r^{\parallel} 
\sim \mu_n B_o \sim  2\times 10^{-13}K$;
on the other hand for junction spins this coupling is  
$\omega^{\parallel}_j = \mu_n B_j
\sim 10^{-8}K$. Notice we have 
ignored any coupling to substrate spins (assume, eg., 
the ring is in
superfluid He-4!), which
might have a much larger coupling to the current. 

At any temperature such that 
$kT \gg \omega_r^{\parallel},\omega^{\parallel}_j$, the typical 
polarisation of these spin baths will be $\sqrt{N_r},\sqrt{N_j}$
respectively, giving a distribution of  
longitudinal bias energies with typical values  
$E_o^j\sim \omega^{\parallel}_j  \sqrt{N_j} \approx 5\times 10^{-4}K$, and 
$E_o^r\sim  \omega^{\parallel}_r \sqrt{N_r} \approx 10^{-4}K$
acting on the tunneling flux
coordinate $\Phi$.
If we now add paramagnetic impurities to the ring, 
with concentration $n_{pm}$, and coupling  
$\omega_{pm}^{\parallel} \sim 2 \times 10^3 \omega_r \sim 4 \times
10^{-10} ~K$ 
to the current, this gives  
a typical longitudinal bias energy  
$E_o^{pm} \sim n^{1/2}_{pm}\times 0.2~K$. This longitudinal term
is obviously bigger
than the nuclear contribution, unless the superconductor is very pure indeed! 

However, there is another much stronger transverse term,
because each spin feels the dipolar fields from the other 
spins. This field is
$\geq 1G$ (much higher near to the paramagnetic spins), and for the 
nuclear spins has an 
associated energy $\omega_k^{\perp} \geq 10^{-7} K$, which is 
$\gg \omega^{\parallel}_r, \omega^{\parallel}_j$. 
Physically, when the SQUID flips, the field on each
nuclear spin hardly changes its direction, being dominated by the 
more slowly varying
(but much stronger) nuclear dipolar field. For 
the paramagnetic spins the analogous coupling $\omega^{\perp}_{pm}$ 
is $> 10^3$ times larger, which 
in the absence of nuclear fields would give an inter-paramagnetic  
"flip-flop" rate $V_{kk'}^{pm} \sim 10^{9} n_{pm}$ Hz, 
except that in pure 
samples these
processes will themselves be blocked by the local dipolar coupling
between the impurity
and nearby nuclear spins (of strength $\omega_{pm}^{\perp} 
\sim 10^{-4} K$); this will 
happen once
$n_{pm} \ll 10^{-3}$.

We can write down an effective Hamiltonian for this system, valid over
timescales considerably greater than $\Delta^{-1}$; we will use this 
later to analyse the effect of the spins on the SQUID dynamics
(section 5.B). We will assume that $\Delta \gg
V_{kk'}$ (the only case of experimental interest); then we can treat the internuclear dipolar fields as
slowly-varying in time. The 
effective Hamiltonian can then be derived, to give \cite{PSsquid}:
\begin{equation}
H{\mbox{\scriptsize eff}}(\Omega_o) = \{ \Delta_o(\Phi_o) \hat{\tau}_+ 
e^{-i\sum_k \vec{\alpha}_k . \hat{\vec{\sigma}}_k} + H.c. \} + 
\xi_H \hat{\tau}_z +
\sum_{k=1}^{N} [ \hat{\tau}_z
\omega^{\parallel}_k \hat{\sigma}_k^z + 
\omega^{\perp}_k\hat{\sigma}_k^x]
\label{cases.3}
\end{equation}
where $\omega_k^{\parallel} = \omega_r$, $\omega_j$ or
$\omega^{\parallel}_{pm}$, depending on the spin, and $\omega_k^{\perp}$ has 
just been discussed; and microscopic analysis shows that 
$\vert \vec{\alpha}_k \vert \sim \omega_k^{\parallel}/\Omega_o$, 
where $\Omega_o$ is the 
Josephson plasma frequency (section 2.B). 
Notice that $\omega^{\parallel}_k/\omega^{\perp}_k \ll 1$, which is the
opposite limit considered to that for the giant spin! Notice 
further that these couplings
are far less than $\Delta_o$ ($E_o^r, E_o^j, E_o^{pm} \gg \Delta_o$ 
only because there are so many
spins involved).
Thus the spin bath 
is no longer "slaved" to the 
central system. 
In section 4 we see how this allows a mapping to an oscillator 
bath, coupled to $\vec{\tau}$ (ie., a spin-boson model).

Finally we note that the external field also acts on the nuclear and
paramagnetic spin dynamics, via the Zeeman coupling (which simply adds to 
$\omega_k^{\parallel}$). This can  
help to suppress decoherence effects, by freezing out the spin 
bath dynamics- for more details see ref.\cite{PSsquid}.

\subsection{General Canonical Models}
\label{sec:2G}

We now recall our assertion that almost any mesoscopic
"central" system, coupled to its environment, may be described at low
energies by (\ref{2.2}), with the environment being written as a sum of an
oscillator bath term (\ref{2.3}) and a spin bath term ((\ref{2.7}), or a
higher spin generalisation). The simplest example is of a single central spin
coupled to both oscillator and spin baths. 
Such a model seems forbidding but in fact a fairly complete analysis has
been given of its dynamics \cite{PS96}- we recall some of the results at the 
end of section 4. 
One can also consider
a much more complicated model in which a {\it macroscopic array}
of central spins $\{ \vec{\tau}_j \}$, at positions $\{ {\bf r}_j \}$, 
couples to both oscillator and spin baths. The effective Hamiltonian is then
an obvious generalisation of what has gone before:
\begin{eqnarray}
H_{\mbox{\scriptsize CS}}(\Omega_o)
& = & \sum_j \left\{ \Delta_j \hat{\tau}_j^{-} \cos
\bigg[ \Phi_j - i\sum_k \alpha_{jk} \vec{n}_{jk} \cdot \hat{\vec{\sigma }}_k
\bigg] + H.c. \right\} \; + \; 
\sum_{i<j} V({\bf r}_i - {\bf r}_j) \hat{\tau}_i^z \hat{\tau}_j^z
\nonumber \\
& + & \sum_j\left\{ 
{\hat \tau }_j^z  \sum_{k=1}^N \omega_{jk}^{\parallel} \:
{\vec l}_{jk} \cdot {\hat {\vec \sigma }}_k  +  \sum_{k=1}^N
\omega_{jk}^{\perp}\: {\vec m}_{jk} \cdot {\hat {\vec \sigma }}_k \right\}
+  \sum_{k=1}^N \sum_{k'=1}^N V_{kk'}^{\alpha \beta }
\hat{\sigma}_k^\alpha
\hat{\sigma}_{k'}^\beta \nonumber \\
& + & \sum_j \sum_q \bigg[ c_{jq}^{\parallel} \hat{\tau}_j^z +
(c_{jq}^{\perp} \hat{\tau}_j^{-} + h.c. ) \bigg] x_q
 + {1 \over 2} \sum_q \left(  {p_q^2 \over m_q} +
m_q \omega_q^2 x_q^2 \right)   ~\;,
\label{QSGl}
\end{eqnarray}
where $ V({\bf r}_i - {\bf r}_j) \hat{\tau}_i^z \hat{\tau}_j^z$ is a 
"high-energy" diagonal coupling between the various "central spin" systems.
If we throw away the spin bath we get a set of 2-level systems coupling to 
an oscillator bath, of which the simplest example is the "PISCES" model
(in which there are only two 2-level systems \cite{dube98,dube98a}). 

Such models seem impossibly complicated, but actually  
one can solve for their dynamics in many 
important regimes! 
What is crucial is the {\it separation} of the 2 baths. 
Often (as 
with nuclear spins) their mutual interaction is very weak
(and can be parametrised by a time $T_1(T)$ which may be very long at low
$T$); in this case this separation is a good one. 
If there {\it are} certain spin bath modes that interact strongly with
the oscillators, then typically we can simply absorb these modes into an 
"augmented" oscillator bath by a canonical transformation. An obvious
example arises with electronic spins in a metallic host
(the Kondo or Kondo lattice problems); one 
rewrites the bath to include the "Kondo
resonance" in the oscillator bath spectrum.

A proof that one may do this in all cases seems
rather difficult- in any case the usefulness of these models tends
to be established by their application. 
Models like (\ref{QSGl}) describe mesoscopic
systems like coupled SQUIDs 
or coupled nanomagnets \cite{PS98,dube98,dube98a}, Quantum Spin Glasses
\cite{cugl98} and low-$T$ dipolar glasses \cite{burin}, as well as 
coupled anisotropic coupled Kondo spins and Kondo lattices, 
coupled nuclear spin systems \cite{IDV}, superconducting arrays, or coupled
defects in solids. They are also useful for analysing purely theoretical
questions about relaxation, dissipation, decoherence and quantum measurements
in quantum systems- 
many questions remain unanswered, having only been 
studied thoroughly in restricted models such as the spin-boson model 
\cite{ajl87,weiss} or the PISCES model \cite{dube98,dube98a}. 
We return to experimental and theoretical applications 
in section 5.

\section{Averaging over the Spin Bath}
\label{sec:3}

To extract useful information from the low-energy canonical models, 
we must calculate their dynamical properties.
Since we are typically not interested in the environment 
(one usually has little control over it), 
one performs a statistical average over
the environment. This procedure is fraught 
with danger, because of  
"memory" effects in the environment, and because  
assumptions such as "self-averaging" in 
the environmental correlation functions may not strictly be valid. 

In this section we show how the spin bath may be "integrated out" by means
of 4 different statistical averages, each involving an integration over a 
particular variable. The end result is a description of the time evolution 
of the "reduced" density matrix for the central system- {\it provided} we can 
ignore memory effects in the environment. 
The starting point is 
no different from that involved in functional averaging over oscillators 
\cite{weiss,cal83,feyv63}; both
begin with a path integral form for the propagator of the reduced central 
system density matrix, written as 
\begin{equation}
K(1,2) \ = \displaystyle \int^{Q_2}_{Q_1} dQ \displaystyle
\int^{Q^{\prime}_{2}}_{Q^{\prime}_{1}}
dQ^{\prime} \ e^{-i/\hbar (S_o[Q] \ - \
S_o[Q^{\prime}])} {\cal F}[Q,Q'] \;,
\label{3.3}
\end{equation}
where $S_o[Q]$ is the free central system action, and
${\cal F}[Q, Q^{\prime}]$ is the famous
``influence functional" \cite{feyv63}, defined in general by
\begin{equation}
{\cal F}[Q,Q'] =  \prod_{k} \langle \hat{U}_k(Q,t)
\hat{U}_k^{\dag}(Q',t) \rangle \;,
\label{neto.2}
\end{equation}
Here the unitary operator $\hat{U}_k(Q,t)$ describes the evolution of the
$k$-th environmental mode, given that the central system
follows the path $Q(t)$ on its
"outward" voyage, and $Q'(t)$ on its "return" voyage; and
${\cal F}[Q,Q']$ acts
as a weighting function,
over different possible paths $(Q(t),Q'(t'))$.

For a central 2-level system,
the paths $Q(t),Q'(t)$ are simple (recall Fig. 2):
\begin{equation}
Q_{(n)}(s)=1-\sum_{i=1}^{2n}\big[ sgn(s-t_{2i-1})+
sgn(t_{2i}-s) \big] \;,
\label{neto.1}
\end{equation}
where $sgn(x)$ is the sign-function, and $n$ is the number of
transitions of the central system, occuring at
times $t_1,t_2, \dots ,t_{2n}$ (for definiteness we assume
trajectories starting and ending in the same state, and use the
convention that $Q = \pm 1$ corresponds to $\tau_z = \pm 1$). The
goal is to find the central spin density matrix; in this article we 
give results for the "return probability" $P_{11}(t)$ for the system
to be in the same state $\vert \uparrow \rangle$ at time $t$ as 
it was at $t=0$. Using (\ref{neto.1}) 
this can be written as an "instanton expansion" over flips
of the central spin (Appendix A): 
\begin{equation}
P_{11}(t)=\sum_{nm}^{\infty} (i\Delta_o)^{2(n+m)}
\int_0^t dt_1 \dots \int_{t_{2n-1}}^t dt_{2n}
\int_0^t dt'_1 \dots \int_{t'_{2m-1}}^t  dt'_{2m} {\cal F}[Q_{(n)},Q_{(m)'}]
\label{neto.22}
\end{equation}
Further 
simplification arises if the environmental modes are uncoupled- then 
${\cal F}[Q, Q^{\prime}]$ factorises, and we can write 
${\cal F}[Q,Q'] = \exp(-i\Phi [Q,Q']) =
\exp(-i \sum_{k=1}^N \phi_k [Q,Q'])$, where the {\it complex} phase 
$\phi_k[Q,Q']$ 
contains both real (reactive),
and imaginary (damping) contributions.

Now for an {\it oscillator bath} one simplifying 
feature is crucial, viz., the very weak
coupling to each oscillator. This allows one to evaluate each $\phi_k[Q,Q']$
up to 2nd order only in these couplings, in terms of a 
spectral function for the {\it unperturbed}
oscillator dynamics
(compare $J(\omega,T)$ in (\ref{3.6})). Even though the paths $Q(t)$ and 
$Q'(t)$ may be complicated, 
the calculation of 
${\cal F}[Q, Q^{\prime}]$ is often tractable \cite{weiss}. 

However because the coupling to each spin bath mode is
{\it not} necessarily weak, it will in general
strongly alter their dynamics, often {\it slaving} them to
the motion of the central system. 
{\it Thus we cannot start from the unperturbed spin bath dynamics}- the problem 
is fundamentally non-perturbative in the $\{ \omega_k \}$. However it is not 
intractable, because one {\it can} rather easily 
deal with the dominant longitudinal terms $\{ \omega_k^{\parallel} \}$.
The other terms can then be dealt with perturbatively (and sometimes even
non-perturbatively). It is the separation of the effects of the various terms
in the Hamiltonian which leads to not one, but 4 different averages. What
is quite remarkable is that these averages can be evaluated  
analytically in most cases (section 4).
 
We begin by explaining the 
4 different averaging integrals required for a general spin bath. This is
done pedagogically, by solving for 4 different limiting cases 
of the central spin Hamiltonian (sections III.A-III.D), each of which requires
only one of the 4 averaging integrals. Then the general
procedure (combining the 4 averages) is given in
III.E. One reason for going through these averages one by one is that each 
corresponds to a different physical mechanism of decoherence- we return to 
this in section 5 (note that more detailed results for 
the 3 limiting cases discussed in sections 
III.B-III.D are given in refs. \cite{PRB,PSchi95}).

\subsection{Phase averaging: Topological decoherence}
\label{sec:3A}

Formally the case of pure topological decoherence applies to the following
special case of $H_{eff}$, in which only non-diagonal terms are included:
\begin{equation} 
H_{\mbox{\scriptsize eff}}^{top} = 2 \Delta_o \left\{ {\hat \tau }_{-} 
\cos \big[ \Phi_o   -i \sum_{k=1}^N
 \alpha_k {\vec n}_k \cdot {\hat {\vec \sigma }}_k  \big] 
 + H.c. \right\} \;,
\label{4.1}
\end{equation}
Since $\omega_k^{\parallel}=
\omega_k^{\perp} = 0$ are zero, 
{\it all} the $2^N$ environmental states are degenerate, 
and there is no exchange of any
energy between ${\vec \tau}$ and the $ \{ {\hat \sigma }_k \}$. The only thing 
that is exchanged is {\it phase};
the phase $\Phi_o$ of ${\vec \tau}$ becomes entangled with that
of the $ \{ {\hat \sigma }_k \}$, during the transitions of ${\vec \tau}$
between $\uparrow \rangle$ and $\downarrow \rangle$,
so that the initial
and final {\it states} of the spin environment are different. Physically 
(\ref{4.1}) would arise if the original high-energy Hamiltonian contained
only "transverse" couplings (to $\{ \sigma_k^{\pm} \}$), which only act while
the central system is making transitions (in the case of a moving particle 
coupled to a spin bath, they would be "velocity couplings", only acting
when the particle is moving).

We wish to determine $P_{11}(t)$ for this case. For pedagogical 
purposes let us begin by assuming that $-i \alpha_k$ is {\it real},
ie., $\alpha_k$ is pure imaginary; then we have added a pure environmental 
phase term to 
the free 2-level Hamiltonian. Then, 
writing $-i\alpha_k \to \tilde \alpha_k$,
the formal solution to this 
problem can be written immediately as \cite{PS93}
\begin{equation} 
P_{11} (t) = {1 \over 2} \left\{ 1 + \langle \cos \big[ 4\Delta_o t 
\cos \big( \Phi_o   + \sum_{k=1}^N
 \tilde \alpha_k {\vec n}_k \cdot {\hat {\vec \sigma }}_k \big)  \big]  
 \rangle \right\}\;,
\label{4.4}
\end{equation}
where the brackets $\langle \cdot \cdot \rangle$ trace over the spin bath.
By writing this as an instanton expansion over central spin flips
(see App. A and refs. \cite{PS93,PSchi95}), we transform it to 
a {\it weighted integration over
topological phase}:
\begin{eqnarray}
P_{11}(t) &= &
\sum_{m=-\infty }^{\infty} F(m)
\int_{0}^{2 \pi} {d\varphi \over 2 \pi } e^{i2m(\Phi -\varphi )}
\left\{ {1 \over 2}+{1 \over 2} \cos (2 \Delta_o ( \varphi ) t) \right\}
\label{4.8a} \\
&= &{1 \over 2} \left\{ 1+\sum_{m=-\infty }^{\infty} (-1)^{m}
F(m)
e^{i2m\Phi } J_{2m} (4\Delta_o t ) \right\} \;,
\label{4.8}
\end{eqnarray}
where $\Delta_o(\varphi )= 2{\tilde \Delta}_o  \cos \varphi $ is a 
{\it phase-dependent} tunneling amplitude, 
$\;\; J_{2m}(z) $ is a Bessel function, and 
\begin{eqnarray}
F(m) = \langle \prod_{k=1}^{N} 
e^{2im \tilde \alpha_k {\vec n}_k \cdot {\hat {\vec \sigma }}_k } 
\rangle 
  = \prod_{k=1}^{N} \cos (2m \tilde \alpha_k ) \;.
\label{4.6}
\end{eqnarray}
For small $\alpha_k$ we may approximate the product in (\ref{4.6}) as 
\begin{equation}
F_{\lambda } (\nu ) = e^{-4 \lambda \nu^2 }\;; \;\;\;\;\;\;\;\;\; 
\lambda  = \sum_{k=1}^N \tilde \alpha_k^2/2 \;.
\label{4.7}
\end{equation}
Notice that $\lambda$ is just the mean number of environmental spins 
that are flipped, each time $\tau$ flips.

Clearly phase
decoherence is important  if $\lambda  >1$, in which case
$F_\lambda( \nu ) = \delta_{\nu ,0 } + \; small\; corrections$.
Then, rather surprisingly, we get a {\it universal form} (shown 
in Fig. 6), in the intermediate
coupling limit, for $P_{11} (t) $:
\begin{equation}
P_{11}(t)  \longrightarrow  {1\over 2}
\biggl[ 1+J_0(4 \Delta_o t) \biggr] \equiv
\int {d\varphi \over 2 \pi } P_{11}^{(0)}(t,\Phi =\varphi) \;\;\;\;\;\;\;\
(intermediate \;\; coupling) 
\label{4.13a}
\end{equation}
with a phase integration over the {\it free} central spin propagator 
$ P_{11}^{(0)}(t,\Phi)$ [cf. eqtn. (\ref{b1.4})]. Thus 
{\it random phases}
arise because successive flips of ${\vec \tau}$ cause, 
in general, a different
topological phase to be accumulated by the spin environment.
In fact, the universal behaviour comes from complete phase
phase randomisation \cite{PS93,sta94}, 
so that all possible phases contribute equally to the 
answer! The final form shows decaying oscillations , with an envelope
$\sim t^{-1/2}$ at long times, which can also be understood 
by noting that the "zero  phase" trajectories contributing to
$P_{11}$
constitute a fraction $(2s)!/(2^{s}s!)^2 \sim s^{-1/2}$ of all 
possible trajectories, where  $s \sim  \Delta_o t$.

In the {\it strong coupling} limit of this model, where the bath spins rotate
adiabatically with the the central spin, one has $\alpha_k \to \pi /2$,
so that $F(m)=(-1)^m$ and 
$P_{11} (t) = {1 \over 2} [1 + \cos (4\Delta_o t\cos \tilde{\Phi } ) ]$,
where $\tilde{\Phi }=\Phi_o +N\pi /2$, i.e., the Haldane/Kramers
phase is now $\tilde{\Phi }$, since the $N$ bath spins are
forced to rotate with $\vec{\tau}$.

The results for complex $\alpha_k$ are given in Appendix A; the basic ideas 
behind them (and the techniques for their calculation) are simple elaborations 
of the above.

\subsection{Average over longitudinal fields: Degeneracy blocking}
\label{sec:3C}

We now consider the effective Hamiltonian 
\begin{equation}
 H_{\mbox{\scriptsize eff}}=2 \Delta_o \tau _x +
 \tau _z \{ \xi_H + 
\sum_{k=1}^N \omega_k^{\parallel} \:{\hat \sigma }_k^z \}\;;
\label{4.27}
\end{equation}
To solve this we will assume the model discussed previously, in which all
$\{ \omega_k^{\parallel} \}$ cluster around a single central value 
$\omega_o$ (cf. eqtns (\ref{4.28}),(\ref{gauss}), and Fig. 4). 
Since the bath now just 
acts as an extra longitudinal field, 
we are dealing with the trivial case of a 
biased two-level system, with bias energy $(\xi_H + \epsilon)$, where  
$\epsilon = \sum_{k=1}^N \omega_k^{\parallel} \sigma_k^z $. The only 
question is how to average over the internal bias- this depends on whether 
we deal with a {\it single} central system, or a statistical ensemble of them
(corresponding to either an average over many measurements on a single 
system, or a single measurement on a 
large number of {\it non-interacting} 
systems- interactions are discussed in section 5.A). 

For a single central spin, the dynamics of $\vec{\tau}$ in this model are
completely trivial- one has $P_{11}(t) = [1 - (\Delta_o^2/E^2) \sin^2 Et]$,
where $E = \xi_H + \epsilon$, ie., resonant tunneling of an isolated spin 
in a longitudinal bias feld (compare (\ref{b1.10})).
 
For an {\it ensemble} of central spins, we must average over the whole 
bias range. In what follows let us assume for definiteness a spin bath at 
some equilibrium temperature $T = 1/\beta$; then the ensemble average is 
just a {\it weighted average over bias}: 
\begin{equation}
\int d\epsilon \; W(\epsilon ) {e^{-\beta \epsilon } \over Z(\beta )}
\label{biasav}
\end{equation}
where $Z(\beta)$ is the appropriate partition function. 
\begin{equation}
P_{11}(t) = \int d\epsilon \; W(\epsilon )
{e^{-\beta \epsilon } \over Z(\beta )}
\bigg[ 1-
{2 \Delta_o^2 \over (\epsilon + \xi_H)^2 + 4 \Delta_o^2 } \left(
1 -\cos (2t\sqrt{(\epsilon + \xi_H)^2+4 \Delta_o^2}  ) \right) \bigg] 
\label{4.32}
\end{equation}
The physical interpretation is obvious \cite{PS93}; only a very small fraction 
$A(\xi_H) \sim \Delta_o/E_o$ 
of central spins in the ensemble in the "resonance window", 
ie.,, having total bias 
$\vert (\xi_H + \epsilon)\vert \leq
\Delta_o$, can make transitions- this selects states with internal bias
around $\epsilon \sim -\xi_H$. All other states lack the 
near-degeneracy between initial and final energies required for resonant 
tunneling- they are "degeneracy blocked" \cite{PS93,PS96}. The resulting 
correlation function is then
\begin{equation}
P_{11}(t) =  1-2 A(-\xi_H ) 
\sum_{k=0}^\infty J_{2k+1}(4 \Delta_o t) \;.
\label{4.33}
\end{equation}
where $A(\xi )=\Delta_o W(\xi)$ (except in the unphysical case 
where the polarisation group linewidths $\tilde \Gamma_M < \Delta_o$. 
For the usual case where all polarisation groups 
overlap, and $W(\epsilon)$ has the Gaussian form (\ref{gauss}), one has 
$A(\xi)/ (2 \pi)^{1/2} = (\Delta_o/E_o) \exp(-2\xi^2/E_o^2)$. It is 
not surprising to find that the spectral absorption function 
$\chi^{\prime \prime} (\omega) = Im \int dt P_{11}(t)$, corresponding to 
(\ref{4.33}), has the "BCS" form 
\begin{equation}
\chi^{\prime \prime }(\omega ) = 
A(-\xi_H) {8 \Delta_o \over \omega \sqrt{\omega^2-16 \Delta_o^2} }
 \eta (\omega -4 \Delta_o )\;,
\label{4.35}
\end{equation}

Finally, let us note that one may imagine a case where one has an ensemble 
of systems in which, although the bath state is not fixed, the polarisation 
group is known to be equal to 
$M$. In this case we must replace (\ref{biasav}) by 
\begin{equation}
\int d\epsilon \; G_M(\epsilon ) {e^{-\beta \epsilon } \over Z_M(\beta )} 
\label{G_Mav}
\end{equation}
in (\ref{4.32}) for $P_{11}(t)$, where $Z_M(\beta)$ is the partition function 
for the $M$-th polarisation group. If we then recalculate 
$\chi^{\prime \prime} (\omega)$ in the same way we 
find almost zero absorption unless $\vert M \omega_o + \xi_H \vert <
Max(\Delta_o, \tilde{\Gamma}_M)$, where $\tilde{\Gamma}_M$ is again 
the linewidth 
of the $M$-th polarisation group (cf. (\ref{G_M}).

\subsection{Average over transverse fields: Orthogonality blocking}
\label{sec:3D}

Until now we have ignored the "transverse field" part 
$ \sum_{k=1}^N
\omega_k^{\perp}\: {\vec m}_k \cdot {\hat {\vec \sigma }}_k$
of the diagonal term
in the effective Hamiltonian (\ref{1.24}). To study this let us consider
again an effective Hamiltonian which has no non-diagonal terms apart
from the "bare" tunneling, but having all diagonal terms:
\begin{equation}
H_{\mbox{\scriptsize eff}} = 2 \Delta_o \tau _x  +
{\hat \tau }_z \omega_o^{\parallel}
\sum_{k=1}^N  \:
{{\vec l}_k \cdot {\hat {\vec \sigma }}_k } +   \sum_{k=1}^N
\omega_k^{\perp}\: {\vec m}_k \cdot {\hat {\vec \sigma }}_k \;,
\label{4.14h}
\end{equation}
We will assume $ \omega_o^{\parallel} \gg \omega_o^{\perp}$, ie., that the
transverse "orthogonality blocking"
part of the diagonal interaction is much smaller than the
longitudinal part. To make things as simple as possible we drop all 
degeneracy blocking effects, ie., we assume all
$\omega_k^{\parallel}$ are equal (ie., $\mu = 0$). It then follows that  
the spin bath spectrum is split by $\omega_o^{\parallel}$ into $2N$
"polarization groups" of degenerate lines, with
$C^{(N+M)/2}_N$
degenerate states in polarisation group $M$ (cf. eqtn. (\ref{density}).

At first glance it seems that the $\{ {\vec \sigma_k} \}$ in (\ref{4.14h})
simply act on the central spin ${\vec \tau}$ as an external field. However 
this is {\it wrong}; it ignores their role as dynamic quantum variables. The 
dynamics come because the $\{ {\vec \sigma_k} \}$ can {\it precess} in the 
fields $\{ {\vec \gamma}_k\}$ acting on them, and these change each time 
${\vec \tau}$ makes a transition (cf. eqtn. (\ref{1.25})). 
Quantum mechanically, the precession caused by $ \omega_k^{\perp}$ is 
equivalent to saying  
that some bath spins are {\it flipped} when the central spin
$\vec{\tau}$ flips (in general a different number
of them during each transition of $\vec{\tau}$).
To see this formally, recall that $\omega_k^{\perp}$
exists when the initial and final fields ${\vec \gamma}_k^{(1)} $ and
${\vec \gamma}_k^{(2)}$ acting on ${\vec \sigma }_k$ are not exactly
equal and opposite (cf. eqtn (\ref{1.25})). Defining the small angle $\beta_k$
by $\cos 2\beta_k = - {\vec \gamma}_k^{(1)} \cdot  {\vec \gamma}_k^{(2)}/
\vert {\vec \gamma}_k^(1) \vert \vert {\vec \gamma}_k^(2) \vert$ 
(recall Fig. 3), 
we see that the initial and final states of ${\vec \sigma }_k$ are related by
\begin{equation}
\mid {\vec \sigma }_k^f \rangle = {\hat U}_k
\mid {\vec \sigma }_k^{in} \rangle =
e^{ -i\beta_k {\hat \sigma }_k^x }  \mid {\vec \sigma }_k^{in} \rangle \;.
\label{b.2}
\end{equation}

Suppose now the initial spin bath state  belongs to 
polarisation group $M$. 
If, when $\vec{\tau}$ flips,
bath spins also flip so that
$M \to -M$, then since
$E_{\Uparrow}( M)= E_{\Downarrow}( -M)$, {\it resonance}
is still preserved, a 
transition is possible-
indeed $\vec{\tau}$ cannot flip at all {\it unless} there is
a change in polarisation state of magnitude $2M$.
For this change in polarisation
of $2M$, {\it at least} $M$ spins must flip; moreover,
for resonant transitions to continue (incoherently),  
the bath polarisation state must change by $\pm 2M$
{\it each time} ${\vec \tau}$ flips. 

Let us therefore define $P_M(t)$ as the correlation function
$P_{11}(t)$ restricted to systems 
for which the bath polarisation is $M$. For a thermal ensemble, 
\begin{equation}
P_{11}(t;T)= \sum_{M=-N}^{N} w(T,M) P_{M} (t) \;,
\label{4.25x}
\end{equation}
with a weighting $w(T,M)= Z^{-1}C_N^{(N+M)/2}e^{-M\omega_o^{\parallel}/k_BT}$,
where $Z$ is the partition function.

In Appendix A.2 we calculate $P_{M} (t)$ 
(see  Eq.~(\ref{b.21b}) as a {\it weighted average over an orthogonality
variable} $x$:
\begin{eqnarray}
P_{M}(t) &= & \int_0^\infty d\!x x\:e^{-x^2} \bigg( 1+ \cos
[4\Delta_\Phi J_M(2x\sqrt{\kappa })t]  \bigg)
\label{4.21o} \\
&= & 2\int_0^\infty d\!x x\:e^{-x^2}
P_{11}^{(0)}(t,\Delta_M(x))\;,
\label{4.21a}
\end{eqnarray}
where $P_{11}^{(0)}(t,\Delta_M)$ is just the usual {\it free}
2-level correlator (Eq.(\ref{b1.2})), but now with an an $x$-dependent
tunneling amplitude $\Delta_M(x) = 2 \Delta_\Phi J_M(2x\sqrt{\kappa })$;
and $\kappa$ is the {\it "orthogonality exponent"}, defined by
\begin{equation}
e^{-\kappa } = \prod_{k=1}^N \cos \beta_k \;\; 
\sim \;\; e^{-(1/2) \sum_k \beta_k ^2}  \;,
\label{k.1}
\end{equation}
The orthogonality blocking term
$\beta_k$ is 
analogous to the topological decoherence term $\alpha_k$. It is important to 
understand why we must introduce the average over $x$. 
Mathematically, it comes from the 
restriction to a single polarisation group (see Appendix A.2). Physically, it 
corresponds to a phase average just like that in (\ref{4.8}) (compare the 
Bessel functions), but now the phase is that accumulated {\it between}
transitions of the central spin (rather than {\it during} these transitions, 
as in topological decoherence). This phase accumulates if 
$\omega_k^{\perp} \neq 0$, because then the field $\vec{\gamma}_k$ on the 
$k$-th bath spin does not exactly reverse when the central spin flips, and so 
this spin must start precessing in the new field. It is random simply because 
the waiting time between flips is a random variable, in the path integral.  

We shall not give full details for the dynamics of this limiting case
(for which see ref \cite{PSchi95}), but just enough to understand the
physics. First, note that
the terms $P_{M}(t)$ are
easily verified to be incoherent, and so
$P_{11}(t) \sim f P_{M=0}(t) + incoherent$, 
where $f=\sqrt{2/\pi N}$. Even the small fraction $f$ of systems in an 
ensemble having $M=0$ will only have coherent dynamics if $\kappa \ll 1$. The 
easiest way to see this is to again calculate the spectral absorbtion function
from (\ref{4.21o}), to get \cite{}:
\begin{equation}
\chi^{\prime \prime}_{M=0}(\omega) = {\pi f \over 2 \Delta_o \kappa^{1/2} }
\sum_j {x_j e^{-x_j^2} \over \vert J_1 (2 \kappa^{1/2} x_j) \vert } 
\bigg\vert_{J_0(2 \kappa^{1/2} x_j)\;=\; \pm (\omega/2 \Delta_o)}
\label{chikappa}
\end{equation}
which leads, as $\kappa$ increases, to an ever-increasing number of square-root
singularities in $\chi^{\prime \prime}_{M=0}(\omega)$. For $\kappa < O(1)$
only a single root $x_1 \sim [(1- \omega/2 \Delta_o)/\kappa]^{1/2}$ enters, 
and $\chi^{\prime \prime}_{M=0}(\omega)
\sim (\pi f /2 \Delta_o \kappa) e^{[(1- \omega/2 \Delta_o)/\kappa]}$ for
$\omega < 2\Delta_o$, ie., a fairly sharp asymmetric peak at the resonant
frequency of the free spin. For larger $\kappa$ the multiple peaks and tails 
mix, and $\chi^{\prime \prime}_{M=0}(\omega)$ shows no obvious peak- moreover, 
its spectral weight is shifted to ever-lower frequencies:
\begin{equation}
\left.
\begin{array}{l}
P_{M=0}(t)=1/2 [ 1+ \cos (4 \Delta_{eff} t) ]\;\; \\
\Delta_{eff} = 2 \Delta_\Phi e^{-\kappa}
\end{array} \right\} \;\; \kappa \ll 1
\label{4.22}
\end{equation}
\begin{equation}
\left.
\begin{array}{l}
P_{0}(t)= 1  -  4 \Delta_{eff}^2t^2
+O( \Delta_{eff}^4 t^4 ) \;\;  \\
 \Delta_{eff} =  \Delta_\Phi /(\pi \kappa)^{1/4}
 \end{array} \right\} \;\; \kappa \gg 1
 \label{4.23}
 \end{equation}
Notice that this reduction of the transition rate is much slower than  
the usual polaronic or Anderson/Hopfield orthogonality catastrophe,
relevant to oscillator baths, which gives exponential suppression
of $\Delta_{eff}$ for strong coupling.  
This is understood as follows. In oscillator
bath models, band narrowing comes essentially without any bath
transitions (most of the polaron "cloud" is in virtual high-frequency
modes) - it is adiabatic. Here, however, roughly $\kappa$ spins
flip each time $\vec{S}$ flips (the probability of $r$ flips is
$\kappa^re^{-\kappa}/r!$, which peaks at $r \sim \kappa$), even though
we only consider $P_{M=0}(t)$, i.e., even though $\Delta M=0$
(just as many bath spins flip one way as the other).

A further examination of the correlation function shows that the structure of 
$P_{M=0}(t)$ 
is exceedingly bizarre- it was described in detail in
ref. \cite{PRB,PSchi95}. In section 4.A and Fig. 8 we will 
return to the physics
of orthogonality blocking, but including other mechanisms as well (see
also Fig. 7 below).

\subsection{Averaging over spin bath fluctuations}
\label{sec:3E}

The previous 3 averages assume no intrinsic spin bath dynamics- the bath
acquired its dynamics from the central spin. Consider now a Hamiltonian
\begin{equation}
H_{\mbox{\scriptsize eff}}= \Delta \tau _x + \xi \tau_z +
 \tau _z \sum_{k=1}^N \omega_k^{\parallel} \:{\hat \sigma
}_k^z + \sum_{k=1}^N \sum_{k'=1}^N V_{kk'}^{\alpha \beta }
\hat{\sigma}_k^\alpha
\hat{\sigma}_{k'}^\beta \;;
\label{Hzener}
\end{equation}
in which we assume $\vert V_{kk'} \vert \ll \omega_k^{\parallel}$, but  
arbitrary $V_{kk'}/\Delta$. The addition of $V_{kk'}$, to what
would have been a simple degeneracy blocking Hamiltonian, gives the spin bath
its dynamics, and causes 2 changes (as noted previously in our 
introductory presentation of the central spin model). First, 
a polarisation group $M$
acquires an "intrinsic linewidth" $\Gamma_o \sim V_o N^{1/2}$, where
$V_o$ is a typical value of $\vert V_{kk'}\vert$ 
for the $N$ bath spins (of course normally
$\Gamma_o \ll \tilde{\Gamma}_M$, unless the $\{ \omega^{\parallel}_k \}$ 
happen to be extremely tightly bunched together).
Second, the transverse
part of $ V_{kk'}^{\alpha \beta }$ causes pairwise flipping amongst bath
spins (eg., transitions $\vert \uparrow_k  \downarrow_{k'} \rangle
\to \vert \downarrow_k  \uparrow_{k'} \rangle$), at a rate
$\sim NT_2^{-1}$, where $T_2^{-1} \sim V_o$. This "spin diffusion" in the
bath causes the internal bias $\epsilon$ to fluctuate in time, inside the
energy range of the polarisation group $M$, with a random walk correlation 
$\langle [\epsilon(t)-\epsilon (t')]^2 \rangle =
\Lambda^3 \vert t-t' \vert$, where $\Lambda^3 =
\tilde \Gamma^2 T_2^{-1}$,  
for timescales $\Lambda \vert t-t' \vert \ll 1$. 

It thus follows that for a given {\it single} central spin, with its 
surrounding spin bath in polarisation group $M$, the problem reduces to 
calculating the dynamics of a 2-level system in a longitudinal bias field 
$(\xi + \epsilon (t))$, where $\xi$ is the applied bias, and the internal 
field is $\epsilon (t) = M\omega_o + \delta \epsilon (t)$. The correlation 
properties of $\delta \epsilon (t)$ are those just described- our task is 
simply to functionally average over these fluctuations in the calculation
of $P_{11}(t)$. In doing this we will make the physically sensible assumption
of "fast diffusion" of 
$\delta \epsilon (t)$, such that the time $\Delta t$ it takes for the bias to 
diffuse across the "resonance window", of energy width $\Delta$, satisfies 
$\Delta t \ll 1/\Delta$. Then 
the system has no time to tunnel coherently, but can only
make an incoherent "Landau-Zener" transition. Since the bias changes by
$\delta \epsilon \sim \delta \omega_o (N/(T_2\Delta))^{1/2}$ in a time 
$1/\Delta$, this formally requires that 
\begin{equation}
\Delta^3 \ll \Lambda^3 \equiv \tilde \Gamma^2 T_2^{-1}
\;\;\;\;\;\;\;\;\;\;\;\; (fast \;\; diffusion).
\label{fdiff}
\end{equation}
This problem is solved in Appendix A, by performing a {\it weighted average
over dynamic bias fluctuations}, with the restriction that these only 
occur inside polarisation group $M$; the relevant average is 
\begin{equation}
\int {\cal D}\epsilon(t)\; e^{-{1 \over 2} \int dt_1 \int dt_2
 K (t_1 - t_2) \epsilon (t_1) \epsilon(t_2)} \;;
\label{T_2fl}
\end{equation}
where $2K^{-1}(t_1 - t_2) = \Lambda^3 (\vert t_1 \vert +
\vert t_2 \vert - \vert t_1 - t_2 \vert)$ is the correlator of the dynamic
spin bath fluctuations (see Appendix A). 
One finds that $P_{11}(t)$ decays as 
a simple exponential $P_{11}(t) = e^{-t/\tau_M(\xi)}$, where 
\begin{equation}
\tau^{-1}_M = 2 \pi^{1/2} {\Delta^2 \over \tilde \Gamma} 
e^{-(\xi + M\omega_o)^2/\tilde \Gamma^2}
\label{xizero}
\end{equation}
This result is easily understood- the bias fluctuations can cause 
the system to 
pass briefly through resonance (allowing a transition of the 
central spin), but only if the net static bias $\xi + M\omega_o$ is not 
greater
than the range $\tilde \Gamma$ of the fluctuations.
By comparing with the case of pure degeneracy blocking we see that 
the important role of the bath dynamics is (i) to {\it unblock}
the central spin dynamics, by helping it to find resonance, now over 
an energy window of width 
$\tilde \Gamma$ (instead of $\Delta_o$) around zero bias (recall Fig. 4), 
and (ii) to change the central spin dynamics from coherent to 
incoherent tunneling.

Note that in a model like (\ref{Hzener})
we have eliminated bath fluctuations between 
different polarisation groups. 
The basic assumption is that any "$T_1$ processes" in the intrinsic 
bath dynamics, which could change $M$ in the absence of the central spin, 
are very slow (At low $T$,
$T_1$ for nuclear or paramagnetic spins does become extremely long). However
this is not always realistic- we return to this point in section 5.A. 
If $T_1$ is short, 
one must make a dynamical average
over 2 kinds of fluctuation, usually with quite different time correlation,
viz., the intra-polarisation group fluctuations, described by (\ref{T_2fl}) and
with correlation time $NT^{-1}_2$, and
the inter-polarisation group fluctuations, occuring on a timescale $T_1$ (cf.
\cite{rose99,rose99a}).

\subsection{Averaging over the Spin Bath: General Results}
\label{sec:33C}

We now turn to the problem of averaging over the spin bath for the general 
form of the Central Spin Hamiltonian given in eqtn. (\ref{1.24}). This can be 
given in the form of a marvellously simple prescription- one simply applies the 
4 averages we have just seen, to the problem of a simple biased 2-level
system! We begin by giving the explicit prescription (whose proof is given
in Appendix B), and make a few comments on it.

The prescription begins with the following 4 averages (all of which we 
have seen in the preceding 4 sub-sections):
\begin{equation}
\hbox{ (a)  A "topological phase average" } \;\;\;\;\;\;
\sum_{\nu =-\infty }^{\infty} F_{\lambda '} (\nu )
\int {d\varphi \over 2 \pi } e^{i2\nu (\Phi -\varphi )} \;;
\label{q.5}
\end{equation}
\begin{equation}
\hbox{ (b) An "orthogonality average" } \;\;\;\;\;\;
2 \int_0^\infty  dx x e^{-x^2} \;;
\label{q.6}
\end{equation}
\begin{equation}
\hbox{ (c) A "bias average" } \;\;\;\;\;\;
\int d\epsilon \;G_M(\epsilon ) {e^{-\beta \epsilon } \over Z_M(\beta )}
\;\;\;\;\;\;\;\;\;\;\;OR\;\;\;\;\;\;\;\;\;
\int d\epsilon \; W(\epsilon ) {e^{-\beta \epsilon } \over Z(\beta )}\sum_M\;;
\label{q.7}
\end{equation}
\begin{equation}
\hbox{ (d) A "bath fluctuation average" } \;\;\;\;\;\;
\int {\cal D}\epsilon(t)\; e^{-{1 \over 2} \int dt_1 \int dt_2
 K (t_1 - t_2) \epsilon(t_1) \epsilon(t_2)} \;;
\label{q.7a}
\end{equation}
As before, we assume a thermal
distribution over spin bath biases, with a corresponding partition function 
and $Z(\beta )$. All
averages are normalized to unity.
The weighting  function 
$F_{\lambda '} (\nu )=e^{-4\lambda ' \nu^2}$ in (\ref{q.5}) is a generalisation
of that in (\ref{4.7}), to allow for arbitrary directions of the unit vector
$\hat{\vec{n}}_k$; we now define 
\begin{equation}
\lambda ={1 \over 2}\sum_k \vert\alpha_k\vert^2 (1-(n_k^z)^2)\;,\;\;\;\;\;\;\;
\lambda '={1 \over 2} \sum_k  \alpha_k^2 (n_k^z)^2 \;,
\label{l.2}
\end{equation}

Now, suppose we want to calculate $P_{11}(t)$. 
The prescription is fairly
obvious in the light of the results given above for the 
4 limiting cases. One follows the following steps: \\

\indent (i) Begin with the quantity
\begin{equation}
P_{11}^{(0)} (t; \Delta_M(\varphi ,x); \epsilon )=
1-{ \Delta_M^2(\varphi ,x) \over E_M^2(\varphi ,x) }
\sin ^2 ( E_M(\varphi ,x) t ) \;,
\label{q.8}
\end{equation}
which is just the {\it free} central spin correlator (cf.
(\ref{b1.10})) having tunneling matrix element $\Delta_M(\varphi
,x)$,
and in an "internal field" 
bias $\epsilon$. The energy splitting $E_M$ is given by
$E_M^2(\varphi ,x) = \Delta_M^2(\varphi ,x) +\epsilon^2$, 
and the matrix element $\Delta_M$ is
\begin{equation}
\Delta_M(\varphi ,x) =2{\tilde \Delta}_o \vert \cos ( \varphi )
J_M(2x\sqrt{\gamma }) \vert  \;,
\label{5.8a}
\end{equation}
\begin{equation}
\gamma =  \left\{
\begin{array}{ll}
\lambda & \;\;\;\;\;\mbox{if
$\lambda  \gg \kappa$ (topological decoherence regime)} \\
\kappa &\;\;\;\;\;\mbox{if
$\kappa \gg \lambda $ (orthogonality blocking regime)}
\end{array} \right.  \;.
\label{q.5b}
\end{equation}
We defined $\kappa$ previously (eqtn. (\ref{k.1}).
We will not give results for the case $\kappa \sim \lambda $; they
are extremely complex, do not appear to add new physics, and
seem unlikely to be realised in practice. \\

\indent (ii) Now carry out the averages over topological phase
[Eq.~(\ref{q.5})] and orthogonality [Eq.~(\ref{q.6})], to give an
expression $P_M(t,\epsilon )$ describing the central spin dynamics
in a bias $\epsilon$, coming from a bath in polarisation state $M$: 
\begin{equation}
P_M(t;\epsilon ) = 2 \int_0^\infty  dx x e^{-x^2}
\sum_{\nu =-\infty }^{\infty} F_{\lambda '} (\nu )
\int {d\varphi \over 2 \pi } e^{i2\nu (\Phi -\varphi )}
\bigg[ 1-{ \Delta_M^2(\varphi ,x) \over E_M^2(\varphi ,x) }
\sin ^2 ( E_M(\varphi ,x) t ) \bigg]               \;,
\label{q.11}
\end{equation}
where the weighting  function is
$F_{\lambda '} (\nu )=e^{-4\lambda ' \nu^2}$
over winding number $\nu$ (recall eqtn. (\ref{4.7})); \\

\indent (iii) Then, carry out the bias average [Eq.~(\ref{q.7})]. We will 
assume in the following for definiteness an ensemble average over all
polarisation groups, thereby ensuring a summation over $M$, to give
\begin{equation}
P_{11 }(t;T ) = 1-
\int d\epsilon W(\epsilon ) { e^{-\beta \epsilon }
\over Z(\beta )} \sum_{M=-N}^N
\left( 1- P_M(t,\epsilon -M\omega_o ) \right)  \;;
\label{6.25x}
\end{equation}
This result summarizes the central spin dynamics in the case where the spin 
bath has no dynamics of its own, and only acquires dynamics through its 
interaction with the central system. In some cases there will be no intrinsic 
bath dynamics, and this will be the final answer. If we wish to apply the 
theory to a single central system, or for some reason we can fix the 
polarisation group to be a definite value $M$, then we drop the summation 
over $M$ in (\ref{6.25x}), and replace $W(\epsilon)$ by $G_M(\epsilon)$. \\

\indent (iv) When the interaction term $V_{kk'}$ plays a role, we apply the 
4th average (\ref{q.7a}) to (\ref{6.25x}), as described in Appendix A 
(cf. also the discussion in ref. \cite{PS96}). This gives
the completely incoherent form
\begin{equation}
P_{11}(t) = \sum_M w(T,M) \int xdx e^{-x^2}
\sum_{\nu=-\infty}^{\infty} \int {\varphi \over 2\pi } F_{\lambda
'}(\nu) e^{i2\nu (\Phi -\varphi )} \big[ 1+e^{-t/\tau_M(x,\varphi) } \big]
\;,
\label{diff.2}
\end{equation}
where the relaxation rate $\tau^{-1}_M(x,\varphi)$ is given by
\begin{equation}
\tau_M^{-1}(x,\varphi) = 2\Delta_M^2(x,\varphi)\int d\epsilon G_{\mu}(\epsilon )
\int_{0}^{\infty}ds e^{i\epsilon s} ~e^{-\Lambda^3 s^3/6} 
= 2\Delta_M^2(x,\varphi)
\int_{0}^{\infty}ds e^{-(\mu \omega_o)^2 s^2/4} ~e^{-\Lambda^3 s^3/6} \;;
\label{diff.7}
\end{equation}
with $\Delta_M(x,\varphi)$ given by (\ref{5.8a}), and 
where $G_{\mu} (\epsilon )$ is a Gaussian of width $\tilde \Gamma = \mu
\omega_o$. Since $ \tilde \Gamma \gg T_2^{-1}$ we have also $\tilde \Gamma \gg
\Lambda$, and so we get 
\begin{equation}
\tau_M^{-1}(x,\varphi) = {2 \Delta_M^2(x,\varphi)\over
\pi^{1/2} \tilde\Gamma}\;,
\label{diff.3}
\end{equation}
This result is the most general one for the dynamics of the central
spin, if all 4 bath averages are included- it is generally valid, with
only the single restriction that the diffusion of the fluctuating bath
bias in energy space be {\it fast} (cf. eqtn. (\ref{fdiff}). In the
absence of such fluctuations we go back to (\ref{6.25x}). 

In essentially all physically realistic cases 
the different polarisation groups strongly
overlap. It is then simpler \cite{PS96} to 
transform the sum over $M$ in (\ref{6.25x}) or (\ref{diff.3}) into an
integral over energy bias $\xi$ , using the change of variables 
$\sum_{M} \rightarrow \int d \xi/2\omega_o$, and then integrate over
$\xi$. One way to do this (using
steepest descents) was
detailed in ref. \cite{PS96} (compare eqtns (4.41)-(4.47) in that paper).
For exact answers one can use the identity
\begin{equation}
\int_{0}^{\infty} xdx e^{-x^2} J_M^2(2x\sqrt{\gamma}) \cos^2 \phi\;\; =\;\;
I_M(2\gamma) e^{-2 \gamma} \cos^2 \phi
\label{bessel}
\end{equation}
to evaluate either (\ref{6.25x}) or (\ref{diff.2}).

In the next section we will evaluate $P_{11}(t)$ and its Fourier
transform for a number of different parameter regimes. But 
even before doing the integrals, the qualitative
behaviour is obvious. Relaxation is only occurring for
central spins which happen to be within a bias $\xi_o$ of exact
resonance. The width $\xi_o$ of this "resonance window" is coming from
the energy which the bath spins can provide to the central spin, by
flipping up to $\sim \gamma$ bath spins; hence $\xi_o \sim \gamma
\omega_o$ (formally this is obvious from the properties of the Bessel
functions in (\ref{5.8a}) and (\ref{bessel}), which fall off very fast
once $M > \gamma$). We show graphically the relaxation of different groups
in Fig. 7(a); again one sees how only groups with $M \leq \gamma$ relax 
quickly. The $T_2$ bath fluctuations help this process by bringing the
a central spin in polarisation group $M$ to its resonance
window (of width $\Delta_M$). Only transitions of systems having $M=0$ 
can show (partial) coherence; all transitions with $M \neq 0$ are 
essentially {\it incoherent}. Note that the resonance window will {\it not}
be visible in a resonant absorption experiment (Fig. 7(b)); higher $M$
groups contribute only a very low frequency contribution to this. This
nicely demontrates that one is very far from any linear-response regime
in the present system (so that, e.g., the fluctuation-dissipation theorem
is somewhat irrelevant here).

Without the bath, transitions of the
central spin would be coherent, but over the far smaller
resonance window of width $\sim \Delta_o$. If we only had $T_2$ bath
fluctuations, but $\gamma \ll 1$ (ie., no bath spins flipped during
the transitions of the central spin), then we would again get
incoherent relaxation, but this time with $\xi_o \sim \tilde \Gamma_{M=0}$ 
(the width of the $M=0$ polarisation group). One can also imagine a situation 
in which $T_1$ is very short, so that all polarisation groups are 
involved in the relaxation, and the resonance window is just the distribution
$W(\xi)$, with $\xi_o = E_o = N^{1/2}\omega_o$. 

These results thus tell us that in the presence of a spin bath, any
ensemble of central spins, initially spread over a range of biases, 
will start relaxing by digging a "hole" of
width $\xi_o$ around zero bias \cite{PS96,PS98}. 
This hole reflects the {\it intrinsic} 
central spin dynamics (ie., it is not being produced by
interaction with some external resonant signal- it should not be
confused in any way with the "spectral hole-burning" done by experimenters
working on glasses or in optics, using an external source). As
discussed in a number of papers \cite{PS96,PS98}, evaluation of
(\ref{diff.2}), using either steepest descents or other means, shows that
the system under most conditions relaxes incoherently with a
relaxation rate (after summing over all polarisation groups, and doing
the orthogonality and phase integrals) given approximately as a
function of bias $\xi$ by
\begin{equation}
\tau^{-1}(\xi)\; =\; \tau_o^{-1} e^{-\vert \xi \vert/\xi_o} 
\; \equiv \; {2 \Delta^2 \over
\pi^{1/2} \tilde\Gamma} e^{-\vert \xi \vert/\xi_o}
\label{xi_o}
\end{equation}
All of this is in complete contrast to how inelastic tunneling
works in the presence of an oscillator bath \cite{ajl87}; there the
relaxation rate typically {\it increases} as one moves away from
resonance, usually as a power in bias ($\tau^{-1}(\xi) \sim \xi$ for
diagonal coupling to phonons, $\sim \xi^3$ for non-diagonal coupling to
phonons, and $\sim \xi^{2 \alpha -1}$ 
for diagonal coupling to Ohmic baths 
like electrons via a
dimensionless coupling $\alpha$). Thus one does not expect hole-digging for 
oscillator bath-mediated quantum relaxation, except over a very 
narrow region of width $\sim \Delta_0$.

Finally, we note that one may also give a formal prescription 
for the case where some central spin couples simultaneously to an 
oscillator bath {\it and} a spin bath. We do not give the details here- they
are discussed fairly exhaustively (along with the results for the central spin 
dynamics) in ref. \cite{PS96} (see also end of section 4).

\section{Dynamics of the Central Spin}
\label{sec:4}

Given the large number of parameters entering into the 4 averages just 
described, we see little point in an exhaustive description of 
$P_{11}(t)$ over the whole parameter domain (for more extensive results 
see refs. \cite{PRB,PS96}). Instead we concentrate on
3 points. First, we show how in the strong coupling regime, coherence 
is destroyed, leaving incoherent quantum relaxation; this regime applies to 
almost all mesoscopic
or nanoscopic magnetic systems, because of their coupling to nuclear spins
and to paramagnetic impurities. 
Second, we discuss the physics of the
weak coupling regime (applicable to, eg., SQUIDs), and how in one 
limit of this regime 
one may 
formally map the spin bath to an oscillator bath. Finally, and very briefly, 
we comment on the results obtained when one couples simultaneously to a 
spin bath and an oscillator bath.

\subsection{Strong coupling regime}
\label{sec:4A}

As already explained, the strong-coupling regime is defined by the
condition $\omega_k^{\parallel}$ and/or $\omega_k^{\perp} \geq \Delta_o$. 
This condition applies to virtually all
situations in which the couplings are hyperfine ones to 
nuclear spins, or exchange couplings to paramagnetic spins; and  
also when one has dipolar couplings 
to paramagnetic impurities or defects.

Almost all interesting physical examples in this regime fall either
into the catagory of ``strong orthogonality blocking" (when $\kappa
\gg \lambda'$) or
strong ``phase decoherence" (when $\lambda \gg \kappa$).
In both cases the central system
makes transitions accompanied by flips in the bath spins- so that even if 
the isolated central system is not in resonance, it can "find resonance" by 
using the flipped bath spins to make up the energy difference. If the 
couplings are such that roughly $\kappa$ bath spins flip, the range of energy 
bias over which transtions can occur is extended to 
roughly $2\kappa \omega_o$. The 
central system is helped in this task by the fluctuations in bath bias caused
by the interspin interactions $V_{kk'}$.

In what follows we concentrate on the physics
of {\it decoherence} in this regime, with an eye to the physics of
"qubits" and of "macroscopic quantum coherence". We also look at 
the form of the relaxation. We begin by 
explaining the results without the bath fluctuations, and then show what 
happens on adding these.

\vspace{4mm}

{\bf (i) Results without bath fluctuations}: In this case we must evaluate
(\ref{6.25x}) and (\ref{q.11}), suppressing either the orthogonality
average or the topological average. In what follows we look at each case 
in turn, focussing particularly on the $M=0$ polarisation group contribution.

\vspace{3mm}

{\it (a) Orthogonality Blocked regime ($\kappa \gg \lambda$)}.  
In this regime only the orthogonality average $2\int xdx e^{-x^2}
$, and the average over bias $\epsilon$, are relevant- the phase average
is approximated by a delta-function. The presence of the $x$-dependent 
transition matrix element
$\Delta_M(x)=2\Delta_o J_M(2x\sqrt{\kappa })$ means that
polarisation groups with $M \sim \kappa$ play a dominant role. Suppose 
however that we are interested in any coherent dynamics of the central spin-
what will be found? It is obvious that transitions with finite $M$ will
be essentially incoherent, so we concentrate on central spins for which 
$M=0$. Thus we simply  
integrate $P_M (t, \epsilon )$ over $\epsilon $, in the weighted
bias average, to get 
\begin{eqnarray} 
P_{11}^{M=0} (t) &=& 1- 4 A  \int dx xe^{-x^2}  
\mid J_0(2x\sqrt{\kappa }  ) \mid  \sum_{k=0}^{\infty} J_{2k+1} 
\big[ 4 \Delta_o  \mid J_0(2x\sqrt{\kappa } ) \mid t  \big] 
\label{5.18} \\
&=& 1-  2 \int dx xe^{-x^2}  2A(x)  \sum_{k=0}^{\infty} J_{2k+1} 
\big[ 2 \Delta_0 ( x ) t  \big] \;,
\label{5.18b}
\end{eqnarray}
where the $x$-dependent spectral weight is 
$A(x) =A \vert J_0(2x\sqrt{\kappa } ) \vert$. 
Notice we have just done an "orthogonality average" over a 
"biased averaged" expression for the free system with $x$-dependent 
tunneling frequency $\Delta_0(x)$. A Fourier transform 
to frequency space (which is
essentially a picture of the relaxation rate, as a function of {\it
energy bias} $\xi$ for this system) gives
the absorption spectrum
\begin{equation} 
\chi_{M=0}^{\prime \prime }(\omega ) = 
{ 1 \over \omega } \int dx xe^{-x^2} 4A(x)
{ \mid \Delta_o(x)\mid \over 
[ \omega^2 - 4 \Delta_0 ^2 (x)  ]^{1/2} } 
\eta ( \omega - 2 \mid \Delta_o  (x) \mid ) \;,
\label{5.19}
\end{equation}

Fig. 8 shows some representative plots for this "coherent" part
of $\chi^{\prime \prime }(\omega )$; it is in fact {\it almost completely 
incoherent}, with total spectral weight
\begin{eqnarray}
\int_{-\infty}^{\infty} (d\omega /2\pi ) 
\chi^{\prime \prime }(\omega )
 & =& 2A  \int dx xe^{-x^2} \mid J_0(2\sqrt{\kappa } x)\mid 
\nonumber \\
&=& {2\Gamma(3/4) \over \pi^{3/2} }{A \over \kappa^{-1/4} } \;;
\label{5.21}
\end{eqnarray}
a result which is very accurate even for $\kappa \sim 0.02 $. Note
that the shape of $\chi^{\prime \prime}(\omega)$ will change 
once we include all other (incoherently relaxing) 
polarization sectors $M \ne 0$, and its total
weight increases - in fact the total weight is $\sim A \kappa^{1/4} $ for
large $\kappa$, since $\sim \kappa^{1/2}$ different polarization sectors
contribute. The absorption in $\chi^{\prime \prime }(\omega )$
from these higher $M$ groups will be concentrated at frequencies $\ll
\Delta_o$ (compare also Fig 7(b)). Note however that relaxation itself
(not described by the linear response function 
$\chi^{\prime \prime}
(\omega )$)
will be spread incoherently over a frequency range
$\xi_o \sim 2\kappa \omega_o$, ie., the 
"hole-digging" in the relaxation occurs over a window of width $\xi_o \sim 
2\kappa \omega_o$.

{\it (b) Phase decoherence regime ($  \lambda \gg \kappa$)}.
Let us now suppose the transverse field terms $\{ \omega_k^{\perp} \}$ 
are negligible compared to the $\{ {\vec \alpha_k} \}$. We then  
deal with an effective Hamiltonian 
\begin{equation}
H_{\mbox{\scriptsize eff}}  = 2 \Delta_o \left\{ {\hat \tau }_{-}
\cos \big[ \Phi   -i \sum_{k=1}^N
 \alpha_k {\vec n}_k \cdot {\hat {\vec \sigma }}_k  \big]
 + H.c. \right\}
+ {\hat \tau }_z \sum_{k=1}^N  \omega_k^{\parallel} 
  \: {\hat \sigma }_k^z  \;.
\label{5n.12}
\end{equation}
Since $\omega_k^{\parallel} \gg \Delta_o$ by assumption, 
energy conservation requires that environmental spins 
flip with the central spin, just as in our
discussion of pure orthogonality blocking. Thus in this case {\it we must
also keep the orthogonality average}, to enforce this constraint, ie., we must
perform the full average embodied in eqtns. (\ref{q.11}),(\ref{6.25x}).
The full answer, including both the real and imaginary
parts of $\alpha_k$, is rather complicated, and
is presented in Appendix B. Here we will consider the more
transparent answer one gets when $\alpha_k$ is purely imaginary and
adds to directly as a random variable to the central spin phase.

Let us again start with only the $M=0$ contribution to $P_{11}(t)$.
Then we have 
\begin{equation}
 P_{11 }^{M=0}(t ) = \int d\epsilon W(\epsilon ) 
{ e^{-\beta \epsilon } \over Z(\beta )} P_0(t,\epsilon )  \;;
\label{5n.13}
\end{equation}
with $P_0(t,\epsilon )$ given by (\ref{6.25x}) with $M=0$. We may now carry
out the integration in (\ref{5n.13}), assuming that $W(\epsilon )$ is given
by the usual Gaussian form (\ref{gauss}), to get 
\begin{equation}
P_{11}^{M=0} (t) = 1 -2 \int_0^\infty  dx x e^{-x^2} 
\sum_{\nu =-\infty }^{\infty} F_{\lambda^\prime } (\nu ) 
\int {d\varphi \over 2 \pi } e^{i2\nu (\Phi -\varphi )} 2
A(\varphi ,x) \sum_{k=0}^{\infty} J_{2k+1} 
\big[ 2 \Delta_0 (\varphi , x ) t  \big] \;;
\label{6.25z}
\end{equation}
with $A(\varphi ,x) =A \cos \varphi \: J_0(2x\sqrt{\lambda })$.
The corresponding absorption $\chi^{\prime \prime }(\omega )$ is 
\begin{equation}
\chi_{M=0}^{\prime \prime }(\omega ) = 
{ 2 \over \omega } \int dx xe^{-x^2} 
\sum_{\nu =-\infty }^{\infty} F_{\lambda^\prime }(\nu ) 
\int {d\varphi \over 2 \pi } e^{i2\nu (\Phi -\varphi )}\: 
{ A(\varphi ,x) \Delta_0 (\varphi , x ) \over 
[\omega^2 - 4 \Delta_0^2 (\varphi , x )]^{1/2} }  \;.
\label{6.27z}
\end{equation}

It is possible to write analytic expressions starting from (\ref{6.27z}), but
in this somewhat pedagogical presentation we simply discuss the case 
when $\mu =0$, i.e., zero degeneracy blocking, when $\omega_k =\omega_o$ for 
all nuclei. The integration over bias  is then
absent (since $W(\epsilon )$ is now just a set of
$\delta$-function peaks, ie.,  $W(\epsilon ) \rightarrow \sum _{M=-N}^{N}
C_N^{(N+M)/2} \delta (\epsilon - M \omega_o)$), and we get
\begin{equation}
 P_{11}(t ) = \sum_M w(T,M)   P_M(t ) 
\;;~~~\;\;\;\;\;\;\; w(T,M)=C_N^{(N+M)/2} e^{-M\omega_o/T}/Z(\beta ) \;,
\label{5n.17}
\end{equation}
where $P_M(t )$ now 
describes the dynamics in zero bias; it is given by exactly the same weighted 
average over phase as in (\ref{4.8}):
\begin{eqnarray}
P_{M}(t) &=& \int dx x e^{-x^2} \sum_{m=-\infty }^{\infty} 
F_{\lambda^\prime}(m) \int {d\varphi  \over 2\pi }
e^{i2m(\Phi -\varphi ) }  \left\{ 1+ 
\cos [4 \Delta_o t J_M(2x\sqrt{\lambda }) \cos \varphi ] \right\} 
\label{5.6a} \\
& =& \int dx x e^{-x^2}  \left\{ 1+ \sum_{m=-\infty }^{\infty} (-1)^{m}
F_{\lambda^\prime}(m)
e^{i2m\Phi } 
J_{2m} [4\Delta_o t J_M(2x\sqrt{\lambda })] \right\} \;,
\label{5.6}
\end{eqnarray}
This can be interpreted {\it either} 
as an orthogonality-blocked expression, with 
frequency scale
$\Delta_M(\varphi ,x)=2\Delta_o  \cos ( \varphi )
J_M(2x\sqrt{\lambda })$  
which is then {\it averaged over $\varphi $}, to give phase randomisation;
{\it or} as an integration $\int dx$ 
over an {\it already topologically decohered}
function having frequency scale  
$\Delta_M (x) = 2\Delta_o J_M(2x\sqrt{\lambda })$. It is intuitively obvious
(and easily demonstrated) that only
$P_0(t)$ may behave coherently, with a fractional     
weight $\sim \sqrt{2/\pi N}$ in an ensemble.

There are various interesting cases of (\ref{5.6}) for $M=0$. 
If $\lambda =0$ (i.e., ${\vec n}_k $ is parallel to 
${\hat {\vec z}}$), then we go 
back to pure topological decoherence - the projection operator then commutes
with the cosine operator. On the other hand if $\lambda^\prime = 0$, we
have pure orthogonality blocking as stated earlier, and in fact 
when $\lambda^\prime = 0$, the parameter $\lambda $ plays 
the role of $\kappa$ in (\ref{4.21o}). Notice that whereas the case $\lambda
=0$ can only occur accidentally, $\lambda^\prime = 0$ is 
quite common -  indeed it pertains to the model in Eqs.
(\ref{2.8}).

We really begin to see the analogy between orthogonality blocking 
and topological decoherence when
 $\lambda ,\:\lambda^\prime \gg 1$;
just as with pure topological decoherence, $F_{\lambda^\prime}(m)$ collapses
to a Kronecker delta, and we get the {\it universal} projected
topological decoherence form:
\begin{equation}
P_{11}^{M=0} (t)  \longrightarrow \int dx x e^{-x^2}
 \biggl[ 1+J_0[4 \Delta_o tJ_0(2x\sqrt{\lambda })] \biggr] \;\;\;\;(\mu =0)\;;
\label{5.9}
\end{equation}
\begin{equation}
\chi_{M=0} ^{\prime \prime } (\omega ) 
\longrightarrow \sqrt{ {2 \over \pi N }} \int dx x e^{-x^2}
  {4 \over [16 \Delta_o^2 J_0^2(2x\sqrt{\lambda }) -\omega^2]^{1/2}}\: 
  \eta (4 \Delta_o \mid 
J_0(2x\sqrt{\lambda}) \mid -\omega ) \;,
\label{5.10}
\end{equation}
which generalizes the result of (\ref{4.13a}) for pure topological
decoherence. Eq.(\ref{5.9}) should be compared to (\ref{4.21o}).

We show in Fig.9 some results for 
$\chi ^{\prime \prime } (\omega )$ for selected values of 
$\lambda $. The results are startling; even a very small value of 
$\lambda $ significantly  washes out pure topological
decoherence; but for any large value of $\lambda^\prime $, we never get 
back the pure orthogonality blocking spectrum. 

The results in the case where $\alpha_k$ is real, and $\hat{\vec n}_k$ is
along the $\hat{\vec{x}}$-direction (see Eq.~(\ref{z.29})) are 
obtained by simply converting the Bessel functions $J_m$ to Bessel
functions of imaginary argument $I_m$. 

Finally, note again that the above discussion of the $M=0$
polarisation group is irrelevant to the real experimental
lineshape- an evaluation of the full expression (\ref{6.25x}), summing
over all $M$, for
the $\lambda, \lambda' \gg 1$ regime, simply gives incoherent
relaxation \cite{PS96,PS98}, spread over a frequency range 
$\xi_o \sim  \lambda \omega_o$.

\vspace{5mm}

{\bf (ii) Including Bath Fluctuations:} 
The modification of the above results, occasioned by the intrinsic spin bath 
fluctuations, was given in detail in Prokof'ev and Stamp \cite{PS96}. The
fluctuations in bias allow the central system 
to cycle rapidly through the whole 
range of biases within a given polarisation group (transitions between 
{\it different} polarisation groups can occur through $T_1$ 
processes- usually much slower). Here we simply 
recall the main result, which is obtained by summing the relaxation forms
from each polarisation group in an ensemble (cf. eqtns. (\ref{6.25x}) and 
(\ref{diff.2})), and assuming that the spin bath $T_1$ is longer than all
experimental times scales.

For a single central system, coupled to a spin bath in equilibrium at
temperature $kT \gg \omega_o$, one finds that after an initial short-time
transient, the relaxation is roughly
logarithmic over a very long period; in fact one finds for the strong
coupling regime that
\begin{equation} 
1-P(t,\xi_H=0) \sim  \sqrt{ {1 \over2\pi  N} }
 { \ln ( t/ \tau_o)  \over 
 \ln \bigg[ {1 \over e\sqrt{\gamma} } \ln 
( t/ \tau_o)\bigg] } 
\;; \;\;\;\;\;\;\;\;\;\;\;(t\ll t_c)  \;,
\label{JLTP.446}
\end{equation}
for times $t \ll t_c$, where $t_c \sim \tau_o (2 N/e^2 \gamma)^{\sqrt{2N}}$, 
and $\tau_o= 2 \Delta^2/\pi^{1/2} \tilde \Gamma$ is the relaxation
time of the $M=0$ polarisation group (compare eqtn. (\ref{diff.3})); thus
$t_c$ is extremely long!
For $t \gg t_c$ the system settles down to a rather different
behaviour \cite{PS96}. 
This logarithmic 
behaviour can be {\it roughly} understood \cite{PS96} as coming from a 
distribution of barrier heights, for the different polarisation groups, 
which are then summed over- as discussed in ref. \cite{PS96}, section 4.3(b), 
the final result looks basically the same as that shown in Fig. 7(a). 
The fastest relaxation 
comes from the those polarisation groups in the "resonance window" (recall
the discussion at the end of section 3). 

Two cautionary notes are in order here. First, (\ref{JLTP.446}) applies to a 
{\it single} relaxing system- but in the case of nanomagnetic systems, 
all experiments until now have been done on large numbers of nanomagnets,
coupled together via 
long-range dipolar forces, which drastically changes the relaxation (see
section 5.A below).
Second, (\ref{JLTP.446}) should
not be applied uncritically to experiments, even on single quantum systems.
This is because in a real experiment there will also be (i) couplings to 
oscillator baths, and (ii) the relaxation will change once $t \geq T_1$. 
In superconductors or metals, electronic oscillator baths will often 
dominate the relaxation \cite{ajl87},
even at short times (for their effect on {\it coherence},
see section 5.C below). Even in insulating systems,
one eventually expects the coupling to phonons 
to take over at long times \cite{PS96}, since this causes exponential
decay- even if very slow, this will eventually become faster than the 
spin bath-mediated logarithmic decay in (\ref{JLTP.446}).

\subsection{The weak coupling regime; relation to the oscillator bath}
\label{sec:4B}

A question of considerable theoretical 
(and practical) interest is the transition to
the weak coupling regime, where the perturbation on the central system
dynamics by a single bath spin is small (even though the net effect of all bath
spins, measured by parameters like $\lambda$ or $\kappa$, may still be large
if $N$ is very large).
The weak-coupling regime is thus defined by the condition 
$\omega_k \ll \Delta_o$, ie., both $\omega_k^{\perp}$ {\it and} 
$\omega_k^{\parallel}$ are $\ll \Delta_o$. Again, a variety of cases is 
possible depending on how large are ratios like $V_{kk'}/\Delta_o$ and 
$\omega_k^{\perp}/\omega_k^{\parallel}$, or parameters like $\lambda$ and
$\kappa$. In the following we will not be exhaustive, but simply 
consider two theoretically interesting cases, in which $V_{kk'}$ is 
assumed negligible, and we look at the limiting behaviour 
arising when either $\omega_k^{\perp}/\omega_k^{\parallel} \ll 1$
or $\omega_k^{\perp}/\omega_k^{\parallel} \gg 1$. We will also assume  
$N \gg 1$, otherwise the problem is trivial (the bath has little effect at 
all).

There are 2 ways to solve for the dynamics in this regime. One is to use the 
averaging already developed above- this simplifies considerably in the weak-
coupling regime. The other is to map the problem onto an oscillator bath one,
and then use standard techniques to solve this. We demonstrate the 2 methods by
solving one problem with each.

\vspace{3mm}

(i) {\it Longitudinally dominated case} 
($\omega_k^{\perp}/\omega_k^{\parallel} \ll 1$):
We assume the same Hamiltonian as in the discussion of the phase decoherence
regime (eqtn. (\ref{5n.12})), but now we can drop the orthogonality average-
the projection to a single polarisation group is not required since
$\omega_k^{\parallel} \ll \Delta_o$.
Again, for simplicity we consider the case when
$\alpha_k $ is imaginary.

We shall solve this using the techniques previously developed; we let
$x=0$ in (\ref{5.8a}), and hence use a matrix element 
$\Delta (\varphi) = 2 \Delta_o \cos \varphi $ (which is 
{\it independent} of $M$). All polarisation groups overlap, and so we 
simply average over bias and topological phase:
\begin{eqnarray} 
P_{11}(t) &=& 1 - 
\int d\epsilon W(\epsilon ) { e^{-\beta \epsilon }
\over Z(\beta )}
\sum_{m=-\infty }^{\infty}  F_{\lambda }(m) 
\int {d\varphi \over 2 \pi } e^{i2m(\Phi -\varphi )}
\left\{ 
1-{\Delta_0 ^2 (\varphi ) \over \epsilon^2 +  \Delta_0(\varphi )} 
\bigg( 
1 -\cos \big[ 2t\sqrt{\epsilon^2+ \Delta_0^2(\varphi )} \big] \bigg) \right\}
\nonumber \\
&=& 1-    \sum_{m=-\infty }^{\infty} F_\lambda(m) 
\int {d\varphi \over 2 \pi } e^{i2m(\Phi -\varphi )}
2 A( \varphi )  \sum_{k=0}^{\infty} 
J_{2k+1} [ 2 \Delta_0 ( \varphi )t ]  \;,
\label{u.2}
\end{eqnarray}
with $\Delta_0 ( \varphi ) = 2\Delta_o
\cos \varphi $ as before, and $A( \varphi ) =A \cos \varphi $. This gives 
an absorption form
\begin{equation} 
\chi^{\prime \prime }(\omega ) = 
{ 2A \over \omega } \sum_{m=-\infty }^{\infty} F_\lambda(m) 
\int {d\varphi \over 2 \pi } e^{i2m(\Phi -\varphi )}
 { \cos^2 \varphi \over 
[ (\omega /4 \Delta_o )^2 - \cos^2 \varphi ]^{1/2} } 
\eta ( \omega /4\Delta_o -
\mid  \cos \varphi \mid ) \;,
\label{u.3}
\end{equation}
which for large $\lambda $ simplifies to 
\begin{equation} 
\chi^{\prime \prime }(\omega ) = 
{ 2A \over \omega }  \int {d\varphi \over 2 \pi } 
 { \cos^2 \varphi \over 
[ (\omega /4\Delta_o )^2 - \cos^2 \varphi ]^{1/2} } \eta ( \omega/ 4 \Delta_o -
\mid  \cos \varphi \mid ) \;,
\label{u.5}
\end{equation}
and can be expressed in terms of Elliptic functions.

Notice that since $\omega_k \ll \Delta_o $, the number 
$N$ of environmental spins must be very large to have a noticeable  
effect, i.e., for $\lambda$ to be
appreciable. Thus if $\lambda \sim N\alpha_k^2 \sim 1$, since 
$\alpha_k \sim \omega_k/\Omega_o$, we have $N \sim (\Omega_o /\omega_k )^2
\gg (\Omega_o /\Delta_o )^2$. This not only implies that 
$\mu =N^{1/2} \delta \omega_k  /\omega_o \gg 1$ (ie., very strongly overlapping
polarisation groups) but also a Gaussian  
half-width $N^{1/2} \omega_o \gg \Delta_o $, so that the  
internal bias $\epsilon \gg \Delta_o $. 
The reason why the two mechanisms (topological
decoherence and degeneracy blocking) are so easily combined
is just because this bias is produced by {\it all} of the environmental
spins, whereas only a few of them are actually flipped and their
contribution to the bias field is small.

\vspace{3mm}

(ii) {\it Transverse dominated case} 
($\omega_k^{\perp}/\omega_k^{\parallel} \gg 1$):
Consider now the case described by
\begin{equation}
H{\mbox{\scriptsize eff}}= \Delta_o \hat{\tau}_x + 
\sum_{k=1}^{N} ( \omega^{\perp}_k 
\hat{\sigma}_k^z + \hat{\tau}_z
\omega^{\parallel}_k \hat{\sigma}_k^x)\;, 
\label{neto.0}  
\end{equation}
with $\omega^{\parallel}_k /\omega^{\perp}_k \ll 1$.
One can map this problem to the spin-boson problem, in a way similar to that 
given in the paper by Caldeira et al. \cite{neto93}. 
We begin by noting that the
time evolution operator for the $k$-th bath spin, under the influence of
a central spin moving along a path $Q_{(n)}(t)$, is just 
\begin{equation}
\hat{U}_k(Q_{(n)},t)=T_{\tau} \exp \left\{ -i\int_0^t ds 
\big[  \omega^{\perp}_k  \hat{\sigma}_k^z + 
Q_{(n)}(s) \omega^{\parallel}_k \hat{\sigma}_k^x \big] 
\right\} \;,
\label{neto.3}
\end{equation}
Weak coupling means we can expand the time-ordered exponent to second order in
$\omega^{\parallel}_k$; completing the average in
Eq.~(\ref{neto.22}), and exponentiating the answer, one derives
the influence functional for this problem \cite{neto93}. The "high
temperature" result $kT \gg \omega^{\perp}_k $ is readily found:
\begin{equation}
{\cal F}[Q,Q']  = \exp \bigg\{ -\sum_k {(\omega^{\parallel}_k)^2 \over
2} \bigg\vert  \int_0^t ds 
e^{2i\omega^{\perp}_k s }~[Q_{(n)}(s)-Q'_{(m)}(s)] \bigg\vert ^2
\bigg\} \;,
\label{neto.4}
\end{equation}


The evaluation of this depends on the distribution of couplings. In the
"strong decoherence" 
case where the bath states are spread over an energy range $E_o$
(defined as usual by 
$E_o^2= 2\sum_k (\omega^{\parallel}_k)^2$) which is 
large (ie., $E_o \gg \omega^{\perp}_k,~\Delta_o$), then ${\cal F}[Q,Q']$ is
simply approximated by its form for a single  
kink - anti-kink trajectory. With kink at $t = t_1$ and anti-kink at
$t_2$, one only has a contribution if 
$\omega^{\perp}_k (t_2-t_1) \ll 1$, which gives 
${\cal F}[Q,Q']  = e^{- {1 \over 4}E_o^2(t_2-t_1)^2}$, ie., 
a decay on a time scale
$E_o^{-1}$.
It then follows that kink/  anti-kink
transitions are bound in closed pairs, with phase correlations decaying over
a time $\ll \Delta_o^{-1}$, and the leading terms sum
to give exponential relaxation, ie.,  $P_{11}(t)=1/2(1-e^{-t/\tau_R})$, 
with a relaxation rate  
\begin{equation}
\tau_R^{-1} = 2\Delta_o \int_0^{\infty}dt  e^{- {1 \over 4}E_o^2t^2}
\equiv 2{\sqrt{ \pi} \Delta_o^2 \over  E_o} \;.
\label{neto.7}
\end{equation}
On the other hand if $E_o \ll \Delta_o$, we will get coherent
oscillations of the system over a time scale $\tau_{\phi} \sim 
\Delta_o/E_o^2$, ie., over roughly $\Delta_o^2/E_o^2$ oscillation
periods.
We will use this result below, in discussing the possibility of seeing
true mesoscopic quantum coherence 
effects in SQUIDs and nanomagnets (section 5.C). 

It is of course obvious from the original discussion of Feynman and Vernon
that in the weak-coupling regime, where an expansion to 2nd order in all bath
couplings is sufficient, a mapping to an oscillator bath model must be 
possible \cite{feyv63}. The method given above is of course not the only one-
others are explained in refs. \cite{weiss,PS96,neto93,hanggi98}. 

\vspace{3mm}

{\bf Central Spin coupled to oscillator and spin baths}: For reference we also 
briefly note results for the case where a central spin is coupled 
simultaneously to a spin bath and an oscillator bath; the Hamiltonian is 
that in eqtn. (\ref{QSGl}), but without the sum over $j$, (and of
course dropping the coupling term $V({\bf r}_i - {\bf r}_j)$ between different
central spins). Some of the dynamic properties of this model 
were studied in 
ref. \cite{PS96}- a complete study is still lacking. Quite generally one 
expects that at very short times the dynamics will be controlled by the 
spin bath, but at longer times incoherent oscillator bath- mediated transitions
will take over. However these transitions still contain spin bath effects,
via an integration over spin bath bias distribution, to give a result 
\begin{equation}
P_{11}(t,\xi) =  A Z^{-1}(\beta) \int d\epsilon W(\epsilon- \xi) 
e^{-\beta(\epsilon - \xi)} \{ f(T,\epsilon) 
+ [1 - f(T,\epsilon)] e^{-t/\tau(\epsilon,T)} \} 
\label{CSosc}
\end{equation}
where we assume that the ensemble of central spins is in equilibrium
with a bath of oscillators at temperature $T = 1/\beta$, in external bias $\xi$ 
(for more general cases see
ref. \cite{PS96}). In this case $f(\epsilon) = 
e^{-\beta \epsilon}/2 \cosh (\beta \epsilon)$, and $\tau(\epsilon;T)$ is the oscillator 
bath- mediated relaxation
time; $A$ is a "renormalisation" constant $\sim O(1)$ coming from the
spin bath dynamics. Most features of the results are obvious from 
(\ref{CSosc}). Thus, the oscillator bath {\it unblocks} 
transitions for a central spin way off resonance (ie., having  
$\xi \gg E_o$); these spins relax much as they would without the 
spin bath. When $\xi /leq E_o$ there is a wide range of relaxation times,
and the physical relaxation depends in a complex way on the oscillator bath
spectrum as well as the bias. In any experiment one would see a crossover
between purely spin-bath mediated relaxation at short times and this more complex behaviour- this crossover has not yet been studied in any detail.

\section{Physical application of spin bath models}
\label{sec:5}

In this section we briefly review some recent work, mainly
experimental, in which interactions
with a spin bath play a role. We begin with magnetic systems where 
such interactions are strong, and where clear evidence exists for their
influence on tunneling phenomena. We then discuss superconducting and normal 
systems. Finally, we discuss the difficult and controversial 
problem of decoherence and the mechanisms which govern it in nature. In
this discussion the results of the Central Spin model are crucial- they
describe the effect of a spin bath on a "qubit", or on a SQUID, or a
magnetic macromolecule, which is trying to show coherent oscillations.

\subsection{Magnetic Systems}
\label{sec:5a}

(i) {\it Magnetic Solitons}: Magnetically ordered systems 
support a wide variety of soliton excitations,
depending on the symmetry of the order parameter. These couple to various 
environmental excitations, which strongly affect their dynamics. 
These include both linear and non-linear couplings to 
``quasiparticle" excitations such as magnons
\cite{IJMPB92,sta91}, phonons \cite{dube97}, electrons \cite{tatara}, 
and photons \cite{sta91}, all oscillator baths. More serious effects 
come from any localized modes coupling to the 
soliton, most notably paramagnetic impurities and nuclear spins. Both of 
these can be understood theoretically using a model in which a moving particle 
couples to a spin bath \cite{dube97}, as in section 2.C.

Most experimental work in this area has looked at 
domain wall tunneling in ferromagnets. Early experiments looked at the 
dynamics of multi-wall systems \cite{multiwall,paulsen}. 
More recently attention has focussed on 
the tunneling of {\it single} domain walls, whose position can be monitored
in various ways \cite{hong,mangin,wernwall}. 
These walls propagate along magnetic nanowires (of
diameter 300-600 \AA)and are thus mesoscopic objects (the material of 
choice is usually Ni, for which the domain wall thickness $\lambda_w \sim 
700 \AA$). 
Although some predictions of the 
tunneling theory have been verified (eg.,
the square root dependence \cite{sta91} of the 
tunneling exponent on the pinning
field $H_c$; cf. \cite{paulsen,hong}), 
the crossover temperatures
to tunneling are usually much higher than calculated values \cite{hong}. A
clue to the reason for this may be found in the microwave resonance 
experiments of Hong \& Giordano \cite{hong}, which show extremely broad
resonances, even down to 
$1.5~K$. The coupling to phonons or electrons is far too 
small to give this; however, Oxygen impurities, which
will act as paramagnetic impurities, would act as a strong time-varying
potential on any wall, and broaden the line \cite{dube97}. 
It is clear that 
much more experimental work will be necessary to really 
understand the quantum dynamics of magnetic domain walls. 

\vspace{3mm}

(ii) {\it Magnetic Macromolecules}: On the other hand, 
a number of spectacular quantum effects 
have unquestionably been seen in 
crystalline arrays of magnetic macromolecules, in one of the most 
important developments in magnetism in recent years. Experiments on 
tunneling phenomena have had striking success in
the "Mn-12" and "Fe-8" molecules (each of which behaves at low $T$ as 
a "giant spin" of spin 10). There have also been highly-publicised 
experiments on the large "ferritin" molecule, which claim the observation of 
"macroscopic quantum coherence"; these are discussed in section 5.C below. 

The early molecular work demonstrated 
resonant tunneling (coming when spin levels of each molecule 
are brought into resonance), at relatively high temperatures \cite{activated}.
More recent experiments \cite{sang97,ohm98,luc98,ww1,ww2,ww3} have gone 
into the quantum regime, in which only the 2 lowest levels of each giant spin
are occupied. The classical-quantum  crossover is clearest in the case of 
the Fe-8 system (which is very conveniently described by the 
easy axis/easy plane model discussed in section 2.F);
below a temperature $T_c \sim 0.4~K$, the dynamics is completely independent of
$T$ (the energy gap to the next spin level in 
this system is $\sim 5K$). In Mn-12 there are effects arising from
"rogue" molecules which make things more complicated \cite{ww3,christou}. 
At such low temperatures 
phonons are utterly irrelevant (experiments have now been pursued to 
$T \sim 30~mK$), and the only dynamic environment left is the nuclear spin
bath. The system thus seems at first to be an ideal 
realisation of the central spin model discussed in sections 3 and 4.
The main complication is that experiments are done on 
a crystalline array of molecules, which interact via  
strong magnetic dipolar interactions. However, once one knows that the 
dynamics of a single central spin is going to be incoherent relaxation
(cf. sections 3.E, 4.A), these interactions are rather easy to deal with
\cite{PS98}. The system is then described by an obvious generalisation of the 
central spin Hamiltonian, viz.
\begin{equation}
H= \sum_j H_j^{CS} 
+ \sum_{ij}  V({\bf r}_i -{\bf r}_j ) \hat{\tau}_i^z  \hat{\tau}_j^z   \;,
\label{Hsqrt}
\end{equation}
in zero applied field (compare eqtn. (\ref{QSGl}), without the
oscillators). Here $ H_j^{CS}$ is the central spin 
Hamiltonian (\ref{1.24}) for the 
molecule at lattice position ${\bf r}_j$, and 
$  V({\bf r}_i -{\bf r}_j )$ is the magnetic dipolar coupling 
between molecules $i$ and $j$.

To solve for the dynamics of such an interacting array of systems, 
one begins \cite{PS98} by defining a distribution function
$P_\alpha (\xi , {\bf r}, t)$ for a molecule at 
position ${\bf r}$ to be in a longitudinal bias
$\xi $, with polarisation $\tau_z = \alpha = \pm 1$. It is then trivial to 
write down a BBGKY-like hierarchy of kinetic equations for 
$P_{\alpha} (\xi , {\bf r}, t)$ and its multimolecular generalisations
$P^{(2)}(1,2) \equiv 
P^{(2)}_{\alpha_1,\alpha_2}(\xi_1,\xi_2; \vec{r}_1,\vec{r}_2;t)$, and
$P^{(3)}(1,2,3)$, etc., of which 
the first member is 
\begin{eqnarray}
\dot{P}_{\alpha } (\xi ,\vec{r} )= & &
- \tau_N^{-1}(\xi ) [P_{\alpha } (\xi ,\vec{r} )
-P_{-\alpha } (\xi ,\vec{r} ) ] \nonumber \\
& -& \sum_{\alpha '} \int {d\vec{r}\: ' \over \Omega_0 } \int
{ d\xi ' \over  \tau_N (\xi ' )}
\bigg[ P_{\alpha \alpha '}^{(2)} (\xi  , \xi ';\vec{r},\vec{r}\: ')
 -  P_{\alpha \alpha '}^{(2)}
(\xi -\alpha \alpha ' V(\vec{r} -\vec{r}\: ')
, \xi ';\vec{r},\vec{r}\: ') \bigg] \;,
\label{kinetic}
\end{eqnarray}
in which relaxation is driven by the nuclear spin-mediated relaxation rate
$ \tau_N^{-1}(\xi ) \sim (\Delta^2/\tilde \Gamma) e^{-\vert \xi \vert/\xi_o}$
(cf eqtn (\ref{xi_o}), in conjunction with the dipolar interactions.
Under general conditions we must also 
solve for $P^{(2)}$ in terms of $P^{(3)}$, etc.; but if the initial 
experimental state is either {\it polarised} or {\it annealed} then 
$P^{(2)}$ factorises at $t=0$ (ie., $P^{(2)}(1,2) = P(1)P(2)$), and 
(\ref{kinetic}) can be solved. This led to the following 
predictions \cite{PS96,PS98}:

i) Relaxation should only occur for molecules having 
$\vert \xi \vert \le \xi_0$; consequently a
``hole" rapidly appears in $M(\xi ,t) = P_+(\xi,t) - P_-(\xi,t)$ 
with time, with initial width $\xi_o$
determined entirely by the nuclear dynamics (and of course typically
$\xi_o \gg \Delta$). In fact microscopic calculations of $\xi_o$ for
the $Fe$-8 molecule \cite{rose99,rose99a} can be done (recall Fig. 5(b)), 
since the hyperfine couplings are
essentially dipolar (the relevant nuclei include 120 protons, 8 $Br$
nuclei, and 18 $N$ nuclei). One finds that even the weak
molecular dipolar fields strongly distort the
hyperfine fields, mixing up the different polarisation groups, and
also giving a large value of $\kappa$. The final value 
$\xi_o$ will obviously depend sensitively on any nuclear isotopic
substitution \cite{rose99,rose99a}.
The subsequent evolution of the hole depends on sample 
shape; this has been studied theoretically using Monte Carlo simulations \cite{PS98}.

(ii) The short-time relaxation for the total magnetisation 
$M(t)= \int d\xi M(\xi ,t)$ should have a "square-root" time form
\begin{equation}
M(H_0,t) =M_0 [1-( t/\tau_Q(H_0))^{1/2} ] \;,
\label{Msqrt}
\end{equation}
where $M_0$ is the initial magnetization (or with appropriate modifications 
for other protocols, such as a zero-field cooling followed by relaxation
in a field \cite{ww1,ww3}), and 
$\tau_Q^{-1} (H_0) = c (\xi_o/E_D) 
\Delta^2 M(\xi = -g\mu_B SH_0, t=0)$, where $W_D^{-1}$ is the "density of
states" of the distribution. The constant $c$ is
dimensionless, depending
on sample shape, and can be evaluated analytically or numerically. Notice that
the existence of the square root does not depend on sample shape, and indeed 
its persistence over fractional relaxations of $\sim 0.1$ is clearly
demonstrated in Monte Carlo simulations for different shapes 
\cite{PS98,ohm98,vill99}.

This result implies that by varying $H_0$, one can measure $M( \xi )$, by 
extracting $\tau_Q^{-1} (H_0)$ at successive
values of $H_0$. If one knows $M(\xi )$ 
(as one does in an annealed sample - it should be
Gaussian) then one may then extract $\Delta $ 
from measurements of $\tau_Q^{-1}$.
The $\sqrt{t}$ dependence of $M(t)$ has since been reported in quite a few 
experiments, on 
both Fe(8) crystals
\cite{ohm98,ww1,ww2} and Mn(12) crystals \cite{luc98} 
(although the situation in 
Mn-12 is seriously complicated by "impurities" \cite{ww3}). 
The Fe-8 experiments have produced remarkable "maps" of $M(\xi )$, and 
its time variation \cite{ohm98,ww1}.
The "hole-digging" has been found in both Fe-8 and Mn-12 \cite{ww1,ww3}, 
with an "intrinsic" short-time intrinsic linewidth which is roughly that 
expected from the hyperfine interactions, provided one takes account of the 
effect of internal fields \cite{rose99a}, 
which make $T_1$ very short (so that $\kappa$ is 
effectively $\sim \sqrt{N}$).

Wernsdorfer et al \cite{ww1} also used this technique to extract the
value of $\Delta$ for Fe-8. In a remarkable experiment,
Wernsdorfer and Sessoli \cite{ww2} extended these measurements of 
$\Delta $ to include a
{\it  transverse} applied field $H_{\perp}$; as noted in section 2.F, 
the topological giant spin phase
$\Phi$ should vary with field $H_\perp$ \cite{garg93,boga92}, 
producing Aharonov-Bohm oscillations in 
$\Delta (H_\perp )$. These oscillations were found, both as oscillations 
in the relaxation rate $\tau^{-1}_Q (H_\perp )$,
{\it and} using a quite different AC absorption ("Landau-Zener") technique;
these independent techniques agreed rather well in the measurement of 
$\Delta$. Notice, incidentally, that neither method can 
properly measure $\Delta $ near its nodes 
(ie., where $\Delta \to 0$); this is because of the distribution of  
internal transverse fields. Incidentally, we
should strongly emphasize that these experiments (even the Aharonov-Bohm ones)
do {\it not} demonstrate coherent tunneling- indeed 
they show exactly the opposite!
This is because the experiments are inherently {\it relaxational} 
(this is why
all rates are $\sim \vert \Delta \vert^2$, and not $\sim \Delta$). 
Readers puzzled 
about how an Aharonov-Bohm effect can occur in a relaxation rate, are 
encouraged to think about other examples in physics where phase interference
shows up in irreversible quantum phenomena. In the present case we may
loosely define a "decoherence time" $\tau_{\phi}$ for the molecular spins
(one should not push this too far, given the non-exponential nature of
the central spin relaxation!), and one finds that $\tau_{\phi} \Delta_o
\ll 1$ (no coherent oscillations possible) but $\tau_{\phi} \Omega_o 
\gg 1$ (ie., coherence is maintained during a single very rapid tunneling
transition).

In a very recent experiment \cite{sess99}, deliberate modification of the 
nuclear isotopes in an $Fe$-8 crystal has shown a dramatic modification of the 
bulk magnetic relaxation- this is the most direct evidence so far for the
role of the nuclear spins in mediating the quantum relaxation. It will be 
interesting to see a quantitative comparison with the calculated 
results for different isotopes \cite{rose99,rose99a}. 

\vspace{3mm}

(iii) {\it Quantum Spin Glasses}: The work on
nanomagnets probes our understanding of collective phenomena in many
magnetic systems, ranging from spin chains to quantum spin glasses. This latter
example is very closely related to the molecular nanomagnets just described,
since the model Hamiltonian is just (\ref{Hsqrt}), with the dipolar interaction
replaced by a set 
$\{ J_{ij} \}$ of "frustrating" interactions, usually long-ranged. A typical 
experimental example is provided by the 
disordered dipolar spin system LiHo$_x$Y$_{1-x}$F$_4$ (where the 
Ho moments interact via dipolar interactions 
\cite{rosen}, and the $\Delta $ are induced by a transverse field). Until now
most theory has ignored the environment in this problem, but from our 
discussion above, this is 
clearly a mistake if one wishes to discuss dynamics. This is 
also the view expressed in recent papers of Cugliandolo et al. 
\cite{cugl98}, who have included coupling to an oscillator bath environment. 
This is presumably correct in metallic glasses, where the coupling to 
electrons dominates (compare the discussion following (\ref{1.35})), 
but in insulators
the coupling to nuclear spins dominates 
(in Li$_{1-x}$Ho$_x$F$_4$  this is
clear- the hyperfine coupling even modifies the phase diagram!). 
An observation of strong hole-digging in $M(\xi ,t)$ in a spin glass 
would be consistent with this,
since oscillator baths typically give most rapid relaxation {\it away} 
from resonance ($cf.$ discussion after eqtn. (\ref{xi_o}). 
Note also that one of
the main theoretical questions concerning quantum spin glasses, viz., 
what happens when $\Delta \sim J_{ij}$,
could be examined experimentally 
using molecular magnets like Fe-8, or the 
LiHo$_x$Y$_{1-x}$F$_4$ system, in a strong transverse field, where one might 
reasonably expect to see coherent propagation of tunneling events from one site
to another if $\Delta$ is sufficiently large.

\subsection{Conductors and Superconductors}
\label{sec:5c}

In mesoscopic conductors the standard weak localisation theory \cite{altsh}
evaluates a "decoherence time" 
$\tau_{\phi}(T)$ in terms of the electron-electron scattering rate,
which is itself strongly influenced at low $T$ by elastic 
impurity scattering. This is essentially an oscillator bath problem- the 
relevant oscillators being diffusons, Cooperons, and phonons.
Curiously, given the known importance of scattering off dynamic
"2-level" fluctuators in these systems \cite{TLS},
there has been little theory on the effect of
these on $\tau_{\phi}$, apart from the pioneering work of Altshuler and Spivak,
and Feng et al.
\cite{feng}. This is of course a spin bath
problem, with the spins representing defects, paramagnetic impurities, etc.,
in the environment. It will be interesting to see if the "saturation" in
$\tau_{\phi}(T)$ reported at low $T$
\cite{mohanty} may at least be partially explained by such scattering. The  
physics of this saturation depends on the more fundamental 
question of how decoherence behaves in the low-$T$ limit
in conductors, and has caused considerable debate in the
recent theoretical literature \cite{mohanty,decT0}.

In the case of superconductors the situation is similar, in that almost all
theoretical work on dissipation and other environmental effects has looked
at the effects of electronic quasiparticle modes or photons, ie., 
delocalised modes which can be mapped directly to oscillator modes. The 
pioneering papers \cite{ajl87,cal83,amb85,ajl84} led to a massive
subsequent literature, both experimental
\cite{voss,washburn,clarke,lukens} and theoretical 
\cite{kagajl92,SQUIDth}. In most work on tunneling there is no question that 
theory and experiment correspond very well \cite{clarke}. However the situation is more delicate for coherence, discussed below.
In spite of the extensive theory of spin 
impurity effects in superconductors 
we are aware of only 2 theoretical papers \cite{sta88,seattlePS} 
(and no experiments) examining 
their effect on the flux dynamics (in particular
tunneling) of SQUIDs.

\subsection{Coherence and decoherence; and "qubits"}
\label{sec:5D}

An understanding of decoherence mechanisms is central to the exploitation of 
mesoscopic systems in quantum devices, as well as to general questions about 
how quantum mechanics applies on the large scale, \cite{ajl88,ajlLH}, and the 
quantum measurement problem \cite{qmmt}. It has become particularly
important now that efforts are being made to construct "qubit" devices,
with a view to making quantum computers.

The last 15 years have seen
a total transformation in how such questions are discussed- instead of 
vague analyses in terms of "measurements" by the environment, we now have
precise and generally applicable models, which can be 
tested in many experiments.
The basic issues are (i) whether phase coherence can be preserved in the 
reduced density matrix of the system of interest and (ii) what are the 
decoherence mechanisms destroying it. Here we briefly review studies 
of superconducting and magnetic systems, and 
then examine things from a more general theoretical standpoint.

\vspace{3mm}

(i) {\it Decoherence in superconductors}: This  has been discussed 
intensively ever since the 
theoretical predictions of Leggett et al. \cite{ajl87,ajl88} 
concerning "macroscopic 
quantum coherence" in SQUIDs, and subsequent proposals for experimental
searches \cite{tesche}.
To date no experimental success has been reported (although there is good 
evidence for resonant "one passage" tunneling transitions between 
near degenerate levels in 2 wells \cite{lukens,naka}). Almost all microscopic
analyses of this problem have assumed environments of electronic
excitations which can be mapped onto oscillator baths 
(see, eg., \cite{kagajl92,amb85,ajl88,SQUIDth}).
In our opinion, as discussed in section 4.B, the basic problem is simply
that the main source of decoherence in most systems (including SQUIDs)
at low $T$ will not be any oscillator bath, but the spin bath of 
paramagnetic and nuclear spins. As discussed in section 2.G and 4.B, the 
low-energy scale of this spin bath means it will not usually have a big effect
on SQUID tunneling, but its effect on macroscopic coherence or on
superconducting qubits will be rather large. 

Although a superconducting qubit has not
yet been built, experiments may be getting rather close
\cite{naka}. To see how big spin bath effects on coherence 
might be, let us recall that the effect
of paramagnetic impurities is to create a Gaussian
multiplet of spin bath states of width 
$E_o \sim \mu_B B_o \sqrt{N_{pm}}$ in energy for a SQUID containing a
total of $N_{pm}$ paramagnetic impurities interacting with the
supercurrent, where $B_o$ is the change in field on each paramagnetic
impurity caused by the change in flux state of the SQUID. To see
coherence it is necessary, from the discussion of section 4.B, that 
$\Delta_o \gg E_o$, because the decoherence time coming from the spin
bath is $\tau_{\phi} \sim \Delta_o/E^2_o$; 
this essentially sets a lower bound for $\Delta_o$. As discussed in some
detail in a recent paper \cite{PSsquid}, this turns out to be a rather
stringent requirement on real SQUIDs; in fact in the experiments of the 
Lukens group \cite{lukens} one infers a value $E_o \sim 0.4~K$ from
their resonant linewidths. Obviously this value could be reduced 
a great deal by careful attention to the nuclear and paramgnetic spin
impurity composition in the system (as well as to the sample geometry).

\vspace{3mm}

(ii) {\it Decoherence in Magnets}: The most dramatic claims for the 
observation of macroscopic 
coherence have been made 
by Awschalom et al. \cite{Awsch2}, working on randomly oriented dilute 
ensembles of ferritin macromolecules (which order antiferromagnetically,
but carry an excess moment of somewhat random size; the antiferromagnetic
"N\'eel" moment is $\sim 23,000 ~\mu_B$). In an effort to make the molecule size
as uniform as possible, these authors filtered them magnetically.
They also artificially 
engineered molecules of smaller size. The essential result was the 
observation of an absorption peak at $MHz$ frequencies, whose 
frequency varies approximately exponentially with the size of the molecules.
This was interpreted as a signature of coherent tunneling 
between "up" and "down" states of the Neel vector. There have been widespread 
objections to this interpretation, both on theoretical and experimental
grounds \cite{rebuke}, and so far no other group has succeeded in confirming
the experiments. Note that the 
Awschalom group saw similar resonances (also with an 
exponential dependence of 
resonant frequency on size) in large FeCo$_5$ particles, 
but did {\it not} 
attribute this to tunneling \cite{Awsch1}.

As we discussed 
in the previous sub-section, there is now 
very extensive evidence that nanomagnetic molecules  
in macroscopically ordered crystals tunnel {\it incoherently}
in the low-$T$ quantum regime \cite{sang97,ohm98,luc98,ww1,ww2,ww3}. 
There is thus 
an apparent contradiction between the ferritin work and that done in the 
Mn-12 and Fe-8 systems (particularly since the ferritin molecules are much 
larger and certainly contain a lot of spin disorder). 
Thus in the very well-characterised $Fe$-8 molecular 
crystals used by the Florence
and Grenoble groups,
the parameter $\kappa$ characterising decoherence from 
the orthogonality blocking mechanism varies 
\cite{rose99,rose99a}, even in an ideal
sample, 
between $\kappa \sim 6-15$ in zero applied field 
(depending on the annealing-dependent spread $W_D$ in the intermolecular  
dipolar fields), to $\kappa \sim 80$ when
$H_x \sim 0.2~T$ (where the first zero in $\Delta$ 
is supposed to occur \cite{ww3}). Recalling from sections 3.C and 4.A that
coherence is practically eliminated unless 
$\kappa \ll 1$, it is hard to see how
experiments on these particular molecules in low fields 
will stand much chance of seeing it. 

On the other hand it is clear that future experiments on {\it single}
nanomagnets in the quantum regime might have a chance of seeing coherence 
{\it iff} 
one could raise $\Delta_o$
to values $\gg$ the hyperfine couplings $\omega_k$ (presumably using a large 
external transverse field), thereby making  
$\kappa \ll 1$ and so removing decoherence (and also reducing the problem
to a straightforward spin-boson model- see section 4.B). Another
possibility, which could be realised in, eg., the LiHo$_x$Y$_{1-x}$F$_4$ system
at high transverse fields, would be to see coherent propagation of spin
flips (ie., spin waves) in a lattice of spins, by making 
$\Delta > W_D$ (here one could also make
$\omega_k \gg kT$, thereby {\it freezing} the nuclear dynamics!). Conceivably
the same could be done in an $Fe$-8 crystal (now with the inequality
$\Delta > W_D \gg \omega_k$ operating). 
We see no reason why such experiments could not be done in the next few
years.

\vspace{3mm}

(iii) {\it Decoherence as} $T \rightarrow 0$: Let us now consider the general 
question of decoherence effects at low $T$. 
Decoherence is often (particularly in conductors) characterised by a
"decoherence time" $\tau_{\phi}$, for the phase dynamics of the degree of
freedom of interest.
If and when $\tau_{\phi}$ is meaningful, it may be much shorter than the energy
relaxation time $\tau_E$ ({\it cf.} the 
example of a single oscillator coupled to an oscillator bath \cite{cal85}, or
the examples of topological decoherence given for the spin bath in sections
3.A and 4.A above)).
Coherence exists if $\tau_{\phi} \Delta_o  \ll 1$, where $\Delta_o$ is
the characteristic frequency of the system's phase dynamics. 

Notice that what allows us to discuss this problem with any generality at all
is the assumption, discussed in sections 1 and 2, that a
few canonical models describe the low-$T$ behaviour of most 
physical systems. Extensive study of the relevant canonical oscillator
bath models (in particular, the spin-boson model \cite{ajl87,weiss,ajl88}
and the "oscillator on oscillators" model \cite{cal83,grabert88}) show  
that with a power-law form $J(\omega) \sim \omega^n$, decoherence disappears
as $T=0$ for $n > 1$; for the Ohmic form $J(\omega) = 
\pi \alpha  \omega$ decoherence 
is finite at $T = 0$, but can be made small if $\alpha \ll 1$. 
If the electronic spectrum is gapped the Ohmic dissipation falls off
exponentially in the low $T$ limit (thus for superconductors 
\cite{amb85,SQUIDth} one has 
$J(\omega,T) \sim \omega e^{-\Delta_{BCS}/kT}$ for 
$\omega < 2\Delta_{BCS}$, 
and
for magnetic solitons \cite{IJMPB92,sta91,dube97} one has 
$J(\omega,T) \sim
\omega (kT/\Delta_m)e^{-\Delta_m /kT}$ for $\omega < 3\Delta_m$, 
where $\Delta_{BCS}$ and 
$\Delta_m$ are the BCS and magnon gaps respectively). Thus, if one believes 
the oscillator bath models, coherence ought to be easily observable at low $T$; 
the condition $\tau_{\phi}(T) \Delta_o \ll 1$ is clearly satisfied for 
temperatures well below the gap energy.

If we examine the canonical spin bath models we find a 
very different story \cite{PS93,sta94,PRB,PSchi95,PS96}. 
Consider first the central spin model as $T \rightarrow 0$; we will go to 
such a low temperature that all of the spins in the spin bath order in the
field of the central spin (ie., $T \sim 1 \mu K$ in some cases), ie., all 
intrinsic fluctuational dynamics of the bath is frozen out. Does 
decoherence disappear? No, because the mechanisms of topological decoherence
(induced bath spin flip) and orthogonality blocking (precession of the 
bath spins in between central spin flips)
still exist- the bath can still acquire dynamics from the central spin. 
We emphasize here that this physics cannot be described by an oscillator
bath model. From the results in section 4 we see  
there is a residual constant decoherence as $T \rightarrow 0$, coming from 
the spin bath, which in an experiment would be signalled by a saturation
of $\tau_{\phi}$ once oscillator bath effects had disappeared.
The extent of this decoherence is characterised by the
behaviour of $P_{M=0}(t)$, (cf. section 4.A), and we saw that unless both
$\kappa$ and $\lambda$ were $\ll 1$, decoherence was strong. In the case where 
all bath spins are polarised by the central spin, transitions are blocked
anyway- there are no spins in the $M=0$ polarisation group!

We conclude that for any system described by the central spin model (ie., 
where the central system reduces to a 2-level system at low energies), a 
general consequence of the coupling to a spin bath will be a loss of coherence,
via either the topological decoherence or orthogonality blocking mechanisms, 
even in the $T \rightarrow 0$ limit.
The residual coherence (if any) will depend on the strength of the 
couplings to the spin bath, in a way discussed quantitatively in sections 
4 and 5. 

We may generalise these considerations to models in which a "particle" moves
through a spin bath (the same model also describes a network  
of spins, or of mesoscopic superconductors, etc., coupled to a spin bath), 
and get the same result. Consider, eg., 
a particle hopping from site to site on a $D$-dimensional
hypercubic lattice, whilst coupled to a spin bath \cite{SBlatt}, and  described by a Hamiltonian  
\begin{eqnarray}
H^{latt}(\Omega_o) & = & \Delta_o \bigg\{ \sum_{<ij>}  \bigg[ c_i^{\dagger}c_j 
\cos [  \Phi_{ij} + \sum_k \vec{V}_k^{ij}. \vec{\hat{\sigma}}_k] 
+ H.c. \bigg] \nonumber \\
& + & \sum_k \bigg[ \;^{\parallel}\omega_k^{ij} (c_i^{\dagger}c_i - 
c_j^{\dagger}c_j) \vec{\hat{\sigma}}_k^z + \omega_k^{\perp} 
\hat{\sigma}_k^x \bigg] \bigg\}
+ \sum_{k,k'} V_{kk'}^{\alpha \beta }\hat{\sigma}_k^{\alpha} 
\hat{\sigma}_{k'}^{\beta} \;;
\label{SBlatt}
\end{eqnarray}
where $<ij>$ sums over nearest neighbour sites. The couplings
$^{\parallel}\omega_k^{ij}$ and $\vec{V}_k^{ij}$ to the bath spin 
$\vec{\sigma}_k$ usually depend upon which "lattice rung" $(i,j)$ the particle 
happens to be, because the coupling normally has a finite range and the spin
$\vec{\sigma}_k$ has some position with respect to the lattice (one can also 
add a longitudinal coupling to the $\{ \hat{\sigma}_k^z \}$ on each site, as in eqtn. (\ref{QSGl})). 
In the continuum limit (\ref{SBlatt}) reduces to 
(\ref{2.6}) and (\ref{2.7}), after dropping the external 
field $\vec{h}_k$ and the 
dependence of $F_k^{\perp}(P,Q)$ on $Q$. To isolate the decoherence effects
let us assume $V_{kk'}=0$, ie., we now 
study the analogue of the
strong-coupling regime in section 4.A. Without loss of generality 
we may then concentrate on the "phase decoherence" regime, ie., on an 
effective Hamiltonian $H_{eff} = \sum_M w_M H_M^{eff}$, where 
\begin{equation}
H_M^{eff} = \Delta_o \sum_{<ij>} [ c_i^{\dagger}c_j \hat{\cal P}_M
e^{-i\sum_k \alpha_k \hat{\sigma}_k} \hat{\cal P}_M \;+ \; H.c.]
\label{Hproj}
\end{equation} 
This is just a generalisation of (\ref{5n.12}) to the lattice; 
the dependence of the coupling
on the lattice position is dropped because it is inessential to what follows.

Coherence, if it exists, will appear in the function $P_{n0}(t)$
(the probability to start at site $0$, and be at site
$n$ a time $t$ later). 
Perfect coherence (ie., no bath) yields
\begin{equation}
P_{\vec{n} 0}^{(0)}(t) = 
\sum_{\vec{p} \vec{p}'} e^{i[(\vec{p} - \vec{p}').\vec{n} 
- (E_{\vec{p}}-E_{\vec{p}'})t]} \;=\; 
\prod_{\mu=1}^D J_{n_{\mu}}^2(2 \Delta_o t)  
\label{ballistic}
\end{equation}
for which $P_{00}^{(0)}(t) \sim 1/(\Delta_o t)^D$ at 
long times; moreover, the 2nd moment 
$\langle (n(t)^2 \rangle  
= \sum_{\vec{n}} n^2 P_{\vec{n}0}^{(0)}(t) \sim D(\Delta_o t)^2$
at long times. Both are characteristic of coherent "ballistic" band motion. 

If we now go to the interacting case one finds, by a straightforward 
generalisation of the calculations in sections 3 and 4, some rather 
interesting results \cite{SBlatt}. Consider first a bath with all
spin states equally populated. Then
at long times $\langle (n(t)^2 \rangle \rightarrow D(\Delta_o t)^2$ but
$P_{00}^{(0)}(t) \rightarrow (1/\Delta_o t)$, {\it independent} of $D$!
Moreover, if we start with an initial Gaussian wave-packet of width
$\Delta n (t=0) = R_o$, one finds $P_{00}^{(0)}(t) 
\rightarrow (1/R_o \Delta_o t)$ as $t \rightarrow \infty$. These results show 
that the naive inference of ballistic propagation from the second moment result
is wrong- in reality one has strongly anomalous diffusion (with an 
energy-dependent diffusion coefficient, demonstrated by the $R_o$-dependence
of the results). In fact the probability function $P_{\vec{n} 0}^{R_o}(t)$ 
decays like $R_o/(\Delta_o t)\vert \vec{n} \vert^{(D-1)}$ at long times 
(ie., $\Delta_o t \gg R_o$) and distances $\vert \vec{n} \vert \gg R_o$, out 
to a "ballistic" distance $l(t) \sim \Delta_o t$, for dimension $D > 1$. Thus
there is an advance "ballistic front" which decays in amplitude with increasing 
distance/time from the origin, but only as a power law; and it is followed by 
a much larger anomalously diffusive (and of course incoherent) contribution.
For sufficiently long times (in fact for $\lambda\Delta_o t \gg 1$) this 
result is {\it independent of} $\lambda$.
Similar results apply for the orthogonality-
blocked case as a function of $\kappa$. 

Now let's take the limit $T \rightarrow 0$, meaning that we allow the bath spins
to order in the field of the particle. It is clear that if  
$\omega_k^{\parallel} \gg \Delta_o$ for all spins, {\it and} provided there
are no non-diagonal momentum couplings to the bath, we cannot get decoherence
by the same mechanism as above, since there is only one state in the relevant 
polarisation group- the particle will then move freely without disturbing the 
spins in any way. However in any other case phase will be exchanged with the 
bath in the same way as above, with or without dissipation (which will 
certainly  arise in the weak-coupling limit
$\omega_k^{\parallel} < \Delta_o$). 

We therefore conclude that a finite 
decoherence in the  $T \rightarrow 0$ limit is a generic consequence of the 
existence of spin bath environments.

\vspace{3mm}

(iv) {\it Qubits and Quantum Computation}: What is the impact of these results 
on hopes for quantum computation? A Quantum computer is an information
processing device which can be imagined as an assembly of 2-state qubits, 
these being none other than the spin-boson and Central spin models discussed 
in this article \cite{feyn96,berman98}. Such a computer has yet to be
built, and papers and books on this topic tend
to divide into 2 classes. The first simply ignores  
decoherence (apart from occasionally referring to it as the main stumbling
block preventing the construction of a quantum computer!), whereas the second 
regards decoherence as the crucial problem, and either tries to treat it
theoretically (eg., \cite{unruh97}) 
or maintains that a quantum computer will never be
built because of it (eg., \cite{haroche}).  

The basic problem here is the lack of serious theory on the effects of decoherence, starting from realistic models which can be tested quantitatively by experiment. The analysis of the Central Spin model in sections 3 and 4 
can be used for a single qubit- it will be clear that the main task is to 
reduce diagonal couplings to the spin bath (and also any oscillator bath) 
as far as possible. However this 
is only the beginning of the problem- the operation of a 
quantum computer involves multi-qubit wave-function entanglement, and thus
one wants to understand the 
behaviour of a decoherence time $\tau_{\phi}^{(M)}(\xi_1,\xi_2,..\xi_M)$,
governing loss of $M$-spin phase correlations, in the presence of coupling 
to spin and oscillator baths. We are aware of no studies of
this problem (or even recognition that it is a problem) in the literature.
Studies of mutual coherence and 
decoherence in the problem of 2 spins coupled to an oscillator bath
\cite{dube97} give some feeling for what might happen, as do the studies of 
lattice systems. However we still have no answer to, for 
example, the question of how $\tau_{\phi}^{(M)}$ behaves for large $M$.
If it ends up having a generic decrease $\sim e^{-aM}$ then the whole quantum
computing enterprise will be in very serious trouble!  

It is clear that in the near future research on quantum computers 
will have to proceed on both 
practical designs (with serious attention given to the decoherence characteristics of the relevant materials), 
and also on general studies 
of decoherence for systems of coupled qubits, themselves coupled to spin and 
oscillator baths. This promises to be one of the great challenges in 
condensed matter physics during the next decade.

\section{Acknowledgements} 

We would like to thank NSERC and the CIAR in Canada, and the Russian
Foundation for Basic Research (grant 97-02-16548) and European Community
(grant INTAS-2124), for financial
support of our research. We also thank 
I. Affleck, Y. Aharonov, B. Barbara, J. Clarke, 
M. Dub\'e, W. Hardy, A.J.Leggett, 
T. Ohm, C. Paulsen, R. Penrose, G.
Rose, I.Tupitsyn, W.G. Unruh, 
I.D. Vagner, and W. Wernsdorfer for helpful discussions
about the spin bath,
going back to 1987.

\section{Note Added in Proof (Feb 22, 2000)}

A number of papers touching upon the present subject (particularly on 
decoherence) have appeared since 
this article was written. 

On the highly controversial question of zero-temperature decoherence
in mesoscopic conductors \cite{mohanty,decT0} 
(which according to some authors throws the
entire conventional theory of metals into doubt), a large number of suggestions
have appeared. A list of them appears in a short experimental review by
Mohanty \cite{moh99}, which concludes that only the original suggestion, of 
electronic coupling to zero-point fluctuations of the EM field 
\cite{mohanty,decT0,zaik99}, can 
explain the experimental saturation of $\tau_{\phi}$. This conclusion is
hotly disputed by other authors, both on experimental and 
theoretical grounds \cite{alein99,vav99}. The idea that the 
decoherence might be coming from "two-level systems" in the sample (ie., from 
a spin bath) has been explored by Imry et al. \cite{imry99} and Zawadowski 
et al. \cite{zaw99}, but rejected by Mohanty, mainly because it would imply 
very large low-frequency noise levels in the sample. In our opinion this 
requires further work- some of the decoherence mechanisms discussed in  
the present article do not appear in these papers, and would not 
necessarily show up in the low-frequency noise. In any case this controversy 
shows clearly how the theory of transport, 
including weak localisation 
theory, depends crucially on a correct understanding of decoherence mechanisms.

The most interesting recent experimental progress concerning the  
mechanisms of decoherence may be work in magnetic systems. The most 
recent work by
Wernsdorfer et al. \cite{sess99}, which looked at the dependence of tunneling
rates and "hole widths" in the resonant quantum relaxation of crystals
of $Fe$-8 molecules, has been supplemented by further work on the same system 
\cite{ww00}. Taken together these experiments give rather strong evidence
that the tunneling is mediated and controlled by the nuclear spins in the 
system. No direct test has yet appeared of the theory of the decoherence 
coming from these same nuclear spins, however (which depends on calculations
of the parameter $\kappa$; cf. section 2.6, including Fig. 5, and sections 
3.3, 4.1, 5.1, and 5.3, including Fig. 8, and refs. \cite{rose99,rose99a}). 
Experiments looking for coherence in $Fe$-8 have also been done at higher 
fields \cite{barco}; however we are unable to see why the reported 
results give
evidence either for or against coherence, since they simply show very broad and
rather weak peaks in the ESR spectrum as a function of field. We emphasize a 
point here which has often been made, viz., that any demonstration 
of coherence requires direct observation of combinations of multi-time 
correlation functions- even a very sharp peak in, say, the AC absorption is 
not enough to demonstrate coherence. In this connection the reader is referred 
to Figs. 7-9 in the present paper, which show peaks in 
$\chi^{\prime \prime}(\omega)$ even when there is no coherence at all. 
Several reviews of the many different experiments in tunneling nanomagnets have
also recently appeared \cite{barb99,gatt99,tupBB99}. One topic not covered in these is the LiHo$_x$Y$_{1-x}$F$_4$ system (section 5.1.3); some 
interesting new results on this compare the thermal and quantum annealing, showing the efficiency of the 
latter in quantum optimisation\cite{aeppli}. 

More general discussions of coherence and decoherence appearing recently 
include several theoretical reviews \cite{zur99,kilin99,zeh99,hang99,gurv99} 
(although these do not really discuss spin bath environments). There has also been interesting experimental work on systems other than 
superconductors and magnets;
see, eg., 
Wiseman et al. \cite{wise99} for coupled quantum dots, and Myatt et al. 
\cite{myatt99} for 
decoherence in quantum optical systems.

\appendix 

\section{ Derivations for limiting cases}

In this appendix we give the derivations of some key formulae in sections 
3 and 4. 
Instanton methods are used to handle the spin environment. 
In Appendix A.1 we give details of the fairly trivial calculations
required to deal with averaging over bias, topological phase, and fluctuations
in bias. Then in Appendix A.2 we discuss 
the more lengthy derivation of the expression 
(\ref{4.21o}) involved in 
orthogonality blocking.

\subsection{Topological phase, bias, and bias fluctuation effects}

We wish to evaluate 
$P_{11} (t)$, the probability for the central spin to return to an
initial state $\vert \uparrow \rangle$ after time $t$, in the presence of
a static bias $\xi$ and a noisy bias $\epsilon (t)$. We use
ssh physics.ubc.ca
standard instanton techniques \cite{ajl87,coleman}, because they easily
generalise to include the spin bath. 

We begin by ignoring the topological
phase of the central spin since its effects are trivial to add.
Then the amplitude for a 2-level system to 
flip in a time $dt$ is  $i\Delta_o dt$, and the return probability is given
by summing over even numbers of flips:
\begin{equation}
P_{11}^o (t) = {1 \over 2} \left\{ 1+ \sum_{s=0}^{\infty}
{ (2i\Delta_o t)^{2s} \over (2s)! } \right\} \; 
= \; {1 \over 2} [1+\cos (2\Delta_ot )]\;,
\label{b1.2}
\end{equation}

Now consider the modification introduced by a {\it longitudinal}
static bias $\xi$.  
In instanton language 
we begin with the return {\it amplitude} $A_{11}(t,\xi)$,
and Laplace transform it; using
the action  $e^{\pm i\xi dt }$ in bias $\xi$, over time $dt$, we
get
\begin{equation}
  A_{11}(t, \xi) = \sum_{n=0}^{\infty} (-i\Delta_o )^{2n} 
\int_0^t dt_{2n} \dots 
\int_0^{t_2} dt_1 e^{i(\xi (t-t_{2n})-\xi (t_{2n} -
 t_{2n-1}) + \dots \xi t_1 )} \; = \; \int_{-i\infty}^{i\infty} e^{pt}
 A_{11}^{^{TLS}}(p, \xi ) \;,
\label{b1.6}
\end{equation}
\begin{equation}
 A_{11}^{^{TLS}}(p, \xi ) = {1 \over p-i\xi } 
\sum_{n=0}^{\infty} \left(
{ (-i\Delta_o )^2 \over p^2+\xi^2 } \right)^{n} = 
 {1 \over p-i\xi } {   p^2+\xi^2 \over p^2+E_o^2 }\;.
\label{b1.8}
\end{equation}
where $E_o = [\xi^2 +\Delta_o^2 ]^{1/2}$ ; 
then the standard answer for 
$P_{11}(t, \xi )$ is just:
\begin{eqnarray}
P_{11} (t,\xi )&=& \int_{-i\infty }^{i\infty} dp_1 dp_2 
{ e^{(p_1+p_2)t}  \over (p_1 -i
\xi) ( p_2 -i\xi)} \sum_{n=0}^{\infty}\sum_{m=0}^{\infty} 
\left( { (-i\Delta_o )^2 \over p_1^2+\xi^2 } \right)^{n}
\left( { (-i\Delta_o )^2 \over p_2^2+\xi^2 } \right)^{m} \nonumber \\
&=& [1-{\Delta_o^2 \over 2E^2 } (1- \cos 2Et)] \; = 
\; [1- {\Delta_o^2 \over E^2 }\sin^2 Et]
\label{b1.10}
\end{eqnarray}
This representation avoids the difficulty in the usual representation coming 
from the square root $E = [\xi^2 +\Delta_o^2 ]^{1/2}$ in the cosine. 

We now add a fluctuating bias $\epsilon (t)$ to $\xi$, with the correlation 
$\langle [\epsilon(t)-\epsilon (t')]^2 \rangle =
\Lambda^3 \vert t-t' \vert$ for short times. Noise averages are then
given by the Gaussian average
\begin{equation}
\langle F [\epsilon (t)] \rangle \;=\; \int {\cal D}\epsilon (t) 
\; F [\epsilon] e^{-{1 \over 2} \int dt_1 \int dt_2
\epsilon (t_1) K (t_1 - t_2) \epsilon (t_2)}
\end{equation}
with a noise correlator $2K^{-1}(t_1 - t_2) = \Lambda^3 (\vert t_1 \vert +
\vert t_2 \vert - \vert t_1 - t_2 \vert)$.
Averaging over a phase function 
$F(t_1,t_2)=e^{i\int_{t_1}^{t_2} ds \epsilon (s) }$ then gives the standard
result
\begin{equation}
\langle F(t_1,t_2) \rangle =e^{i\epsilon (t_1) (t_2-t_1)}
~ e^{-\Lambda^3(t_2-t_1)^3/6}  \;,
\label{diff.6}
\end{equation}
In applying this to (\ref{b1.6}) we assume fast diffusion (see text), and
thus only expand to $\sim O(\Delta^2)$. This gives $P_{11} (t) =
e^{-t/\tau(\xi)}$, with
\begin{equation}
\tau^{-1}(\xi)\; =\; 2\Delta^2 \int d\epsilon
G_{\mu}(\epsilon)
\int_{0}^{\infty}ds e^{i(\epsilon + \xi) s} ~e^{-\Lambda^3s^3/6}
\;\;=\;\; 2 \pi^{1/2} {\Delta^2 \over \Gamma} e^{-(\xi/\Gamma)^2}
\label{LZener}
\end{equation}
where $G(\epsilon) = (2/\pi \Gamma^2)^{1/2} e^{-2(\epsilon/\Gamma)^2}$ is
the probability $\epsilon (t)$ takes value $\epsilon$
(ie., it is the lineshape of the polarisation group).

Adding the the topological phase $\Phi_o = \pi S$ to these calculations 
changes  the flip amplitude to 
$i\Delta_o \exp \{ \pm i\Phi_o  \} dt $; one then sums over all
paths
with an even number of flips {\it and} over all combinations of $\pm$
(clockwise
and counterclockwise) flips; thus (\ref{b1.2}) becomes
\begin{equation}
P_{11}^{(0)}(t) = {1 \over 2} \left\{ 1+ \sum_{s=0}^{\infty}
{ (2i\Delta_o t)^{2s} \over (2s)! } \sum_{n=0}^{2s}
{(2s)! \over (2s-n)!n! } e^{i\Phi_o(2s-2n) } \right\} \;
= \; {1 \over 2} [1+\cos (4\Delta_o \cos \Phi_o)t ]\;,
\label{b1.4}
\end{equation}
with similar obvious modifications to include bias. 
The generalization to include the phase from the bath spins
(topological decoherence) is now obvious for both imaginary and full
complex
$\alpha_k$ (see text, section 3.A).

\subsection{Orthogonality blocking effects}

We consider the situation  described in section \ref{sec:4}.B., where  
the "initial" and "final" fields on the bath spin ${\vec \sigma }_k$ are 
${\vec \gamma}_k^{(1)} $ and 
${\vec \gamma}_k^{(2)} $, related by an angle $\beta_k$, which is 
assumed small, and is defined by  
$\cos 2\beta_k = - {\vec \gamma} _k^{(1)} \cdot {\vec \gamma} _k^{(2)}/
\vert {\vec \gamma} _k^{(1)} \vert \vert {\vec \gamma} _k^{(2)} \vert$.
We choose axes in spin space such that the initial and 
final spin bath wave-functions 
are related by  
\begin{equation}
\mid \{ {\vec \sigma }_k^f \} \rangle = \prod_{k=1}^N {\hat U}_k 
\mid \{ {\vec \sigma }_k^{in} \} \rangle = 
{\hat U}  \mid \{ {\vec \sigma }_k^{in} \} \rangle \;.
\label{b.3}
\end{equation}
where ${\hat U}_k =  e^{ -i\beta_k {\hat \sigma }_k^x }$ (compare eqtn. 
(\ref{b.2}).

In general the initial spin bath state will belong to some 
polarisation group $M_o$ (not necessarily $M_o =0$), 
where the polarisation is defined along some direction defined by the 
central system (for example, in a nanomagnetic problem, one could define 
it as the direction of initial orientation of the nanomagnetic spin). As
explained in the text, during a central system transition 
energy conservation requires 
the polarisation to change from 
$M_o$ to $M_o -2M$ (and back, for further transitions); for "pure"
orthogonality blocking (ie., when no other terms are involved in the 
Hamiltonian), $M_o = M$. 
In what follows we calculate the correlation function $P_{M_o,M} (t)$,
the central spin correlator defined under the restriction that the 
spin bath transitions
are between subspaces defined by $\langle \hat{\cal P} \rangle 
\equiv \langle \sum_{k=1}^N {\hat \sigma }_k^z  \rangle  = M_o$ and 
 $\langle \hat{\cal P} \rangle = M_o -2M$ subspaces, which are 
supposed to be in resonance.
The statistical weight
of states with $M_o > N^{1/2}$ is negligible, so we will
assume that $M_o, M < N$.

We enforce the restriction to a polarisation group $M$ using  
 the projection operator
\begin{equation}
{\hat \Pi}_M 
= \delta (\sum_{k=1}^N {\hat \sigma }_k^z -M) =
\int_0^{2\pi} {d\xi \over 2\pi} e^{ i\xi ( 
\sum_{k=1}^N {\hat \sigma }_k^z -M) } \;.
\label{b.4}
\end{equation}
We can now write down an expression for the 
{\it amplitude} (not the probability!) $A^{11}_{M_o,M}(t)$ for the central spin
${\vec \tau}$ to stay in state $\vert \uparrow \rangle$ during a time $t$: 
\begin{equation}
A^{11}_{M_o,M} (t) =  \left\{
\sum_{n=0}^\infty {(i\Delta_o(\Phi ) t)^{2n} \over (2n)!}
\prod_{i=1}^{2n}  \int {d\xi_i \over 2\pi } 
e^{-i M_o (\xi_{2n}+\xi_{2n-1}+\dots +\xi_1 )} 
e^{2iM(\xi_{2n-1}+\xi_{2n-3}+\dots +\xi_1 )}
{\hat T}_{2n} \right\} 
\mid \{ {\vec \sigma }_k^{in} \} \rangle \;,
\label{b.5}
\end{equation}
where  ${\hat T}_{2n}$ is
\begin{equation}
 {\hat T}_{2n} =\bigg[ e^{i\xi_{2n}\sum_{k=1}^N {\hat \sigma }_k^z} 
{\hat U}^{\dag } 
e^{i\xi_{2n-1}\sum_{k=1}^N {\hat \sigma }_k^z}
{\hat U} \dots {\hat U}^{\dag } 
e^{i\xi_{1} \sum_{k=1}^N {\hat \sigma }_k^z} 
{\hat U} \bigg] \;.
\label{b.6}
\end{equation}
>From (\ref{b.5}) we can now
write the full correlation function  $P^{11}_{M_o,M} (t)$ as
\begin{eqnarray}
&&P_{M_o,M}(t) \equiv  \langle R^{*}_{M_o,M} (t) R_{M_o,M} (t) 
\rangle \nonumber \\ 
 &= &
\sum_{n=0}^\infty \sum_{m=0}^\infty {(i \Delta_o(\Phi ) t)^{2(n+m)} \over (2n)!(2m)!}
\prod_{i=1}^{2n} \prod_{j=1}^{2m} \int {d\xi_i \over 2\pi } 
\int {d\xi_j^{\prime} \over 2\pi }
e^{-i M_o (\sum_i^{2n}\xi_{i} - \sum_j^{2m}\xi_{j}^\prime )} 
e^{2iM( \sum_{i=odd}^{2n-1} \xi_{i} - \sum_{j=odd}^{2m-1} \xi_{j}^\prime )}
\langle 
{\hat T}_{2m}^{\dag} {\hat T}_{2n} \rangle \;.
\label{b.7}
\end{eqnarray} 

We now use the assumption  that the $\beta_k$ are small; 
more precisely we assume that the orthogonality exponent $\kappa $,
defined previously by
$e^{-\kappa } = \prod \cos \beta_k$ (cf. eqtn (\ref{k.1})), 
can be approximated by the perturbative expansion
$\kappa  \approx  { 1\over 2}  \sum \beta_k^2$.
This assumption makes it much easier to calculate the average
in (\ref{b.7}). We consider first the problem with only one environmental spin
${\vec \sigma }_k$, and calculate the average  
$\langle {\hat T}_{2m}^{\dag} {\hat T}_{2n} \rangle_k $ in this case;
since ${\hat T}_{2m}^{\dag} {\hat T}_{2n}$ is a product of operators acting
separately on each $\vec{\sigma}_k$, the average over all spins is also
the product of single spin results.

We only need consider processes with 
$0,\; 1$, or $2$ flips of the environmental spin, i.e., we expand in powers 
of $\beta_k$, and stop at $\beta_k^2$.
Then it is clear that, if the initial state of ${\vec \sigma }_k$ is 
$\mid \uparrow_k \rangle $
\begin{eqnarray}
{\hat T}_{2n}^{(k)} \mid \uparrow_k \rangle & = & 
e^{i\xi_{2n} {\hat \sigma }_k^z} e^{ -i\beta_k {\hat \sigma }_k^x } \dots 
e^{ -i\beta_k {\hat \sigma }_k^x } e^{i\xi_{1} {\hat \sigma }_k^z} e^{ i\beta_k {\hat \sigma }_k^x }
\mid \uparrow_k \rangle \nonumber \\
 & = & e^{i\sum_{i=1}^{2n}\xi_i} \bigg[
(1-n\beta_k^2)\mid \uparrow_k \rangle +i\beta_k \mid \downarrow_k \rangle
\sum_{l=1}^{2n} (-1)^{l+1} e^{ -2i \sum_{i=l}^{2n} \xi_i } \nonumber \\
 & \; & \;\;\;\;\;\;\;\;\;\;\; -\beta_k^2 \mid \uparrow_k \rangle 
\sum_{l^\prime =l+1}^{2n} \sum_{l=1}^{2n-1} (-1)^{l^\prime -l} 
e^{-2i \sum_{i=l}^{l^\prime -1} \xi_i } + O(\beta_k^3) \bigg] \;,
\label{b.10}
\end{eqnarray}
where the first term arises from the sequence $[ 11 \uparrow
\dots 11 ]$, the second from the sequence 
$[ 11 \uparrow \dots \uparrow \downarrow \downarrow
\downarrow \dots \downarrow \downarrow ]$, with a flip when $j=l$; and so on.
In the same way we find 
\begin{eqnarray}
\langle \uparrow_k \mid ({\hat T}_{2m}^{(k)})^{\dag} 
{\hat T}_{2n}^{(k)} \mid \uparrow_k \rangle  &= & 
e^{i(\sum_{i=1}^{2n} \xi_i - \sum_{j=1}^{2m} \xi_j^\prime )}
\bigg[ 1 - \beta_k^2 \big[ (n+m) + \sum_{l^\prime =l+1}^{2n} \sum_{l=1}^{2n-1}
(-1)^{l^\prime -l}  e^{-2i \sum_{i=l}^{l^\prime -1} \xi_i } \nonumber \\ 
&+ &
\sum_{p^\prime =p+1}^{2m} \sum_{p=1}^{2m-1}
(-1)^{p^\prime -p}  e^{2i \sum_{j=p}^{p^\prime -1} \xi_j^\prime } \nonumber \\
 &- & 
\sum_{p =1}^{2m} \sum_{l=1}^{2n}
(-1)^{l+p}  e^{-2i (\sum_{i=l}^{2n} \xi_i - \sum_{j=1}^{p-1} \xi_j^\prime )}
\big] \bigg] \;,
\label{b.11}
\end{eqnarray}
to order $\beta_k^2$. The sequence 
$\langle \downarrow_k \mid ({\hat T}_{2m}^{(k)})^{\dag} 
{\hat T}_{2n}^{(k)} \mid \downarrow_k \rangle$
will have a similar expression, but with reversed signs coming from the 
$e^{i\xi_j {\hat \sigma }_k^z}$ factors.

We now observe that the state with polarisation $M_o $ consists of 
$(N+M_o)/2$ spins up and $(N-M_o)/2$ spins down.
Consequently, for each ${\vec \sigma }_k$, we add $\uparrow $ or $\downarrow$ 
averages like (\ref{b.11}), and then take  the product
\begin{equation}
\langle {\hat T}_{2m}^{\dag} {\hat T}_{2n} \rangle =\prod_{k=1}^{N_{\uparrow}}
\langle ({\hat T}_{2m}^{(k)})^{\dag} {\hat T}_{2n}^{(k)} \rangle\;
\prod_{k^\prime =1}^{N_{\downarrow}}
\langle ({\hat T}_{2m}^{(k)})^{\dag} {\hat T}_{2n}^{(k)} \rangle
\label{b.12}
\end{equation}
Substituting (\ref{b.11}) into this expression we get 
\begin{equation}
\langle {\hat T}_{2m}^{\dag} {\hat T}_{2n} \rangle = 
e^{i M_o (\sum_i^{2n}\xi_{i} - \sum_j^{2m}\xi_{j}^\prime )}
\exp \big\{ -K^{eff}_{nm} (\xi_i ,\xi_j^\prime , M_o ) \big\}
\;,
\label{b.12b}
\end{equation}
where the "effective action" $K^{eff}_{nm} (\xi_i ,\xi_j, M_o )$
has two contributions $K^{eff} = K_1+K_2$:
\begin{eqnarray}
K_1= 2\kappa (1-{M_o \over N })
 \bigg\{ (n+m) & + & \sum_{l^\prime > l}(-1)^{l^\prime -l}
\cos [2\sum_{i=l}^{l^\prime -1}\xi_i ] + \sum_{p^\prime > p}(-1)^{p^\prime -p}
\cos [2\sum_{j=p}^{p^\prime -1}\xi_j^\prime ] \nonumber \\
 &-& \sum_{p =1}^{2m} \sum_{l=1}^{2n} (-1)^{l+p}  
\cos [2\sum_{i=l}^{2n} \xi_i - \sum_{j=1}^{p-1} \xi_j^\prime ] \bigg\}
\;,
\label{b.12c}
\end{eqnarray}
\begin{eqnarray}
K_2= 2\kappa {M_o \over N }
 \bigg\{ & & \sum_{l^\prime > l}(-1)^{l^\prime -l}
\exp [-2i \sum_{i=l}^{l^\prime -1}\xi_i ] + 
\sum_{p^\prime > p}(-1)^{p^\prime -p}
\exp [2i \sum_{j=p}^{p^\prime -1}\xi_j^\prime ] \nonumber \\
 &-& \sum_{p =1}^{2m} \sum_{l=1}^{2n} (-1)^{l+p}  
\exp [2i \sum_{j=1}^{p-1} \xi_j^\prime -2i \sum_{i=l}^{2n} \xi_i ] \bigg\}
\;,
\label{b.12d}
\end{eqnarray}
We recall now that $M_o \le N^{1/2} \ll N$, which allows us to neglect
the contribution due to $K_2$ and  drop the correction  $\sim M_o / N$ to 
the coefficient $\kappa $ in $K_1$.
 Notice also that the phase factor in front of 
$\exp \{ -K^{eff} \} $ in (\ref{b.12}) cancels exactly the phase 
proportional to $M_o$ in the formula (\ref{b.7}) for $P_{M_o,M}(t)$. 
Thus, quite surprisingly, we find the correlation function to 
be independent of $M_o$ in this limit:
\begin{equation}
P_{M}(t)  =\sum_{n=0}^\infty \sum_{m=0}^\infty {(i \Delta_o(\Phi ) t)^{2(n+m)} \over (2n)!(2m)!}
\prod_{i=1}^{2n} \prod_{j=1}^{2m} \int {d\xi_i \over 2\pi } 
\int {d\xi_j^{\prime} \over 2\pi } 
\exp \big\{2iM(\xi_{2n-1}+\xi_{2n-3}+\dots +\xi_1 )
 -K^{eff}_{nm} (\xi_i ,\xi_j^\prime ) \big\}
\;,
\label{b.13}
\end{equation}

We can render this expression more useful by changing variables; first 
we consider the whole sequence $\xi_\alpha = (\xi_1, \dots , \xi_{2n}, 
-\xi_1^\prime, \dots , -\xi_{2m}^\prime)$ together, and then define new 
angular variables
\begin{equation}
\chi_\alpha = \sum_{\alpha^\prime =
\alpha }^{2(n+m)} 2\xi_{\alpha^\prime} +\pi \alpha \;,
\label{b.15}
\end{equation}
so that now
\begin{eqnarray}
P_{M}(t)  = & &\sum_{n=0}^\infty \sum_{m=0}^\infty {(i\Delta_o(\Phi ) t)^{2(n+m)} \over (2n)!(2m)!}
\left( \prod_{\alpha =1}^{2(n+m)} \int {d\chi_\alpha \over 2\pi } \right) 
\nonumber \\ 
& \times & 
\exp \bigg\{iM \sum_{\alpha} (-1)^{\alpha +1}\chi_{\alpha} 
-2\kappa \big[ (n+m) +\sum_{\alpha^\prime > \alpha } 
\cos (\chi_\alpha - \chi_{\alpha^\prime} ) \big]   \bigg\} \;.
\label{b.16}
\end{eqnarray}
Thus we have mapped our problem onto the partition function of a rather
peculiar system of spins, interacting via infinite range forces, with
interaction strength $2\kappa$.

To deal with this partition function , we define "pseudo-spins"
${\vec s}_\alpha= (\cos \chi_\alpha, \sin \chi_\alpha )$ and 
${\vec {\cal S}}$, such that 
\begin{equation}
{\vec {\cal S}} = \sum_{\alpha =1}^{2(n+m)} {\vec s}_\alpha
\;, \;\;\;\;\;\;\;\;\;\;\;\;
\sum_{\alpha^\prime , \alpha } 
\cos (\chi_\alpha - \chi_{\alpha^\prime} ) = {\vec {\cal S}}^2 \;,
\label{b.18}
\end{equation}
We can think of ${\vec s}_\alpha $ as rotating in our fictitious angular space
defined by the projection operator (\ref{b.4}). Now consider the term 
$G({\vec {\cal S}} )$ in (\ref{b.16}) defined by 
\begin{eqnarray}
G({\vec {\cal S}} ) & = & \left(
\prod_{\alpha =1}^{2(n+m)} \int {d\chi_\alpha \over 2\pi }  
e^{iM(-1)^{\alpha +1}\chi_{\alpha} } \right) \: 
\exp \big\{ -\kappa \sum_{\alpha^\prime , \alpha } 
\cos (\chi_\alpha - \chi_{\alpha^\prime} ) \big\} \nonumber \\
& =&  \left(
\prod_{\alpha =1}^{2(n+m)} \int {d\chi_\alpha \over 2\pi } 
e^{iM(-1)^{\alpha +1}\chi_{\alpha} } \right)  \: 
e^{ -\kappa {\vec {\cal S}}^2} \;.
\label{b.19}
\end{eqnarray}
This is easily calculated, viz.,
\begin{eqnarray}
G({\vec {\cal S}} ) &=& \int d{\vec {\cal S}} e^{ -\kappa {\vec {\cal S}}^2}
\prod_{\alpha =1}^{2(n+m)} \int {d\chi_\alpha \over 2\pi }
e^{iM(-1)^{\alpha +1}\chi_{\alpha} }
\delta ({\vec {\cal S}} - \sum_\alpha {\vec s}_\alpha ) \nonumber \\
&=& \int {d{\vec  z } \over 2\pi } 
\int d{\vec {\cal S}} e^{ -\kappa {\vec {\cal S}}^2 +i{\vec  z } \cdot 
{\vec {\cal S}}}\: \left( \int_0^{2\pi } {d\chi_\alpha \over 2\pi }
e^{-i{\vec  z} {\vec s}_\alpha +i M \chi_{\alpha }} \right) ^{2(n+m)} 
\nonumber \\
 &=& {1 \over 2\kappa } \int d z  z e^{- z^2/4\kappa }
J_M^{2(n+m)}( z ) \;,
\label{b.20}
\end{eqnarray}
where $J_M(\lambda )$ is the $M$th-order Bessel function. Using
\begin{equation}
\sum_{l=0}^{\infty} {\delta_{2(n+m),2s} \over (2m)! (2n)! } =
{  \delta_{s,0} +2^{2s} \over 2(2s)! }  \;,
\label{b.21}
\end{equation}
to reorganize the sum over $n$ and $m$ in (\ref{b.16}) and changing 
the integration variable
$z \to 2x \sqrt{\kappa }$, we then find
\begin{eqnarray}
P_{M}(t)&=& 2\int_0^\infty dx x\:e^{-x^2} \; 
{1 \over 2}  \left( 1+ \sum_{s=0}^{\infty} {
[2it \Delta_o(\Phi ) J_M(2x\sqrt{\kappa })]^{2s} \over (2s)! } \right) 
\nonumber \\
& = &  P_{M}(t)= \int_0^\infty dx x\:e^{-x^2} \big( 1+ \cos 
[2 \Delta_o(\Phi ) J_M(2x\sqrt{\kappa } )t]  \big)  
 \equiv 2\int_0^\infty dx x\:e^{-x^2} P^{(0)}_{11}(t,\Delta_M(x)) \;.
\label{b.21b}
\end{eqnarray}
\begin{equation}
\Delta_M(x) =  \Delta_o(\Phi ) J_M(2x\sqrt{\kappa } )
\;.
\label{b.21c}
\end{equation}

Here we come to the crucial point in our derivation. Eq.(\ref{b.21b}) gives 
the final answer as a {\it superposition  of 
non-interacting} correlation functions for  
effective tunneling rates $\Delta_M(x)$  with the proper weighting.
For $M=0$ this is the form quoted in Eq.(\ref{4.21o}) of the text.

It is worth noting that non-zero $M$ enters this calculation as the overall
phase factor which we can follow from (\ref{b.7}) up to (\ref{b.20}), where 
we finally integrate over $\{ \chi_\alpha \} $ to produce the Bessel function
of order $M$. This observation allows one to generalise any calculation done
for $M=0$ to finite $M$ by simply replacing $J_0 \to J_M$ in the final answer -
the prescription which we use in other Appendices.

\section{ Derivations for the generic case}

We outline here the derivations for section 4, in which
topological decoherence, degeneracy blocking, and orthogonality blocking are 
all simultaneously incorporated (the average over bias fluctuations being 
essentially trivial- see Appendix A.1).
We have demonstrated in section 3 and 
Appendix A how each different term in the effective Hamiltonian (\ref{1.24})
influence the central spin dynamics. From these limiting cases we learned 
that static (or diagonal) terms in the Hamiltonian can be partly 
absorbed into a redefinition of the transition amplitude
between states with equal initial and final energies. If we now deal with 
the full central spin Hamiltonian, we can still write 
the instanton expansion in central spin transitions in the form 
(see (\ref{b1.10})): 
\begin{equation}
P_{M}(t) =\int_{-i\infty }^{i\infty} dp_1 dp_2 { e^{(p_1+p_2)t}  \over (p_1 -i
\epsilon ) ( p_2 -i\epsilon )} \sum_{n=0}^{\infty}\sum_{m=0}^{\infty} 
\left( { (-i\Delta_o )^2 \over p_1^2+\epsilon^2 } \right)^{n}
\left( { (-i\Delta_o )^2 \over p_2^2+\epsilon^2 } \right)^{m}
B_{nm}(M) \;,
\label{y.1}
\end{equation}
\begin{equation}
B_{nm}(M) = \sum_{\{ g_l=\pm \} } e^{i\Phi \sum_{l=1}^{2(n+m)}g_l }
\langle {\hat T}_{2m}^{\dag}(M, g_l) {\hat T}_{2n}(M, g_l) \rangle \;,
\label{y.2}
\end{equation}
where the sum over $\{ g_l=\pm \} $ with $1 \le l \le 2(n+m)$ describes all
possible clockwise and anticlockwise transitions, and the operator product 
is defined as 
\begin{equation}
{\hat T}_{2n}(M, g_l) = \hat{U}_{M}(g_1) \hat{U}_{M}^{\dag}(g_2)
\hat{U}_{M}(g_3) \dots
\hat{U}_{M}^{\dag}(g_{2n}) \;,
\label{y.3}
\end{equation}
\begin{equation}
\hat{U}_{M}(g) = \hat{\Pi}_M  e^{ig\sum_k \alpha_k \vec{n}_k \cdot 
\hat{\vec{\sigma}}_k} 
e^{-i\beta_k \hat{\sigma}_k^x } \hat{\Pi}_{-M}  \;.
\label{y.4}
\end{equation}
Here $\hat{\Pi}_M$ as before projects on the polarisation state 
$\Delta N =M$, and 
$\beta_k$ describes the mismatch between the initial and final nuclear 
spin states.
If all the couplings were equal ($\omega_k=\omega_o \gg \Delta_o$) then 
the above set
of equations would be the complete solution of the $M$ polarisation 
group dynamics.
One may then further average over different grains in the ensemble by
summing over different polarisation groups with proper weigthing. 
If there is a small spread in the nuclear hyperfine couplings, it will
produce an internal bias field acting on the grain, as described in the 
text (section
\ref{sec:4}.C. The final answer for the ensemble of grains is obtained then by
averaging (\ref{y.1}) over the bias field. This bias field is due to
{\it all} environemntal spins interacting with $\vec{S}$, and hardly changes
when a few nuclei (of order $\sim \lambda $) flip with $\vec{S}$. For this
reason there is no back influence of the induced nuclear spin flips 
on the bias field, at least during the time scale set by the damping of 
coherent oscillations (when many environemntal spins are flipped the coherence
is obviously already lost).

The crucial observation  is that if the sum over the clockwise and 
anticlockwise 
trajectories and the average of the operator product in (\ref{y.2}) 
can be presented
as some weighted average and/or sum, of form
\begin{equation}
 B_{nm}(M)=\int dx_1dx_2\dots dx_a \sum_{k_1k_2\dots k_b} 
Z_M(x_1,\dots ,x_a; k_1,\dots ,k_b) R_M^{2(n+m)} 
(x_1,\dots ,x_a; k_1,\dots ,k_b)   \;,
\label{y.5}
\end{equation}
with fixed integer values $a$ and $b$, then the problem may be considered as
solved because the instanton summation then reduces to that of a coherent
(non-interacting) dynamics with the renormalized tunneling amplitude
\begin{equation}
\Delta_o \longrightarrow \Delta_M(x_1,\dots ,x_a; k_1,\dots ,k_b) =
\Delta_o  R_M (x_1,\dots ,x_a; k_1,\dots ,k_b)   \;.
\label{y.6}
\end{equation}
and the final answer acquires a form
\begin{equation}
P_M (t, \epsilon ) =  \int dx_1dx_2\dots dx_a \sum_{k_1k_2\dots k_b} 
Z_M(x_1,\dots ,x_a; k_1,\dots ,k_b) P_{11 }^{TLS}(t,\epsilon , 
\Delta_M(x_1,\dots ,x_a; k_1,\dots ,k_b) \;,
\label{y.7}
\end{equation}
where $P_{11 }^{TLS}(t,\epsilon , \Delta_M )$ is described by Eq.~(\ref{b1.10}).

We have already seen that Eq.~(\ref{y.5}) is indeed valid for the cases
of pure topological decoherence and pure orthogonality blocking-
we now prove that it also holds when we combine
the effects of topological decoherence with the projection on a 
given polarisation
state. Here we evaluate the $M=0$ contribution;
the result for $P_{M}(t)$ then follows from the generalisation 
explained at the end of
the previous Appendix.

Introducing as before the spectral representation for the projection
operator [see Eq.~(\ref{b.4})] we write 
\begin{equation} 
B_{nm}(M) =  \sum_{\{ g_l=\pm \} } e^{i\Phi \sum_{l=1}^{2(n+m)}g_l }
\prod_{\rho=1}^{2(n+m)} \int {d\xi_\rho \over 2\pi } \exp \left\{ -
K^{eff}_{nm}( \{ g_l \} , \{ \xi_\rho \})  \right\}\;.
\label{y.8}
\end{equation}
\begin{equation} 
\exp \left\{ - K^{eff}_{nm}( \{ g_l \} , \{ \xi_\rho \})  \right\} =
\langle e^{i\xi_1 \hat{{\cal P}}} 
e^{ig_1\sum_k \alpha_k \vec{n}_k \cdot \hat{\vec{\sigma}}_k} 
e^{i\xi_2 \hat{{\cal P}}} 
e^{ig_2\sum_k \alpha_k \vec{n}_k \cdot \hat{\vec{\sigma}}_k}  \dots
e^{i\xi_{2(n+m)} \hat{{\cal P}}} e^{ig_{2(n+m)} \sum_k \alpha_k \vec{n}_k 
 \cdot \hat{\vec{\sigma}}_k} \rangle \:.
\label{y.9}
\end{equation}

With the usual
assumption that the individual $\alpha_k$ are small (but not necessarily
$\lambda $), the "effective action" $K^{eff}_{nm}$ 
can be written as (compare Eq.(\ref{b.16}))
\begin{equation} 
K^{eff}_{nm}( \{ \xi_\rho \}) =  \lambda^\prime
 \sum_{\rho^\prime , \rho }^{2(n+m)} g_\rho g_{\rho^\prime}+ 
\lambda  \sum_{\rho^\prime ,\rho }^{2(n+m)}
\cos (\chi_\rho - \chi_{\rho^\prime} ) g_\rho g_{\rho^\prime} \;,
\label{c.5}
\end{equation} 
which generalizes from orthogonality blocking; 
the $\chi_\rho $ are defined as in 
 (\ref{b.15}), and 
\begin{equation}
\lambda = {1 \over 2} \sum_{k=1}^N \alpha_k^2 (1-(n_k^z)^2)\;;\;\;\;\;\;\;\;\; 
\lambda^\prime = {1 \over 2} \sum_{k=1}^N \alpha_k^2(n_k^z)^2  \;.  
\label{c.6}
\end{equation}
as before. We use the same trick of introducing ``pseudo-vectors" 
${\vec s}_\rho =(\cos \chi_\rho ,\sin \chi_\rho )$, and 
${\vec {\cal S}}^2 =  
\left( \sum_{\rho =1}^{2(n+m)} g_\rho {\vec s}_\rho \right) ^2
= \sum_{\rho^\prime , \rho } g_\rho g_{\rho^\prime}
 \cos (\chi_\rho - \chi_{\rho^\prime} ) $, to decouple integrals over  
the new variables $\chi_\rho =\chi_\rho +\pi g_\rho /2 $. After some lengthy, 
but straightforward
algebra we get $P_{11} (t)$ in the form
\begin{equation}
P_{0} (t) = 2\int dx x e^{-x^2} \int {d\varphi \over 2\pi } 
\sum_{m=-\infty}^{\infty} F_{\lambda^\prime}(m)  e^{i2m(\Phi -\varphi )} 
P_{11}^{(0)} (t, \Delta_o (\varphi ,x ))
\label{c.11}
\end{equation}
where 
$\Delta_o (\varphi ,x ) = 2  \Delta_o \cos \varphi J_0(2x\sqrt{\lambda })$
as before.


The case of complex $\alpha_k$ (even assuming $\omega_k^{\perp} =0$ 
in the effective Hamiltonian)
is more subtle technically, but goes through in exactly the same way. 
Here we just outline the key
steps; a more detailed derivation may be found in \cite{PRB}.
The effective action now has the form 
\begin{eqnarray} 
K^{eff}_{nm} &= & \sum_{\rho ,\rho^\prime =1}^{2(n+m)} 
g_\rho g_{\rho^\prime}  \bigg\{ 
\big[ \lambda^\prime -\eta^\prime (-1)^{\rho +\rho^\prime}  
-i\gamma^\prime ((-1)^\rho +(-1)^{\rho^\prime} ) \big] \nonumber \\
& + & \cos (\chi_\rho - \chi_{\rho^\prime }) \big[
(\lambda - \lambda^\prime ) -(-1)^{\rho +\rho^\prime} (\eta - \eta^\prime ) 
-i(\gamma - \gamma^\prime )((-1)^\rho +(-1)^{\rho^\prime} ) \big] \bigg\} \;,
\label{c.25}
\end{eqnarray}
where the constants are  defined by:
\begin{equation}
\lambda ={1 \over 2} \sum_{k=1}^N \alpha_k^2 \;; \;\;\;\; 
\lambda^\prime ={1 \over 2} \sum_{k=1}^N \alpha_k^2(n_k^z)^2 \;; 
\label{c.26}
\end{equation}
\begin{equation}
\eta ={1 \over 2} \sum_{k=1}^N \xi_k^2 \;; \;\;\;\; 
\eta^\prime ={1 \over 2} \sum_{k=1}^N \xi_k^2(v_k^z)^2 \;;
\label{c.27}
\end{equation}
\begin{equation}
\gamma ={1 \over 2} \sum_{k=1}^N \alpha_k\xi_k {\vec n}_k \cdot 
{\vec v}_k \;; \;\;\;\; 
\gamma^\prime ={1 \over 2} \sum_{k=1}^N \alpha_k\xi_k n_k^z v_k^z \;; 
\label{c.28}
\end{equation}

As before we change variables according to $\chi_\rho =\chi_\rho +\pi $ 
when $g_\rho =-1$, to introduce odd and even spin fields
\begin{equation} 
{\vec {\cal S}}_o =\sum_{\rho =odd}^{2(n+m)-1} {\vec s}(\chi_\rho ) 
\;; \;\;\;\;\;
{\vec {\cal S}}_e =\sum_{\rho =even}^{2(n+m)} {\vec s}(\chi_\rho ) \;.
\label{c.29}
\end{equation}
which are used to decouple integrations over $\chi_\rho $ with the final 
goal to get the answer
in the form of Eq.~(\ref{y.5}). This indeed can be done, and the final 
answer reads
\begin{eqnarray}
P_{11} (t) = \int {d\varphi_1  \over 2\pi }\int {d\varphi_2  \over 2\pi } & &
\sum_{m_1 =-\infty }^{\infty}\sum_{m_2 =-\infty }^{\infty} 
\int dx_1 \int dx_2  \nonumber \\
& \times &
{\cal Z}(\varphi_1,\varphi_2,x_1,x_2,m_1,m_2) P_{11}^{(0)} 
[t, {\tilde \Delta}_o (x_1,x_2,\varphi_1,\varphi_2)] \;,
\label{5.29}
\end{eqnarray}
and (\ref{5.29}) has an obvious generalisation to include the bias
integration $\int d \epsilon$. The weight is given by  
\begin{eqnarray}
{\cal Z} &= & e^{2i[ m_1(\Phi -\varphi_1) -m_2
\varphi_2 +4m_1m_2\gamma^\prime ]} 
e^{4(\eta^\prime m_2^2 - \lambda^\prime m_1^2)}
\nonumber \\
 &\times & {x_1x_2 \over 8(ab-c^2)} 
I_0\left( {(a+b) x_1x_2 \over 8(ab-c^2) } \right) 
\exp \left\{ { (a-b+2ic)x_1^2 + (a-b-2ic)x_2^2 \over 16(ab-c^2) } \right\} \;,
\label{5.34}
\end{eqnarray}
and the  renormalized tunneling splitting equals 
\begin{equation}
{\tilde \Delta}_o^2 (x_1, x_2, \varphi_1, \varphi_2 ) = 
4{\tilde \Delta}_o^2 \cos (\varphi_1 + \varphi_2)
\cos (\varphi_1 - \varphi_2 ) J_0(x_1)J_0(x_2) \;,
\label{c.34}
\end{equation}
where 
\begin{equation}
a=\lambda - \lambda^\prime \;;\;\;\;\;
b=\eta - \eta^\prime \;;\;\;\;\;\;
c=\gamma - \gamma^\prime \;.
\label{c.31}
\end{equation}



\begin{center}
{\bf FIGURE CAPTIONS }
\end{center}

{\bf Figure 1 } The flow of a class of effective Hamiltonians 
describing a central system coupled to a background environment,
in coupling constant space, 
as the UV cutoff in the joint Hilbert space is reduced from $E_c$
to $\Omega_o$.
Here we show flow to a fixed point FP, in a simplified 2-dimensional 
space of couplings $\alpha_1, \alpha_2$, but one may also have fixed lines or 
more complex topologies.

{\bf Figure 2 }  A typical path for a 2-level central system 
(solid line) coupled
to environmental modes (wavy lines) as a
function of time, showing the couplings which exist in both the 
spin-boson and central spin models. 
We show both diagonal couplings D to $\tau_z$ 
and non-diagonal couplings ND to $\tau_{\pm}$ (in the central spin model
these are strong enough to lead to multiple excitation of environmental modes).

{\bf Figure 3 } Definition of the longitudinal and transverse parts
of the diagonal coupling to a bath spin 
in the Central Spin Hamiltonian, in terms of the 
initial and final fields $\vec{\gamma}^{(1)}$ and 
$\vec{\gamma}^{(2)}$ acting on this spin- this also defines the angle 
$\beta$, and the mutually perpendicular unit vectors $\hat{l}$ and 
$\hat{m}$  (see text).

{\bf Figure 4} Classifying the 
states of the Central spin Hamiltonian. 
Each level of ${\vec \tau}$
is associated with a $2^N$-fold multiplet of bath states (Fig 6(a)). 
These are classified 
into polarisation groups $\{ M \}$ (where $M$ is 
the total polarisation along ${\hat z}$), separated by energy $\omega_o$ and 
with width $\tilde {\Gamma}_M$; Fig 6(b) shows the density of states $G_M(\xi)$
of the separate groups, and Fig. 6(c) their sum $W(\xi)$.
We show $W(\xi)$ for 2 different values of the 
parameter $\mu = \tilde {\Gamma}_M/\omega_o$; 
in realistic cases $\mu \gg 1$ (ie., the polarisation 
groups strongly overlap), and $W(\xi)$ is Gaussian. Longitudinal transitions 
between 2 different polarisation groups $M_1$ and $M_2$ go at a rate 
$T_1^{-1}$; transitions within a polarisation group at a rate $T_2^{-1}$. 

{\bf Figure 5} Example of the application of the Central Spin model to 
a magnetic macromolecule (the $Fe$-8 molecule, further descibed in section 5).
In (a) we show the effective tunneling matrix element 
$\vert \tilde \Delta_{eff} \vert = 
\vert \Delta \cos (\pi S + i\beta_o {\bf n}_o.{\bf H}_o \vert)$, for this 
easy axis/easy plane nanomagnet in the presence of a 
field ${\bf H}_o = \hat{x} H_x$ 
in the $x$-direction (transverse to the easy axis), assuming an 
angle $\varphi$ between $\hat{x}$ and the magnetic "hard axis" 
(perpendicular to the easy plane).
Aharonov-Bohm oscillations appear when
$\varphi$ is small, so that the action of the 2 relevant paths on the
spin sphere have similar {\it magnitudes}, but almost opposite
phase. For larger $\varphi$, one path dominates over the other and
oscillations are suppressed. In (b) we show a histogram of the 
$\omega_k^{\parallel}$ for this system- the main figure shows the protons and
the lower inset the $N$ and $O$ contributions. The upper inset in (b) shows 
the variation of $E_o$ and $\xi_o$ with $H_x$ (the parameter $\xi_o$ is 
discussed in sections 4 and 5). These figures are adapted from  Ref. [69].  

{\bf Figure 6 } Behaviour of $P_{11}(t)$ in the case of pure 
topological decoherence.
We show $P_{11} (t)-1/2$ for intermediate  coupling, for which 
$P_{11} (t)$ takes the "universal form" discussed in the text. 

{\bf Figure 7 } The effect of relaxation on 
a statistical ensemble of central spins, each interacting with a 
spin bath. In (a) we assume that 
$\lambda = 0, \;\kappa=5$ and $N=1000$, and show the normalised 
time dependence of 3 different 
contributions $P_M(t)$ to the total relaxation function $P_{11}(t)$; they sum 
to give a roughly logarithmic time dependence for the total function
$P_{11}(t)$. The small $M$ contributions relax quickly
(up to $M \sim \kappa$), so the effect on an initial ensemble distributed
over bias $\xi$, at short times, is to  
dig a hole around zero bias, of width $\sim \kappa \omega_o$. 
In (b) we show the spectral absorption function 
$\chi^{\prime \prime}(\omega)$ for $\kappa = 2$, dividing this into the
$M=0$ contribution and the contributions from higher $M$ groups (which
relax more slowly and thus peak at lower $\omega$).

{\bf Figure 8 } The spectral absorption function 
$\chi^{\prime \prime}_{M=0} (\omega )$ for several values of $\kappa$, 
for an ensemble of central spins in the 
$M=0$ polarisation group, in the 
case where orthogonality blocking dominates, and degeneracy blocking effects
(ie., a bias average) are also incorporated. 
Contributions from higher polarisation groups $M \neq 0$ are not shown; they  
are spread over a range $\sim M \omega_o$, up to 
$\sim \kappa \omega_o$. Contributions from groups with $M > \kappa$ are 
negligible. 

{\bf Figure 9 }  Graphs of $\chi^{\prime \prime}_{M=0} (\omega )$  
for "projected 
topological decoherence" (ie., including a bias average over an ensemble in 
which topological decoherence dominates), for several different values of
the parameter $(\lambda - \lambda^{\prime})$. Contributions from
higher polarisation groups, which are spread over an energy range $\sim 
\lambda \omega_o$, are not shown. 


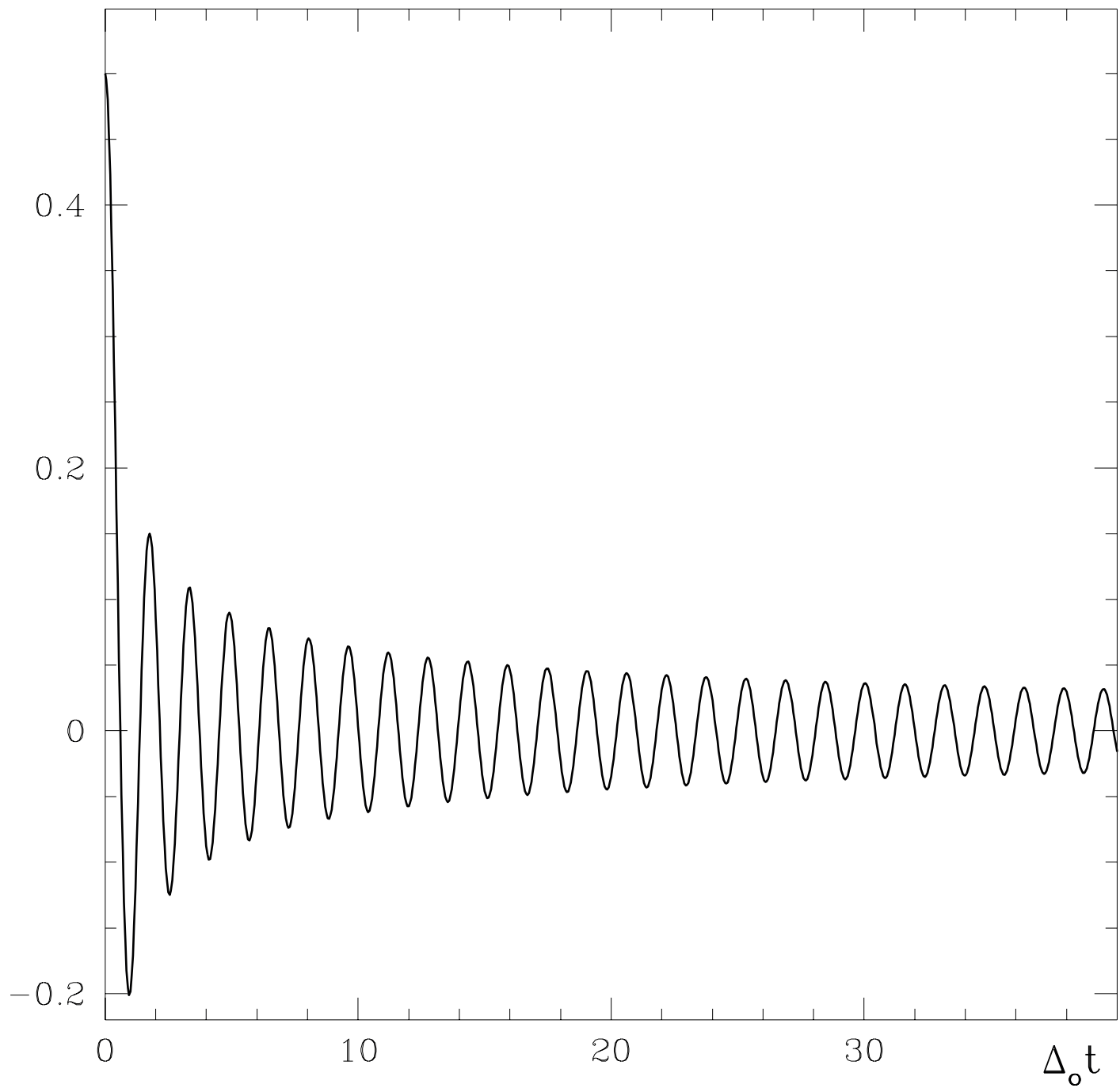
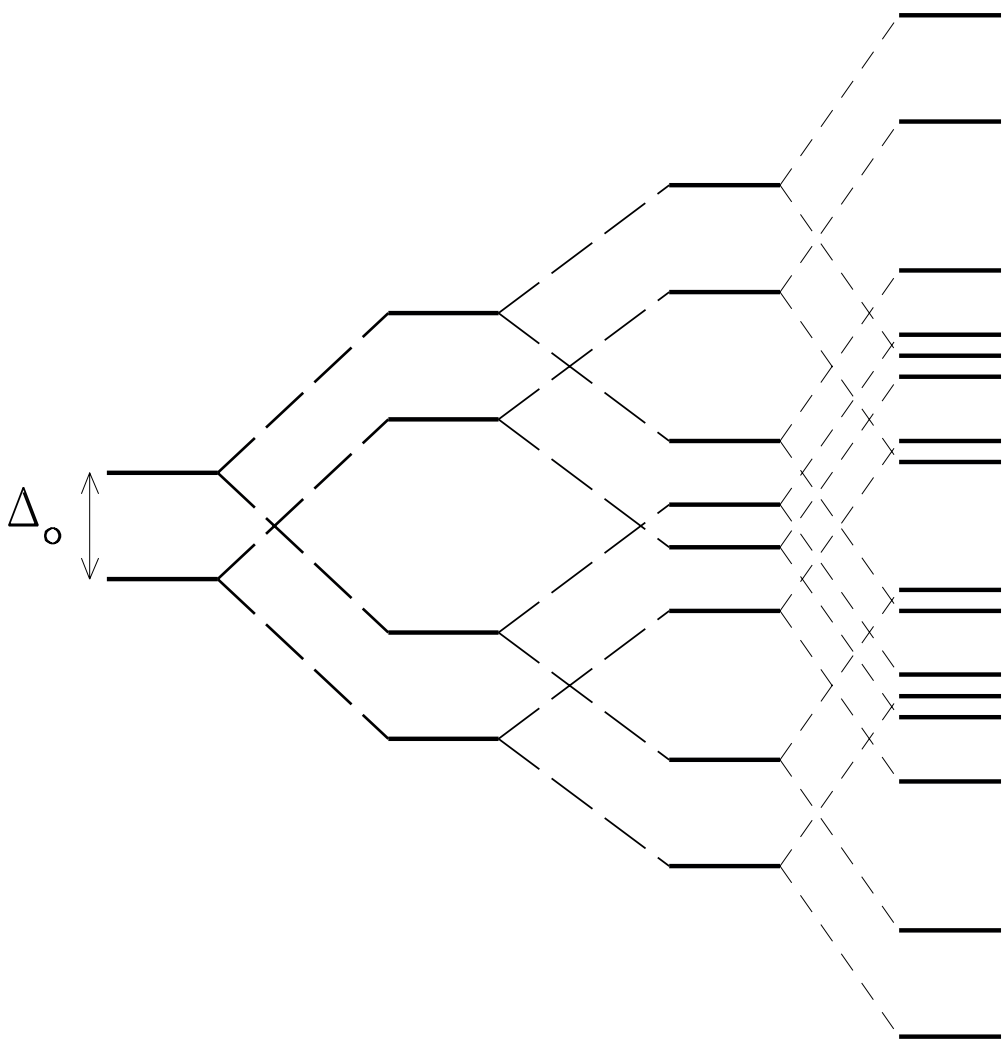
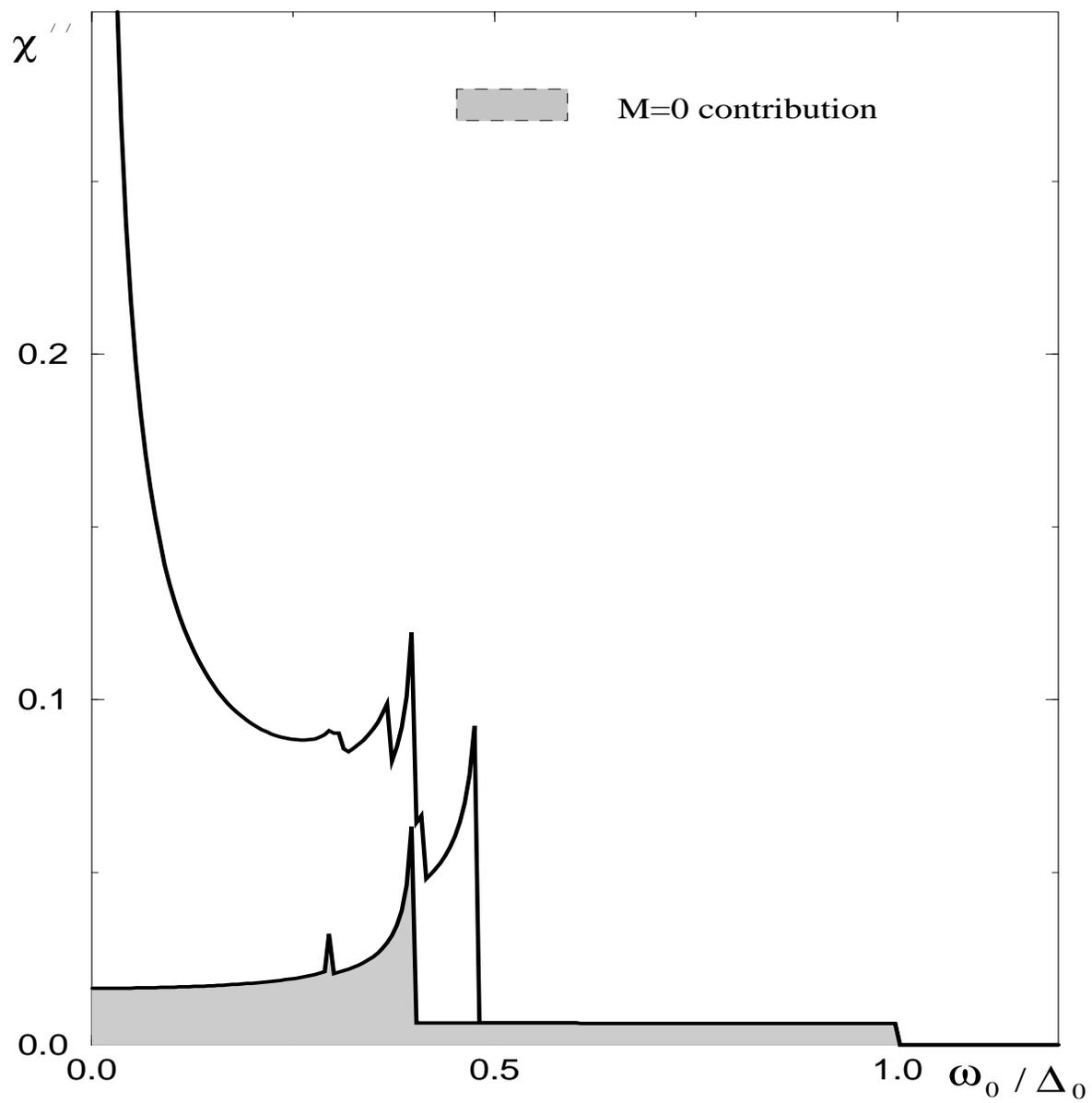
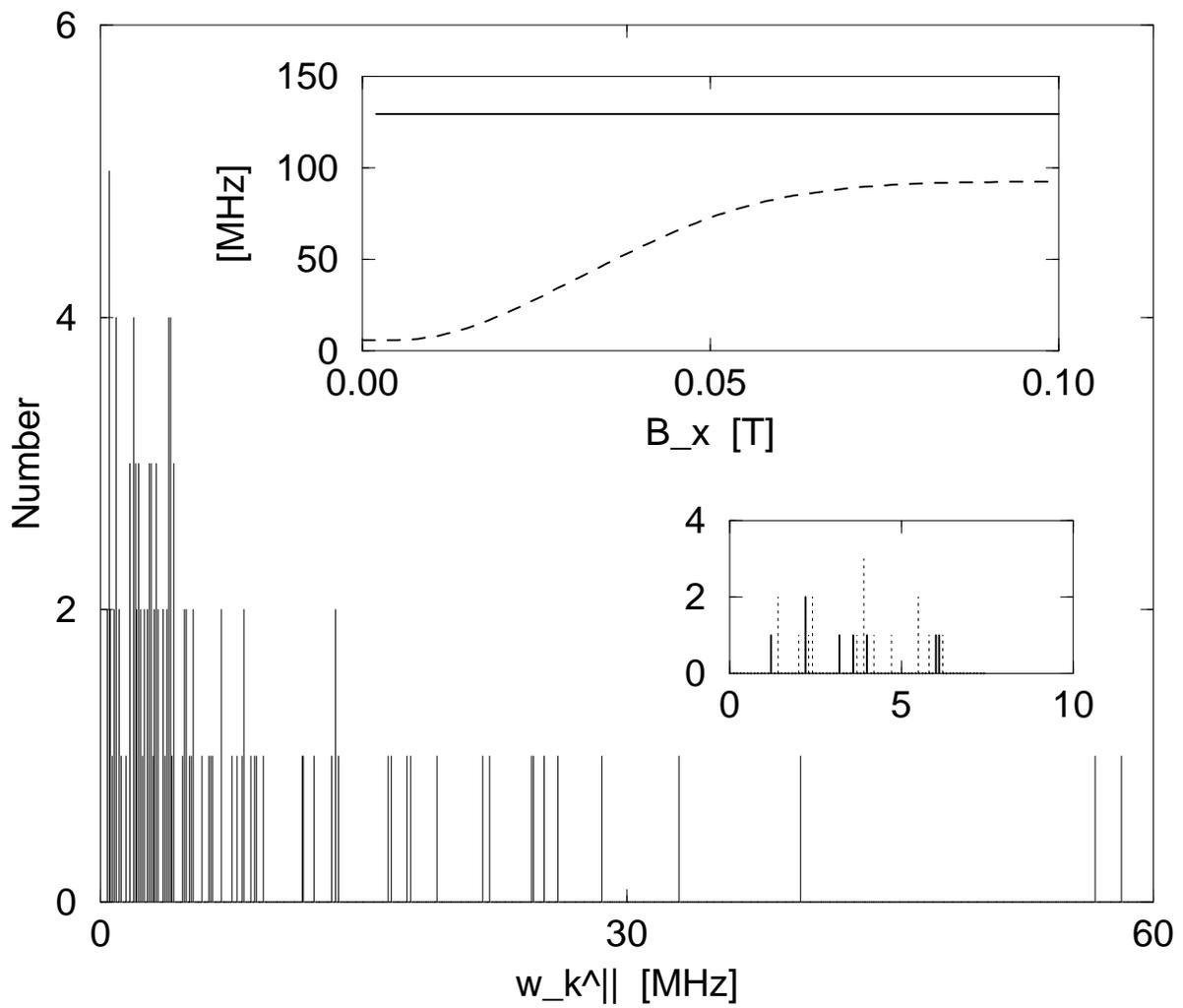
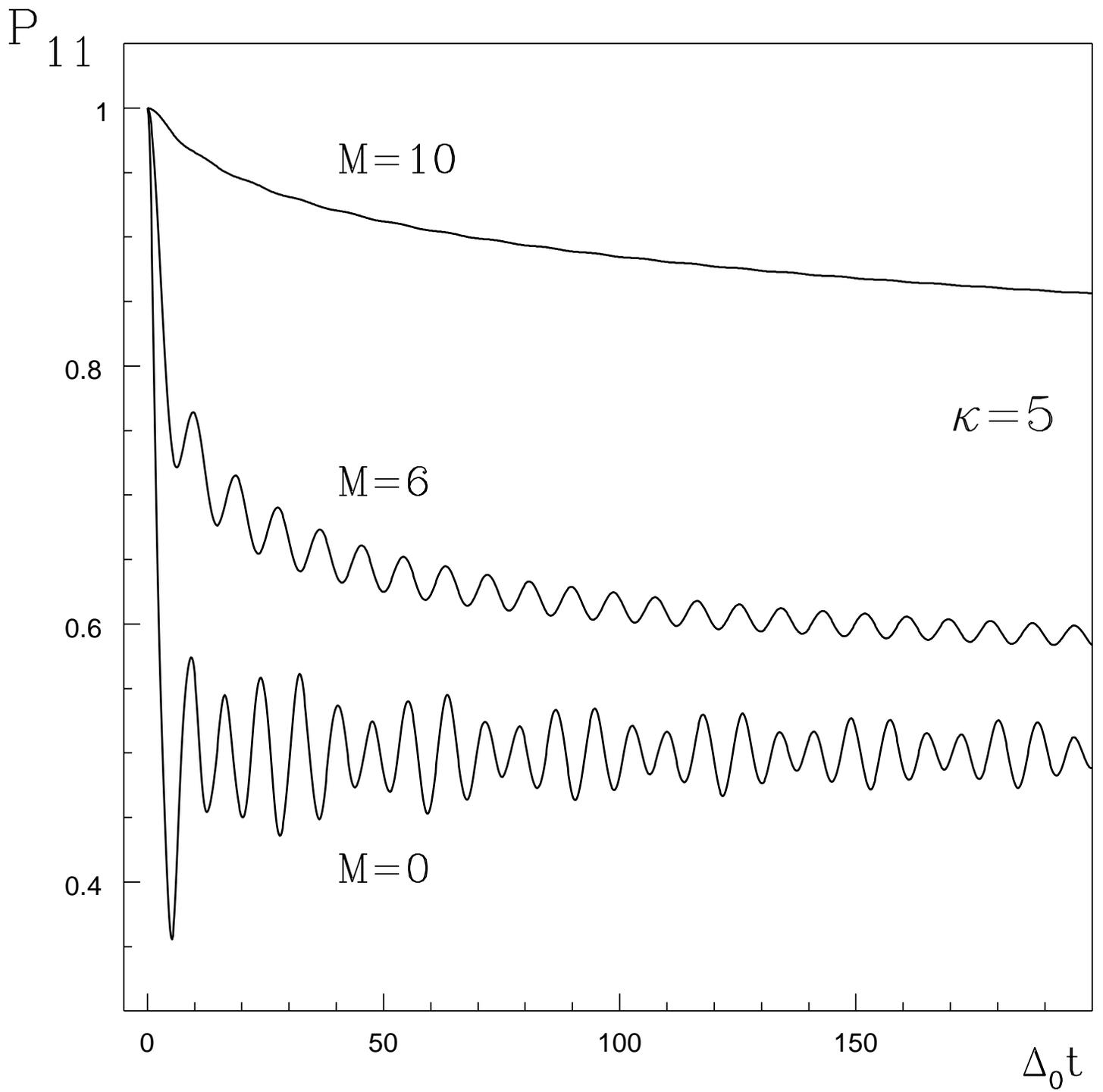
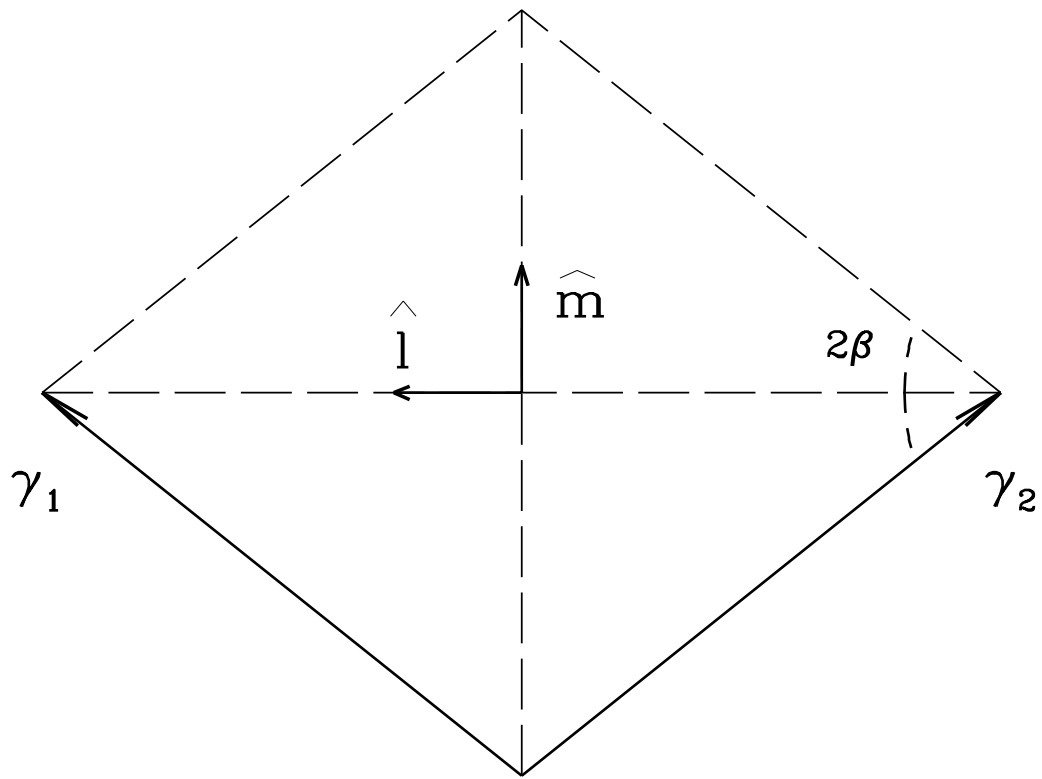
\end{document}